\documentclass[seceq]{ptptex}
\usepackage[pdftex]{graphicx}

\title{
Antisymmetrized molecular dynamics and its applications to cluster phenomena
}

\author{
Yoshiko \textsc{Kanada-En'yo}$^1$, Masaaki \textsc{Kimura}$^2$, and Akira \textsc{Ono}$^3$
}

\inst{
$^1$Department of Physics, Kyoto University, Kyoto 606-8502, Japan\\
$^2$Creative Research Initiative ``Sousei,'' Hokkaido University, Sapporo 001-0021, Japan\\
$^3$Department of Physics, Tohoku University, Sendai 980-8578, Japan
}

\abst{
Structure and reaction studies with a method of antisymmetrized molecular dynamics (AMD)
were reviewed. 
Applications of time-independent and time-dependent versions of the AMD 
were described.
In applications of time-independent AMD to nuclear structure studies, 
structures of neutron-rich nuclei such as Be, C, Ne, and Mg isotopes were described
focusing on cluster aspects. Important roles of valence neutrons were discussed.
The results suggested a variety of cluster structures appear
also in unstable nuclei as well as in stable nuclei.
Deformation and cluster phenomena in $Z\sim N$  nuclei in $p$- and $sd$- shell regions
were also discussed. 
Applications of time-dependent AMD contain various topics such as 
fragmentation in heavy-ion collisions as well as nuclear responses. 
The AMD calculations successfully describe multifragmentation which is one of the remarkable phenomena 
in heavy-ion collisions. 
The approach is able to link reactions and
nuclear matter properties. The studies suggested the important balance between single-particle
motions and correlations to form clusters and fragments.
}

\begin{document}

\maketitle

\tableofcontents

\section{Introduction}
A nucleus is a finite quantum many-body system consisting of protons and neutrons
interacting via nuclear forces. Its ground state has shell structure,
in which nucleons move almost independently in an averaged field (mean field)
analogously to an atomic system.  
The
shell structure and excitation modes associated with single-particle and collective
motions are important facets of nuclear system. 

On the other hand, in nuclear systems, one 
may also find unique characters different from atomic systems. One of the important
differences is that a nucleus is a self-bound system formed by attractive nuclear forces. 
Because of the attraction, spatial correlations among nucleons can be rather strong, 
and therefore, assembling and disassembling of nucleons occur in various ways.  Furthermore,
the saturation property of nuclear system, where binding energy per nucleon and central
density are almost constant independently to the mass number, implies that assembling and
disassembling of nucleons can take place with a small excitation energy. 
As well known, this characteristic of nuclear system  manifests itself as
cluster structures in which a nucleus is divided into several subunits
(clusters) and nucleons are confined within each cluster. 
  The important roles of assembling and disassembling of nucleons
  continue to higher excitation energies as in intermediate-energy
  heavy-ion collisions where a lot of clusters and fragment nuclei are
  produced from a hot source whose excitation energy is typically
  comparable to the binding energy of a nucleus.

In spite of importance of cluster aspect in nuclear systems,  
usual mean-field approaches often fail to describe those cluster phenomena
because they take into account insufficiently 
many-body correlations, which are essential in cluster formation.
The theoretical method of antisymmetrized
molecular dynamics
(AMD) \cite{Ono:1991uz,Ono:1992uy,ENYOabc,ONO-ppnp,ENYOsupp,AMDrev,AMDrev2} has been 
proposed in studies of 
heavy-ion collisions.
AMD describes nuclear many-body systems
by antisymmetrized products of Gaussian wave packets of nucleons
and incorporates quantum effects and nucleon-nucleon collisions.
It has described 
fragment formation in heavy-ion collisions successfully. 
AMD has been proved to be a powerful tool also for
nuclear structure study. All centers of the Gaussian packets are independently treated
as variational parameters in the AMD framework, and it is possible to describe various
cluster structure wave functions as well as independent-particle motion in a mean field
without {\it a priori} assumption. Thus, AMD has been applied to investigate various
phenomena in nuclear structure and reactions. 

In the early days, AMD studies were limited to light systems. This is because  computational cost increases rapidly
in proportion to $A^{4\sim 6}$ ($A$ is the mass number)
due to the non-orthogonality of single-particle wave functions. 
However, AMD calculations have 
developed remarkably toward various nuclear systems owing to the rapid progress of computational facilities. 
For instance, it enabled the reaction
studies by AMD  up to Au+Au collisions ($A\sim 400$)\cite{Ono:1998yd}, the
structure studies up to $pf$-shell nuclei ($A\sim 40$) \cite{Kimura:2004ez}  and variational
calculation after the angular momentum projection that covers up to very high excitation
energy \cite{ENYOe}.
Further high-performance computing will  extend the subjects of  AMD
studies and  enable more sophisticated AMD calculations. 

\begin{figure} 
\centerline{\includegraphics[width=15.0 cm] {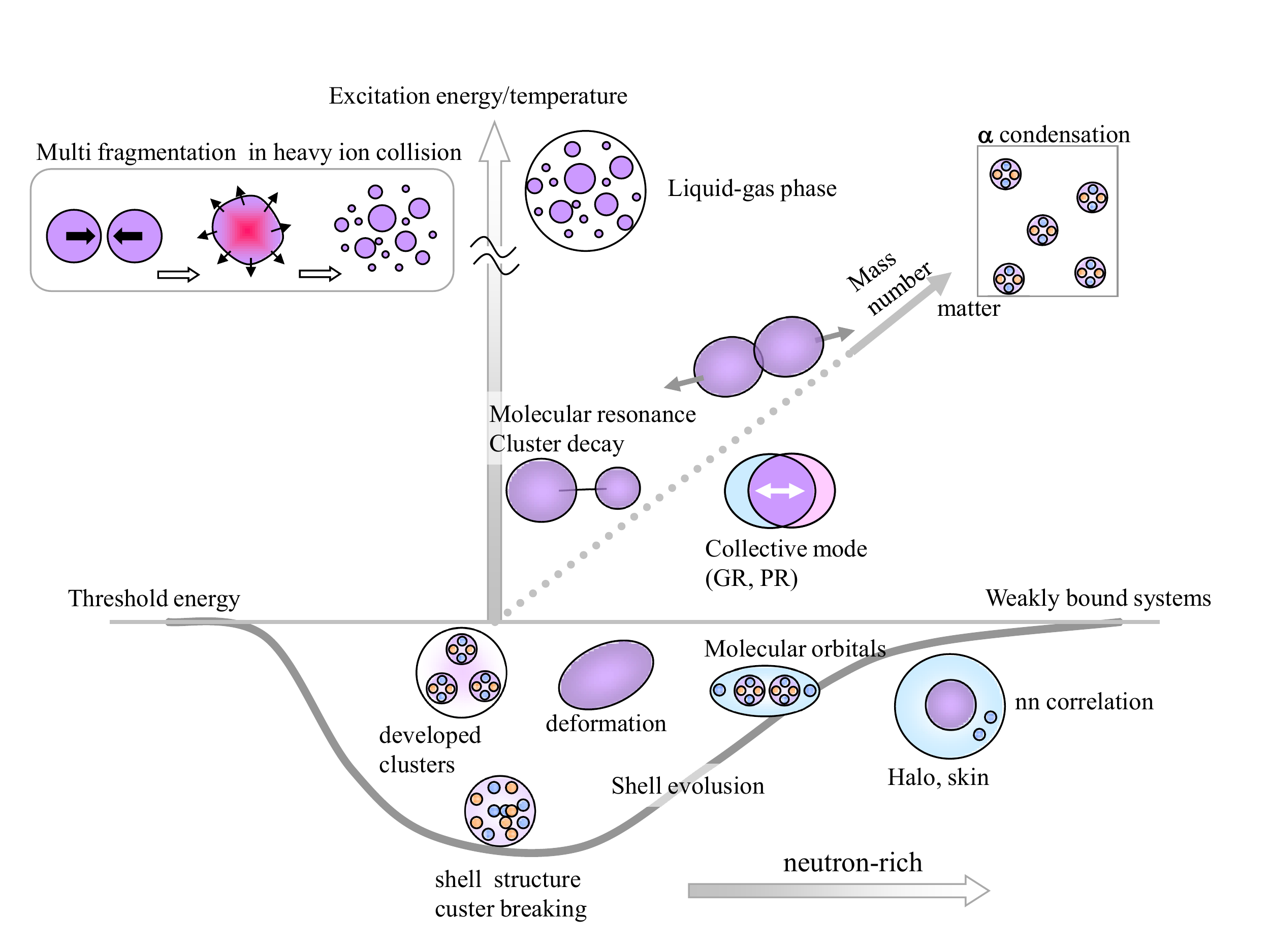}}
 
   \caption{Schematic figure for rich phenomena in nuclear systems.}
   \label{fig:1}
\end{figure}

AMD studies have revealed that 
cluster phenomena emerge widely in various nuclear systems. 
Many exotic and novel features of clustering have been discovered and 
the concept of the clustering has been renewed and extending now.
Today, it is well established that clustering is an essential
aspect of nuclear many-body systems as well as the mean-field aspect. 
Coexistence of 
cluster and mean-field aspects brings out rich phenomena to nuclear many-body systems
as functions of excitation energy and isospin degrees of freedom (Fig.~\ref{fig:1}). 
As the excitation energy increases, one may see transitions from mean-field to cluster structures. 
In deeply bound systems such as low-lying states of stable nuclei, mean-field effects are
rather strong. However, even if a nucleus has a shell-model-like structure in its ground
state,  developed cluster structures appear in excited states near the corresponding
cluster-decay  threshold energy (so-called Ikeda's threshold rule
\cite{ikeda-diagram}). Above the threshold energy, further remarkable cluster phenomena  such
as alpha decays, molecular resonances, and fission etc. are known. The coexistence and
competition between the clustering and mean field have been studied by AMD in a unified way
within a single theoretical framework. When the excitation energy increases further, 
nuclear systems may 
enter the region of nuclear liquid-gas phase transition.  AMD studies
have confirmed the phase transition by obtaining caloric curves for
equilibrium systems.  The link between the phase transition and copious
fragment formation in heavy-ion collisions is now clearer with the
unified description by time-dependent AMD.
Another important degree of freedom in nuclear systems 
is the isospin asymmetry, i.e., neutron (proton) excess, which is one of the major 
directions in recent nuclear physics. 
In the neutron-rich and proton-rich domains, the
saturation law of energy and density is broken. Because of the unbalanced proton-neutron 
ratio, we may encounter many novel cluster phenomena in isospin asymmetric systems. 
Indeed, such exotic clustering phenomena 
as cluster structures in neutron-rich Be isotopes and 
isospin  fractionation/distillation at liquid-gas separation in
fragmentation reactions have been investigated with AMD.


In this paper, we review the AMD approach and its applications to nuclear structures and reactions.
In the next section, the formulation of AMD is described. 
Applications of the time-independent version of AMD to static problems of nuclear structures are explained
in \S\ref{sec:sec3}, and those of the time-dependent version to dynamical phenomena such as
nuclear responses and nuclear reactions are described in \S\ref{sec:sec4}. 
Finally, a summary and perspectives are given in \S\ref{sec:summary}.

\section{Antisymmetrized molecular dynamics (AMD)}\label{sec:formulation}
As mentioned above, cluster aspect is one of the essential features of nuclear systems as well as mean-field
aspect. Originating in coexistence of two kinds of nature, cluster and mean-field aspects, a variety of phenomena arise in nuclear many-body systems.
To investigate rich phenomena concerning cluster aspects such as cluster structures and multifragmentations,
a theoretical framework that can describe both the cluster and mean-field features is required. 
The AMD method has been proved to be one of the powerful approaches in description of those features. 
In this section, we briefly review formulation of AMD. After describing basic formalism of AMD, 
we explain extended methods of AMD which have been applied to nuclear structure 
and nuclear reaction studies.
For detailed formulation of the AMD method, the reader is refereed to Refs.~\citen{ONO-ppnp,ENYOsupp,AMDrev,AMDrev2}.

\subsection{Basic formulation of AMD}
\subsubsection{AMD wave function}
In the AMD framework, an basis wave function for an $A$-nucleon system 
is expressed by a Slater determinant of Gaussian wave packets;
\begin{equation}
 \Phi_{\rm AMD}({Z}) = \frac{1}{\sqrt{A!}} {\cal{A}} \{
  \varphi_1,\varphi_2,...,\varphi_A \},
\end{equation}
where the $i$th single-particle wave function is written by a product of
spatial ($\phi$), intrinsic spin ($\chi$), and isospin ($\tau$) 
wave functions as follows.
\begin{eqnarray}
 \varphi_i&=& \phi_{{\bf Z}_i}\chi_i\tau_i,\\
 \phi_{{\bf Z}_i}({\bf r}_j) &\propto&  
\exp\bigl (-\nu({\bf r}_j-\frac{{\bf Z}_i}{\sqrt{\nu}})^2\bigr ),
\label{eq:spatial}\\
 \chi_i &=& (\frac{1}{2}+\xi_i)\chi_{\uparrow}
 + (\frac{1}{2}-\xi_i)\chi_{\downarrow}.
\end{eqnarray}
The spatial part $\phi_{{\bf Z}_i}$ of the $i$th single-particle wave 
function is represented by a complex variational parameter 
$Z_{i\sigma}$ with $\sigma=x,y,z$ which indicates the center of the Gaussian wave
packet. The spin part $\chi_i$ is parametrized by a complex number 
parameter, $\xi_i$. 
The isospin function $\tau_i$ is fixed to be up (proton) or down (neutron). 
The width parameter $\nu$ takes a common value for all nucleons. It is
chosen to be an optimum value for the studied system.
Accordingly, an AMD wave function
is expressed by a set of variational parameters, ${Z}\equiv 
\{{\bf Z}_1,{\bf Z}_2,\cdots, {\bf Z}_A,\xi_1,\xi_2,\cdots,\xi_A \}$.
These parameters 
indicate centers of localized Gaussians and 
spin orientations, which 
are treated independently for all nucleons. 
That is to say, a system written by a single AMD wave function is specified by the 
configuration of single-nucleon wave packets in the phase space 
and their spin orientations.
In the simplest version of AMD, the spin part $\chi_i$ is sometimes fixed 
to be up or down, and only parameters ${\bf Z}_1,{\bf Z}_2,\cdots, {\bf Z}_A$ for Gaussian centers 
are treated as variational parameters. In time-dependent AMD, intrinsic spin orientations are usually 
fixed. In the case of fixed spins, we redefine 
${Z}\equiv \{{\bf Z}_1,{\bf Z}_2,\cdots, 
{\bf Z}_A \}$ by omitting the spin labels $\xi_i$. 

In the AMD wave function, all single nucleons are treated independently as 
localized Gaussians. Although any constituent clusters are not assumed {\it a priori}, 
multi-cluster structures can be described by grouping of 
single-nucleon Gaussian wave packets in the spatial configuration.
On the other hand, if all the Gaussian centers gather around a certain position, 
the AMD wave function becomes equivalent to a harmonic oscillator shell-model
wave function around the position due to the effect of antisymmetrization. 
Thus, the model space of AMD can describe both cluster and mean-field features
with assembling and disassembling of Gaussian wave packets.
If a system favors a specific cluster channel, such a cluster
structure will be
automatically obtained in energy variation or in dynamics. 

The AMD wave function is quite similar to a wave function of 
fermionic molecular dynamics (FMD)\cite{Feldmeier:1989st,Feldmeier:1994he} where 
more generalized wave functions are adopted. 
In applications of AMD to nuclear structure and reaction studies, 
  the description has been improved by superposing many AMD wave
  functions or by introducing stochastic processes rather than by
  choosing more general single-particle wave functions.

\subsubsection{Equation of motion}
In the time-dependent versions of the AMD method which have been applied to 
dynamics of nuclear systems, 
the time evolution of the variational parameters ${Z}$ are determined by the time-dependent variational
principle. The equation of motion for ${Z}$ derived from the time-dependent
variational principle is
\begin{equation}\label{eq:eqmotion}
i\hbar \sum_{j\rho} C_{i\sigma, j\rho} \frac{dZ_{j\rho}}{dt}=\frac{\partial {\cal H}}{\partial Z^*_{i\sigma}},
\end{equation}
where 
  $\sigma,\rho=x,y,z$ are the labels for the components of
  $\mathbf{Z}_i$ ($i=1,2,\ldots A$).  When the spin wave function is a
  variable, it may be regarded as the fourth component $Z_{i4}=\xi_i$.
The expectation value of the 
Hamiltonian $\hat{H}$ is given by
\begin{equation}\label{eq:hamil}
{\cal H}({Z},{Z}^*)=
\frac{
\langle\Phi_{\rm AMD}({Z})|\hat{H}|
\Phi_{\rm AMD}({Z})\rangle}{\langle\Phi_{\rm AMD}({Z})|\Phi_{\rm AMD}({Z})\rangle}.
\end{equation}
A positive definite Hermitian matrix
\begin{equation}\label{eq:cmat}
C_{i\sigma,j\rho}\equiv 
\frac{\partial^2}{\partial Z^*_{i\sigma}\partial Z_{j\rho}}
\ln \langle\Phi_{\rm AMD}({Z})|\Phi_{\rm AMD}({Z})\rangle
\end{equation}
appears in the
  equation of motion, suggesting that the
  variables $Z$ are not canonical coordinates.

\subsubsection{Energy variation}
To get an optimum solution for the energy minimum state, 
the energy variation is performed.
Namely, the variational
parameters ${Z}$ are optimized to minimize the 
expectation value of the Hamiltonian in the AMD model space.
We introduce the following frictional cooling equation,
\begin{equation}
i\hbar \sum_{j\rho} C_{i\sigma, j\rho} \frac{dZ_{j\rho}}{dt}
=\left(\lambda+i\mu\right)\frac{\partial {\cal H}}{\partial Z^*_{i\sigma}}.
\end{equation}
The parameter $\lambda$ is an arbitrary real number and $\mu$ is an arbitrary
negative real number. It is easily proved that the energy of the system decreases as time develops
due to the frictional term $i\mu$. 
In the energy variation, the matrix $C_{i\sigma, j\rho}$ can be replaced with $\delta_{ij}\delta_{\sigma\rho}$
and then, the frictional cooling method with $\lambda=0$ becomes
equivalent to the steepest decent method.
In both cases of $C_{i\sigma, j\rho}$, we obtain the optimum set of parameters ${Z}$ that
gives the AMD wave function for
the minimum energy state in the model space after enough cooling time (iteration steps).

\subsubsection{Hamiltonian}
 The Hamiltonian for an $A$-nucleon system consists of the kinetic energy, the nuclear and Coulomb
force terms,
\begin{eqnarray}
 \hat{H} = \sum_{i} \hat{t}_i + \sum_{i,j} \hat{v}_{ij} + \sum_{i,j} \hat{v}^{\text{coulomb}}_{ij}  - \hat{T}_g.
\end{eqnarray}
Here $\hat{t}_i$ is the kinetic energy. 
The energy of the center-of-mass
motion $\hat{T}_g$ is subtracted exactly 
because the total wave function can be separated into the internal wave function and the
center-of-mass wave function.
In applications to heavy ion collisions, spurious kinetic energy of the zero-point oscillation of
fragment mass centers is also subtracted from the Hamiltonian. 
For the effective two-body nuclear force $\hat v_{ij}$, 
finite-range forces such as 
Volkov \cite{VOLKOV} forces supplemented by G3RS form spin-orbit forces \cite{LS}, Gogny
\cite{GOGNY,D1S} forces, and Skyrme \cite{SIII,cha98} forces are used.
Finite-range two-body forces with zero-range 
three-body forces such as the modified Volkov forces \cite{MVOLKOV} are also used. 
The Coulomb force $\hat v^{\text{coulomb}}_{ij} $ is approximated by a sum of seven Gaussians. 

These effective forces are phenomenological ones constructed to describe 
low-energy properties of nuclear structure. 
In heavy-ion reaction, residual interactions contribute to nucleon-nucleon collisions 
which are incorporated by stochastic collision process in the AMD framework as explained later. 

\subsection{Basis AMD and its Extensions in applications}
In the early stage of AMD studies, simple versions of the AMD method have been applied to 
reaction and structure studies \cite{Ono:1991uz,Ono:1992uy,ENYOabc}, and later, 
the AMD method has been developed to many extended versions
(Refs.~\citen{ONO-ppnp,AMDrev,AMDrev2}
and references therein). In this section, we explain the formulation of the 
basic AMD method and some advanced versions for structure study and those for reaction study.

An basis AMD wave function is given by a single Slater determinant. 
Generally, many-body wave functions for quantum systems should be expressed by a 
superposition of many Slater determinants. 
Restriction of the model space 
within a single Slater determinant is the limit of a mean-field approximation. 
To incorporate beyond-mean-field effects, superposition of Slater determinants 
is essential. Firstly, 
in structure study, parity and angular-momentum projections, which are done by superposition of Slater determinants,
are essential to describe properties of energy-eigen states.
In second, superposition of Slater determinants is significant to improve wave functions by
taking into account quantum fluctuation around a mean-field, many-body
correlations, and spin-parity projections. 
It is also necessary to describe excited states orthogonal to lower states. 

In applications of the AMD method to structure study, 
the projections and superposition of AMD wave functions are practically performed. 
In applications to heavy-ion reactions, however, 
emergence of multiple reaction channels is introduced by stochastic
branching processes, neglecting quantum interference between different
channels each of which is described by an AMD wave function.

\subsubsection{Projections and superposition of AMD wave functions}

The parity-projected AMD wave function is given as 
\begin{equation}
 |\Phi^{\pm}_{\rm AMD}\rangle \equiv P^\pm |\Phi_{\rm AMD}({Z})\rangle =
 \frac{1\pm \hat P_r}{2} |\Phi_{\rm AMD}({Z})\rangle,
\end{equation}
where $P^\pm=1\pm \hat P_r$ is the parity projection operator.
The angular-momentum projected AMD wave function is written as
\begin{equation}
 |\Phi^{J}_{MK}\rangle = P^{J}_{MK}|\Phi_{\rm AMD}({Z}) \rangle 
 = \int d\Omega D^{J*}_{MK}(\Omega) \hat R(\Omega) |\Phi_{\rm AMD}({Z})\rangle.
\end{equation}
Here $D^J_{MK}(\Omega)$ is the Wigner's $D$ function and $\hat R(\Omega)$ is a rotation operator
with respect to Euler angle $\Omega$. 
As clearly shown, the angular-momentum projected state is
expressed by a linear combination of wave functions rotated from the intrinsic AMD 
wave function $\Phi_{\rm AMD}({Z})$ with the weight function, $D^{J*}_{MK}$.
The matrix element of a tensor operator $\hat T^k_q$, where $k$ is the rank and $q$ is the $z$-component, 
can be calculated to be
\begin{multline}
\langle P^{J}_{MK} \Phi_{\rm AMD}({Z})|\hat T^k_q|P^{J'}_{M'K'} \Phi_{\rm AMD}({Z'}) \rangle
= \frac{8\pi^2}{2J+1} \langle J'M'kq|JM\rangle\\
\times  \sum_{\mu\nu}  \langle J'\mu k\nu|JK\rangle 
 \int d\Omega D^{J'*}_{\mu K'}(\Omega) \langle \Phi_{\rm AMD}({Z})
|\hat T^k_\nu \hat R(\Omega)|
\Phi_{\rm AMD}({Z'}) \rangle.
\end{multline}
In practical calculations, the integrations with respect to $\Omega$ is performed 
by numerical integration on grid points of angles $\Omega=(\theta_1,\theta_2,\theta_3)$.
In calculations of expectation values for observable operators such as 
Hamiltonian, radii, moments, and transitions, 
AMD wave functions are projected to 
parity and angular-momentum eigenstates. 
In the usual AMD calculations for structure study, the parity projection is done before 
energy variation while the angular-momentum projection is performed after the energy variation, 
i.e., variation before projection (VBP).

Superposition of independent AMD wave functions is useful to improve wave functions 
and it is essential in description of excited states to satisfy orthogonality between energy levels.
Let us consider superposition of independent AMD wave functions $\Phi_{\rm AMD}({Z}^{(k)})$ 
$(k=1,\cdots, k_{\rm max})$ ($k_{\rm max}$ is the number of adopted basis AMD wave functions). 
Superposed wave functions for $J^\pm$ states is written as 
\begin{eqnarray}\label{eq:superpose}
 |\Phi^{J\pm}_n\rangle = \sum_{kK} c_{n,kJK}|P^{J\pm}_{MK} \Phi_{\rm AMD}({Z}^{(k)})\rangle,
\end{eqnarray}
where $P^{J\pm}_{MK}\equiv P^J_{MK}P^\pm$. 
Here the values for the coefficients $c_{n,kJK}$ are determined by the variational principle,
\begin{eqnarray}
 \delta\bigl\{\langle\Phi^{J\pm}_n|\hat{H}|\Phi^{J\pm}_n\rangle 
  - \epsilon_n \langle\Phi^{J\pm}_n|\Phi^{J\pm}_n\rangle
\bigr\}=0,
\end{eqnarray}
which is equivalent to diagonalization of the norm and the Hamiltonian matrices
and leads to the Hill-Wheeler equations.
The $K$ sum in Eq.~(\ref{eq:superpose}) stands for the $K$-mixing.  

To adopt efficient AMD wave functions as basis wave functions for the superposition,
constraint methods, which are often used in a generator coordinates method (GCM)\cite{Itagaki:1999um,defAMD,defAMD2,Suhara:2009jb},
or stochastic variational methods \cite{Aoyama:2006dn} are applied in the AMD framework. 
Here we consider a constraint 
$\langle \hat g \rangle=\bar g$. 
By changing the constraint value as $\bar g=\bar g_1, \bar g_2, \cdots, \bar g_{k_{\rm max}}$, 
the minimum energy state $\Phi^\pm_{\rm AMD}(\bar g)$ in the 
AMD model space is obtained for each constraint value
by the constraint energy variation after parity projection. Then the obtained
 wave functions $\Phi^\pm_{\rm AMD}(\bar g_k)$ $(k=1,\ldots, k_{\rm max})$ are superposed, and 
coefficients are determined by the diagonalization.
This method corresponds to a GCM calculation for a generator
coordinate $\bar{g}$ when an enough number of the basis wave functions
for different values $\bar{g}_k$ are taken into account.
 This method (called AMD+GCM) is useful, in particular, for study of excited states.
For the constraints, the nuclear quadrupole deformation parameters are often 
used  \cite{defAMD,defAMD2,Suhara:2009jb}.
In the AMD+GCM method, variation is done before the angular-momentum 
projection.

It is also efficient to perform variation after the
angular-momentum projection as well as the parity projection, especially, for study of excited states. 
Namely, the energy expectation value for a parity and angular-momentum projected AMD
wave function,
\begin{equation}
{\cal H}= 
\frac{\langle P^{J\pm}_{MK}\Phi_{\rm AMD}({Z})|\hat H|P^{J\pm}_{MK}\Phi_{\rm AMD}({Z})\rangle} 
{\langle P^{J\pm}_{MK}\Phi_{\rm AMD}({Z})|P^{J\pm}_{MK}\Phi_{\rm AMD}({Z})\rangle},
\end{equation}
is minimized with respect to the variational parameters ${Z}$
by using the frictional cooling method.
Firstly, for a given spin and a parity $J^\pm$, the wave function 
for the lowest $J^\pm$ state is obtained by the energy variation. 
Then, a wave function for a higher state ($J_n^\pm$) is provided by varying ${Z}$ to
minimize the energy for the component of an AMD wave function 
orthogonal to the lower states $(J^\pm_1,\cdots,J^\pm_{n-1})$ which are already obtained.
In the present paper, we call the variation after spin-parity projection 
``VAP'' \cite{ENYOe}.

\subsubsection{Extensions for single-particle wave functions}
In a basis AMD wave function, a single-particle wave function is written by a
spherical Gaussian. In extended versions of AMD, a single-particle wave function
is written by a deformed Gaussian or superposition of different-range Gaussians 
to improve single-particle wave functions\cite{defAMD,defAMD2,Dote:2005un,Furutachi:2009}. 
In the method of deformed-basis AMD\cite{defAMD,defAMD2}, 
triaxially deformed Gaussians are employed as single-particle wave
packets instead of spherical ones,
\begin{equation}
 \phi_{{\bf Z}_i} ({\bf  r}_j)\propto \exp\bigl\{-\sum_{\sigma=x,y,z}\nu_\sigma
 (r_{j\sigma}-\frac{Z_{i\sigma}}{\sqrt{\nu_\sigma}})^2\bigr\},
\end{equation}
where the width parameters, $\nu_x$, $\nu_y$ and $\nu_z$, take different values for each
direction. They are determined in the energy variation to
optimize energy of a system. By using this
deformed basis, it is possible to successfully describe coexistence (or mixing) 
of cluster and deformed mean-field structures, which are 
essential especially in heavy systems.

Superposing different range Gaussians is another method to improve 
single-particle wave functions of an AMD wave function as is done in 
the FMD method \cite{Dote:2005un,Furutachi:2009}.

Instead of adopting deformed or superposing Gaussians,  
a method of stochastic branching on wave packets is used to describe 
the diffusion and deformation of single-particle motions 
in the time-dependent version of AMD for 
nuclear reaction calculations\cite{Ono:1998yd,Ono:1996rk,ONOj} as explained in the next subsection.

\subsection{Branching in time evolution}

\begin{figure}
\begin{center}
\includegraphics[width=0.5\textwidth]{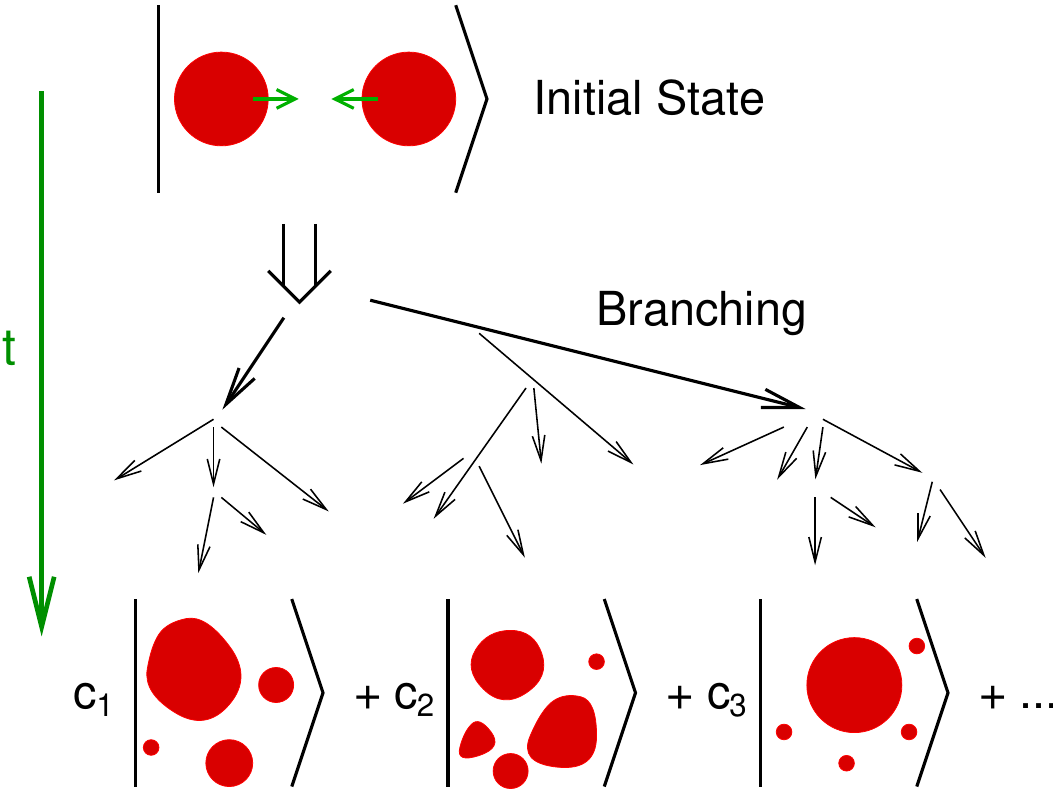}
\end{center}
\caption{\label{fig:multichannel}
A schematic picture of the quantum branching processes for
multichannel reactions.}
\end{figure}

The description of dynamics of excited nuclear many-body systems, such
as in heavy-ion collisions, is a highly quantum-mechanical many-body
problem.  If the many-body time-dependent Schr\"odinger equation is
solved for an initial state which may be roughly approximated by a
single Slater determinant, the intermediate and final states will be a
very complicated states containing a huge number of reaction channels
corresponding to different fragmentation configurations, as
illustrated in Fig.\ \ref{fig:multichannel}.  The AMD model for
reactions respects the existence of channels, while it neglects some
of the interference among them.  Namely, the total many-body wave
function $|\Psi(t)\rangle$ is approximated by a mixed state
\begin{equation}
|\Psi(t)\rangle\langle\Psi(t)|\approx
\int
\frac{|\Phi(Z)\rangle\langle\Phi(Z)|}{\langle\Phi(Z)|\Phi(Z)\rangle}
 w(Z,t)dZ,
\label{eq:AMDensemble}
\end{equation}
where each component is represented by an AMD wave function
$|\Phi(Z)\rangle$ with a time-dependent weight $w(Z,t)$.

Interference between quite different components is not of practical
importance because the matrix elements are negligible for the
Hamiltonian and other operators of our usual interest.  Furthermore,
when we adopt mean field approximation in some way, we need to take
care of the spurious nonlinearity introduced by the approximation.
A many-body state composed of different reaction channels cannot be
described by a single mean field.  Our strategy is to suitably
decomposed the state in such a way that the mean field approximation
is valid in each component.

Let us consider the motion of a nucleon in the system, ignoring the
Pauli principle for the moment.  In mean field approximation, the
one-body density matrix is given by a pure single-particle wave
function $\hat{\rho}(t)=|\psi(t)\rangle\langle\psi(t)|$ at any time.
Under the mean field, the phase space distribution will spread in some
directions and may shrink in other directions.  In contrast, in the
real time evolution, the condition $\hat{\rho}(t)^2=\hat{\rho}(t)$
should no longer hold due to many-body correlations.  Namely,
decoherence should occur on the single-particle state so that it turns
into a mixed state.  In the AMD approach, decoherence is taken into
account by splitting the wave packet in such a way that the spreading
of the distribution in the mean field is respected while the shrinking
is discarded.  By the decoherence of the single-particle states, the
decomposition (or branching) of the many-body state is induced
naturally.  This approach of using compact wave packets is
advantageous for the case with many fragmentation channels because it
is free from spurious coupling of different fragmentation channels.

Instead of directly considering the weight function $w(Z,t)$ in Eq.\
(\ref{eq:AMDensemble}), we solve a stochastic equation of motion for
the wave packet centroids $Z$, which may be symbolically written as
\begin{equation}
\frac{d}{dt}\mathbf{Z}_i
=\{\mathbf{Z}_i,\mathcal{H}\}_\text{PB}
+\mbox{(NN coll)}
+\Delta\mathbf{Z}_i(t)
+\mu\,(\mathbf{Z}_i,\mathcal{H}').
\end{equation}

The first term $\{\mathbf{Z}_i,\mathcal{H}\}_\text{PB}$ is that in the
deterministic equation of motion [Eq.\ (\ref{eq:eqmotion})] derived
from the time-dependent variational principle.

The second term represents the effect of stochastic two-nucleon
collisions, where a parametrization of the energy-dependent in-medium
cross section is adopted.  The collisions are performed with the
``physical nucleon coordinates'' that take account of the
antisymmetrization effects, and then the Pauli blocking in the final
state is automatically introduced \cite{Ono:1991uz,Ono:1992uy}.

The third term $\Delta\mathbf{Z}_i(t)$ is a stochastic term for the
wave packet splitting mentioned above
\cite{Ono:1998yd,Ono:1996rk,ONOj}.  The change of the width and shape
of each wave packet is calculated by solving the Vlasov equation (for
some time period) with the same effective interaction as for the term
$\{\mathbf{Z}_i,\mathcal{H}\}_\text{PB}$.  An essential ingredient
here is how long the coherent single-particle motion is solved before
decoherence.  The properties of fluctuations $\Delta\mathbf{Z}_i(t)$
are determined depending on this time scale called coherence time
$\tau$.  When the wave packet splitting was first introduced into AMD
\cite{Ono:1998yd,Ono:1996rk}, the limit of $\tau\rightarrow0$ was
taken, for which the decoherence effect is maximum.  The choice of a
finite coherence time has been formulated in Ref.\ \citen{ONOj}, and a
reasonable choice may be to consider decoherence for a nucleon when it
collides with another nucleon.  This choice is symbolically denoted by
$\tau=\tau_{\text{NN}}$, though $\tau$ is not a constant number.

The last term $\mu\,(\mathbf{Z}_i,\mathcal{H}')$ is a dissipation term
related to the fluctuation term $\Delta\mathbf{Z}_i(t)$.  This term is
necessary in order to restore the energy conservation that is violated
by the fluctuation term.  The coefficient $\mu$ is given by the
condition of energy conservation. However, the form of this term is
somehow arbitrary.  The variables $Z$ are shifted to the direction of
the gradient of the energy expectation value $\mathcal{H}$ under the
constraints of nine conserved quantities (the center-of-mass variables
and the total angular momentum).  When the number of other nucleons
around the nucleon $i$ within a phase space radius of
$|\mathbf{Z}_j-\mathbf{Z}_i| <2.5$ is more than a certain number
$N_{\text{c9}}$, other twelve global one-body quantities (monopole and
quadrupole moments in coordinate and momentum spaces) are also
included as constrained quantities for the dissipation corresponding
to the fluctuation $\Delta\mathbf{Z}_i(t)$.  The latter constraints
should be imposed because the one-body time evolution has already been
considered by
$\{\mathbf{Z}_i,\mathcal{H}\}_\text{PB}+\Delta\mathbf{Z}_i(t)$.
Empirically, $N_{\text{c9}}$ has been chosen between 5 and 15.

A summary of the complete formulation of AMD for reactions can be found in Ref.\
\citen{ONO-ppnp}.

\section{Applications of time-independent AMD method to nuclear structure} \label{sec:sec3}
We here discuss some topics investigated with the time-independent AMD method
focusing on cluster aspects.

\subsection{Molecular structures in Be and Ne isotopes}
Cluster structure of Be isotopes is one of the fascinating subjects of unstable nuclei. 
A $2\alpha$-cluster core is favored in neutron-rich Be isotopes as well as $^8$Be whose
ground state is a $2\alpha$ resonance state. 
The low-lying states of neutron-rich Be isotopes are 
described well by a molecular-orbital picture based on a 2$\alpha$ core 
and valence neutrons moving around the 2$\alpha$
\cite{SEYA,OERTZEN,OERTZENa,ITAGAKI,DOTE,Ito:2003px,Oertzen-rev}.
In contrast to the molecular-orbital structures in low-lying states, developed 
di-cluster states such as $^6$He+$^6$He  in $^{12}$Be
have been suggested in highly excited states\cite{AMDrev2,Ito:2003px,Descouvemont01,ENYObe12,Ito:2008zza}. There, 
valence neutrons are moving not around the whole system but around one of two $\alpha$ clusters.
It means that a variety of cluster structures coexist in neutron-rich Be isotopes where
valence neutrons play important roles.

The molecular-orbital picture has been extended also to Ne isotopes such as $^{21}$Ne and $^{22}$Ne 
based on an $^{16}$O+$\alpha$-cluster core and  valence neutrons in molecular orbitals
\cite{AMDrev2,OERTZENa,k:kimurane}. 
Di-cluster states like $^{18}$O+$\alpha$-cluster states in $^{22}$Ne is another attractive subject \cite{AMDrev2}. 

\subsubsection{Molecular-orbital structure} \label{sec:results1}

The idea of the molecular orbitals
surrounding a 2$\alpha$ core was
suggested in $^9$Be with a 2$\alpha$+n cluster model\cite{OKABE} in 1970's. 
In 1980's and 1990's, molecular-orbital models
were applied to neutron-rich Be isotopes and succeeded to describe 
rotational bands \cite{SEYA,OERTZEN,OERTZENa,ITAGAKI}.

In a 2$\alpha$ system, molecular
orbitals are formed by a linear combination of $p$ orbits around
two $\alpha$ clusters. In neutron-rich Be isotopes, valence neutrons 
occupy the molecular orbitals around the 2$\alpha$ core.
The negative-parity orbital is called 
'$\pi$ orbital', while the longitudinal orbital with positive parity is 
'$\sigma$ orbital' (Fig.\ref{fig:orbit}). 
Since the $\sigma$ orbital has two nodes along the $\alpha$-$\alpha$ 
direction, it gains the kinetic energy as the $2\alpha$ cluster develops.
The energy gain of the $\sigma$ orbital in the developed $2\alpha$ system
results in the intruder configurations 
of the $^{11}$Be and $^{12}$Be ground states.
In other words, it is the origin of the breaking of 
the neutron magic number $N=8$ in the Be isotopes. 

In analogy to neutron-rich Be isotopes, molecular-orbital structures in Ne isotopes have been
suggested from the experimental systematics\cite{OERTZENa}.  Indeed, AMD calculation
has predicted the presence of the molecular-orbital bands with $^{16}$O+$\alpha$-cluster core
surrounded by two valence neutrons in the $\sigma$ orbital \cite{AMDrev2,k:kimurane}. Different from Be isotopes,
$\sigma$ orbital is a linear combination of $sd$ orbits around $^{16}$O and $p$ orbits around
$\alpha$ and it results in a $pf$-shell like molecular orbital. Another difference is the
parity asymmetry of the core, that produces parity doublet of the molecular bands. 
Similar molecular-orbital structures have been also suggested for F isotopes \cite{KimF10}.
Detailed discussions are given in later sections. 


\subsubsection{Cluster structures in neutron-rich Be isotopes}

Cluster structures of Be isotopes have been
intensively investigated in many theoretical works with 
cluster models\cite{Ito:2008zza,OGAWA,Arai01,Descouvemont02}, 
molecular-orbital models\cite{SEYA,OERTZEN,ITAGAKI,Oertzen-rev,OKABE}, 
and AMD\cite{AMDrev,DOTE,ENYObe12,ENYObe10,ENYObe11}. 
In the cluster and molecular-orbital models,  
the existence
of two $\alpha$ clusters are {\it a priori} assumed.
On the other hand, AMD does not rely on model
assumptions of the existence of cluster cores. Nevertheless, 
the results of AMD calculations indeed indicate the appearance of the 
2$\alpha$ core surrounded by valence neutrons in the molecular orbitals
in low-lying states of Be isotopes.
It means that 
the formation of the 2$\alpha$ core and molecular orbitals 
has been theoretically confirmed by those AMD calculations without assuming clusters.
Here we discuss the cluster
structures of neutron-rich Be isotopes based 
on AMD calculations \cite{AMDrev,ENYObe12,ENYObe10,ENYObe11}.
 
\begin{figure}[th]
\centerline{\includegraphics[width=12cm]{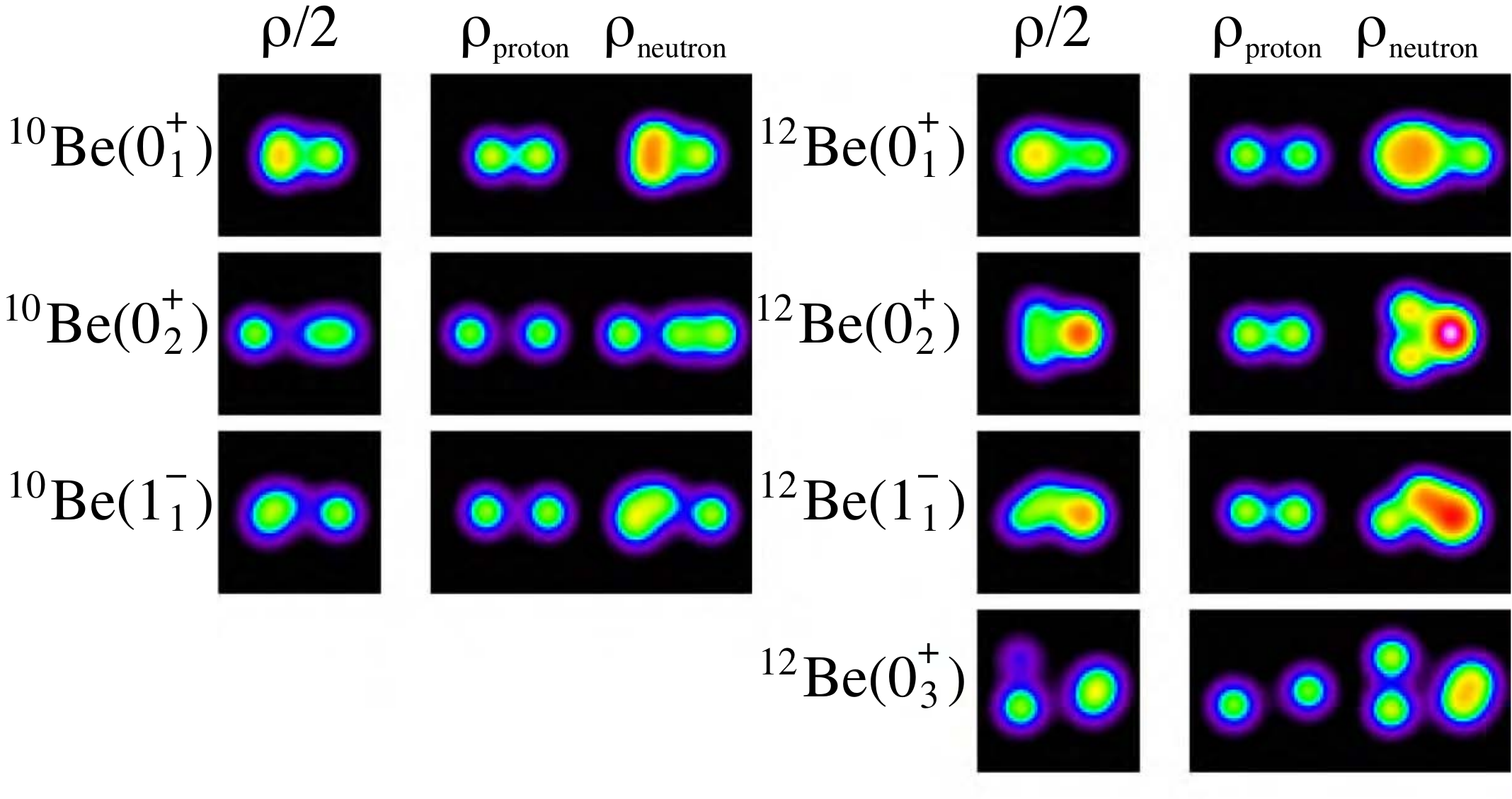}}
\vspace*{8pt}
\caption{\label{fig:be-dense}
Density distributions of the intrinsic states for the band-head 
states of $^{10}$Be and $^{12}$Be obtained 
by AMD-VAP \cite{ENYObe10,ENYObe12}.
The integrated densities of matter, proton and  neutron densities are presented
in the left, middle and right panels.
}
\end{figure}

The systematic study of the ground and excited states of 
Be isotopes was performed with VAP calculations in the AMD model (AMD-VAP).
Many rotational bands having the 2$\alpha$ core 
structure were obtained in the theoretical results.
In Fig.~\ref{fig:be-dense}, density distributions of 
the intrinsic wave functions for the band-head states of
$^{10}$Be and $^{12}$Be are shown.
As is seen, the proton-density distribution
indicates the formation of  
the 2$\alpha$ core, while 
the neutron density distribution exhibits the behavior of 
valence neutrons around the 2$\alpha$ core.

\begin{figure}[th]
\centerline{\includegraphics[width=12cm]{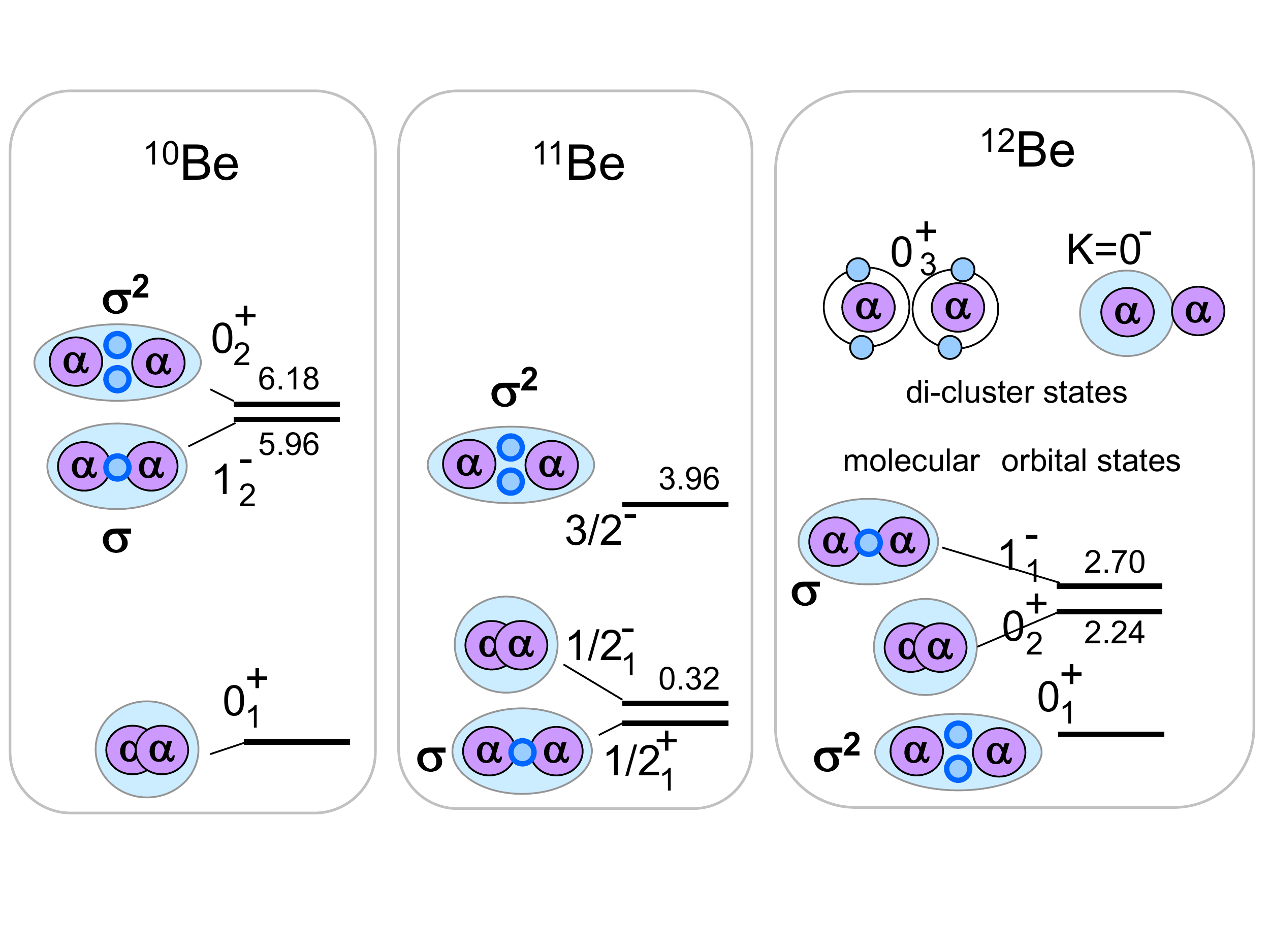}}
\vspace*{-0.5cm}
\caption{\label{fig:be-cluster}
Schematic figures for cluster states
suggested in $^{10}$Be, $^{11}$Be, and $^{12}$Be. 
For the molecular-orbital states, 2 $\alpha$ cores and 
the valence neutrons in the $\sigma$ orbital are illustrated.
The experimental values of the excitation energies are also shown.}
\end{figure}

In analysis of single-particle wave functions in the AMD wave functions, 
valence neutron orbits in low-lying states of Be isotopes were found to be 
associated with the molecular orbitals around the 2$\alpha$ core.
In Fig.~\ref{fig:be-cluster}, we show schematic figures of the cluster states
suggested in $^{10}$Be, $^{11}$Be, and $^{12}$Be. 
In the figure, we show the number of neutrons occupying the $\sigma$-like
orbitals, which have dominant positive-parity components and are 
regarded as the $\sigma$ orbital.
The experimental value for the excitation energies of the 
corresponding states are also shown in the figure.

\begin{figure}[th]
\centerline{\includegraphics[width=10cm]{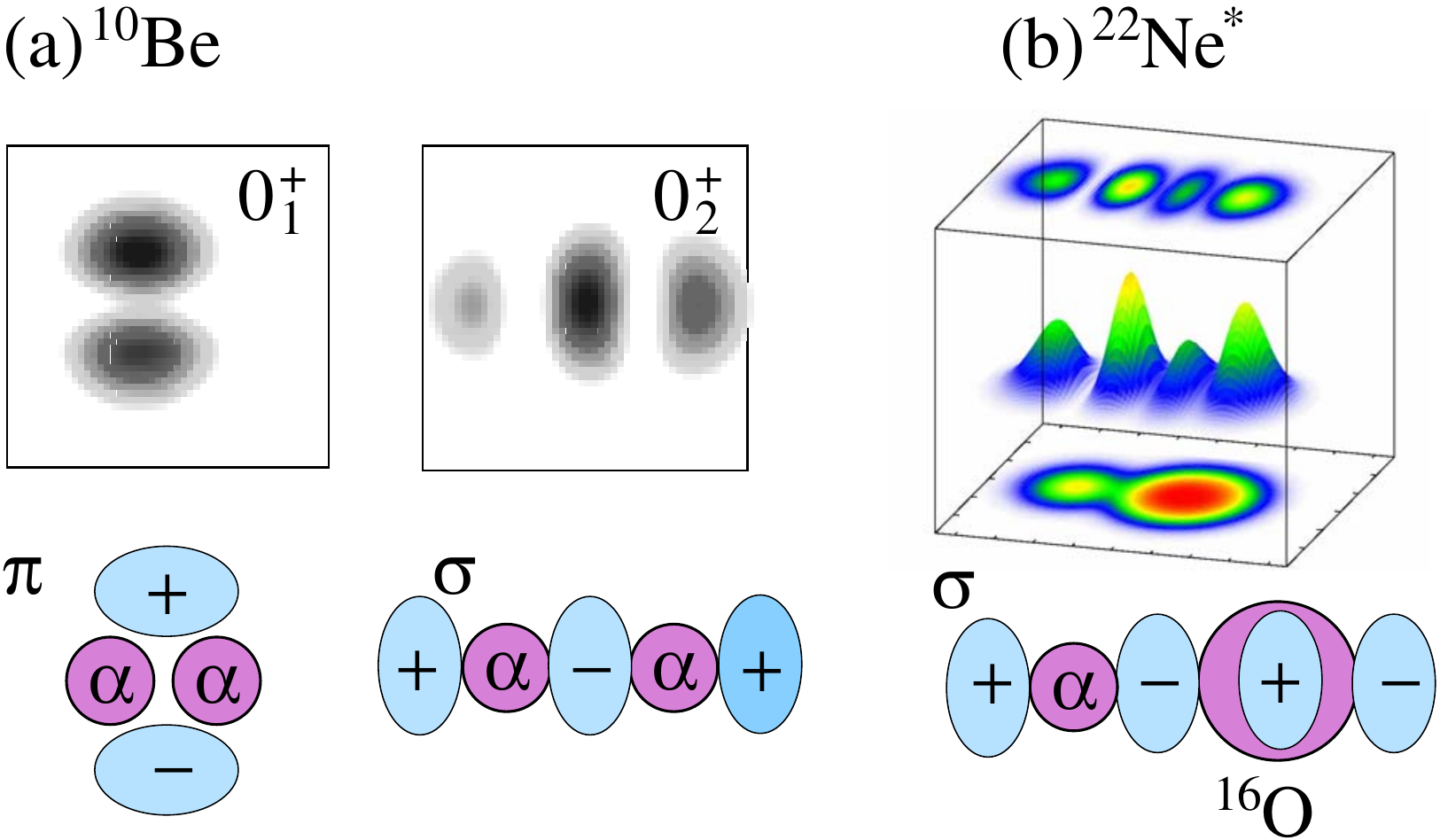}}
\vspace*{8pt}
\caption{\label{fig:orbit}
(a) Density distributions of the  
single-particle wave functions for valence neutrons in 
$^{10}$Be($0^+_1$) and $^{10}$Be($0^+_2$) \cite{ENYObe10}. 
Schematic figure of the molecular orbitals, $\pi$
and $\sigma$ orbitals around the 2$\alpha$ core are also shown at the bottom. 
(b) Density distribution of the excited band ($K^\pi=0^-$) in $^{22}$Ne obtained by AMD \cite{k:kimurane}. 
The middle and top figures show the density distribution of the 
single-neutron wave function of the highest single-particle level.
The matter density of the total system is displayed at the bottom of the box.
}
\end{figure}

In $^{10}$Be,
the valence neutron configurations of 
$^{10}$Be($0^+_1$), $^{10}$Be($1^-$), and $^{10}$Be($0^+_2$) can be regarded as 
$\pi^2$,  $\pi\sigma$, and $\sigma^2$ configurations meaning two neutrons in the $\pi$
orbitals, one neutron in the $\pi$ and the other neutron in the $\sigma$, and 
two neutrons in the $\sigma$, respectively. 
Spatial distributions of the single-particle orbits
of valence neutrons in $^{10}$Be($0^+_1$) and    
$^{10}$Be($0^+_2$) are shown in Fig.~\ref{fig:orbit}(a), where
the $\pi$-like orbital and the $\sigma$-like orbital are clearly seen.
Similarly to $^{10}$Be, it was found that 
the 
$^{11}$Be($1/2^+$),$^{11}$Be($1/2^-$), and $^{11}$Be($3/2^-_2$) states
correspond to $\pi^2\sigma$,  $\pi^3$  and $\pi\sigma^2$ configurations,
while the $^{12}$Be($0^+_1$), $^{12}$Be($0^+_2$), and $^{12}$Be($1^-_1$) states
are roughly interpreted as $\pi^2\sigma^2$,  $\pi^4$  and $\pi^3\sigma^1$ configurations. 

Interestingly, the degree of the $2\alpha$-cluster development 
strongly correlates with the number of valence neutrons in the 
$\sigma$ orbital. Namely, the 2$\alpha$ cluster 
develops as the neutron number in the $\sigma$ orbital increases. 
It is easily understood because 
the single-particle energy of the $\sigma$ orbital decreases because of the 
kinetic energy gain in largely distant $2\alpha$ systems.
The enhancement of the $2\alpha$ cluster with neutrons in
the $\sigma$ orbital is consistent with the arguments in 
Refs.~\citen{OERTZEN,OERTZENa,ITAGAKI}. 
On the other hand, as the neutron number in the $\pi$ orbitals increases, 
the cluster structure tends to weaken. 

Another interesting characteristic in Be isotopes is the breaking of neutron magicity in 
 $^{11}$Be and $^{12}$Be. 
The breaking of the $p$ shell for the neutron magic number $N=8$ in $^{11}$Be
has been experimentally known from the unnatural parity $1/2^+$ ground state, while 
that in $^{12}$Be has been suggested from slow $\beta$ decay\cite{suzuki97}. Those exotic features 
of $^{11}$Be and $^{12}$Be can be understood 
from the molecular orbital picture. 
The ground states of $^{11}$Be
and $^{12}$Be are considered to have dominant intruder configurations with $\sigma$-orbital neutron(s)
instead of normal $0\hbar\omega$ configurations.
The ground $1/2^+$ state of $^{11}$Be corresponds 
to the $\pi^2\sigma$ configuration, while $^{12}$Be($0^+_1$) is the 
intruder state $\pi^2\sigma^2$ in terms of molecular orbitals.
In the one-center shell-model limit, the $\pi$ and the $\sigma$ orbitals
correspond to the $p$ and $sd$ orbits. Therefore, in the 
ground states, $^{11}$Be($1/2^+_1$) and $^{12}$Be($0^+_1$) have dominant  
$1\hbar\omega$ and $2\hbar\omega$ configurations, respectively, indicating the 
vanishing of the $N=8$ magic number in $^{11}$Be and $^{12}$Be. 
The breaking of the neutron shell in neutron-rich Be isotopes is caused by the
lowering $\sigma$ orbital in 
the developed 2$\alpha$ structures as 
discussed in Refs.~\citen{AMDrev,ENYObe12,ENYObe11}.
Again, the $\sigma$ orbital in the 2$\alpha$ structure plays an important role.

In addition to molecular-orbital structures in such low-lying states, the AMD results for $^{12}$Be
suggested
molecular resonant states having di-cluster $^6$He+$^6$He and $^8$He+$\alpha$  structures
in highly excited states  \cite{ENYObe12}.
The result is consistent with the experimental observations of cluster states in He+He break-up reactions
\cite{FREER,SAITO04} and also with theoretical suggestions by cluster model calculations 
\cite{Descouvemont01,Ito:2008zza,Ito:2000de}.

\subsection{Three-body cluster states in $^{12}$C, $^{11}$B, and $^{14}$C.}
One of the typical examples where cluster and shell features coexist is $^{12}$C. 
The ground state of $^{12}$C is an admixture of 3$\alpha$-cluster and $p_{3/2}$-shell closure structures.
One the other hand, a variety of  3$\alpha$-cluster states have been suggested in excited states
in many $3\alpha$ model calculations since 1970's \cite{Fujiwara80}. 
Recently, Tohsaki {\it et al.} have proposed a new concept of cluster structure in the 
second $0^+$ state of $^{12}$C, where three $\alpha$ clusters are weakly interacting like 
a gas \cite{Tohsaki01,Funaki:2003af}. 
Because of the bosonic behavior of $\alpha$ particles in a dilute 3$\alpha$ gas state
this phenomenon has been discussed in relation with Bose-Einstein Condensation 
of $\alpha$ particles in a dilute nuclear matter \cite{Ropke98}. It is a challenging issue to search 
for such dilute cluster gas-like states in other nuclei, for instance, $^{11}$B and $^{13}$C.
Another interesting problem to be solved is a linear-chain 3$\alpha$ structure \cite{morinaga66}. 
It has been a long-standing 
problem whether a linear-chain 3$\alpha$ state appears in excited states of $^{12}$C or not. 
Possibility of linear-chain structures in neutron-rich 
C isotopes is also attracting a great interest as it might be stabilized by valence neutrons.

Applying the AMD method to $^{12}$C, $^{11}$B, and $^{14}$C, 
we have found various three-body cluster structures in their excited states as well as shell-model structures in low-lying states. 
Some are weakly interacting three-cluster states, and some shows rather geometric configurations of clusters.
We here discuss cluster aspects of these nuclei based on the AMD-VAP calculations for $^{12}$C and $^{11}$B\cite{ENYOe,Enyo-c12v2,Enyo-c11}
and the $\beta$-$\gamma$ constraint AMD calculations  for $^{14}$C \cite{Suhara-c14}. 

\subsubsection{Cluster structures of $^{12}$C} 

\begin{figure}[th]
\centerline{\includegraphics[width=12cm]{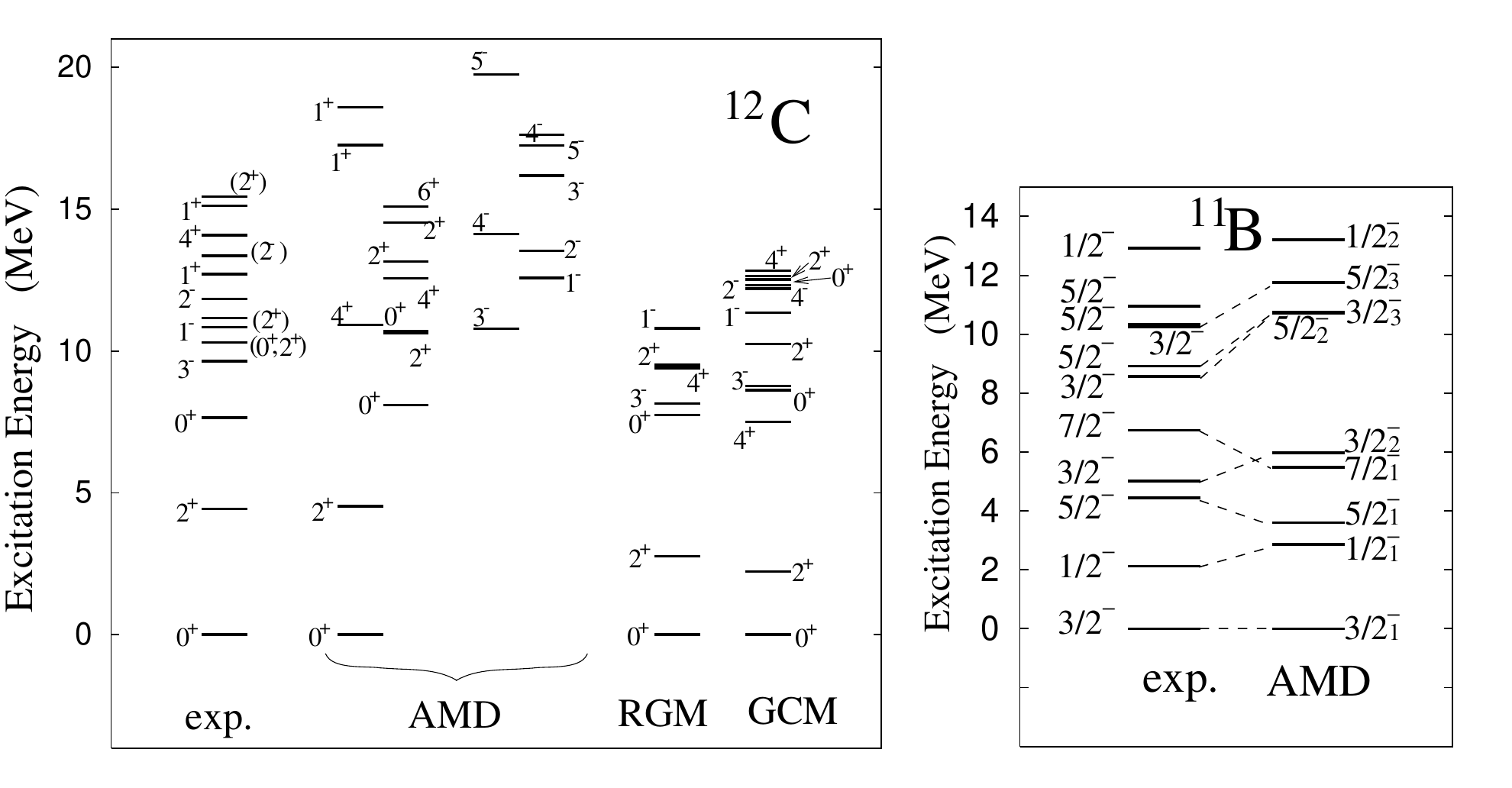}}
\vspace*{8pt}
\caption{\label{fig:c12-b11-spe}
Left: Energy levels of $^{12}$C calculated with AMD-VAP\cite{Enyo-c12v2}.
The theoretical levels of $^{12}$C calculated with the $3\alpha$RGM\cite{RGM} and
$3\alpha$GCM\cite{GCM} are also shown.
Right: Energy levels of $^{11}$B calculated with AMD-VAP\cite{Enyo-c11}. 
}
\end{figure}

\begin{figure}[th]
\centerline{\includegraphics[width=7cm]{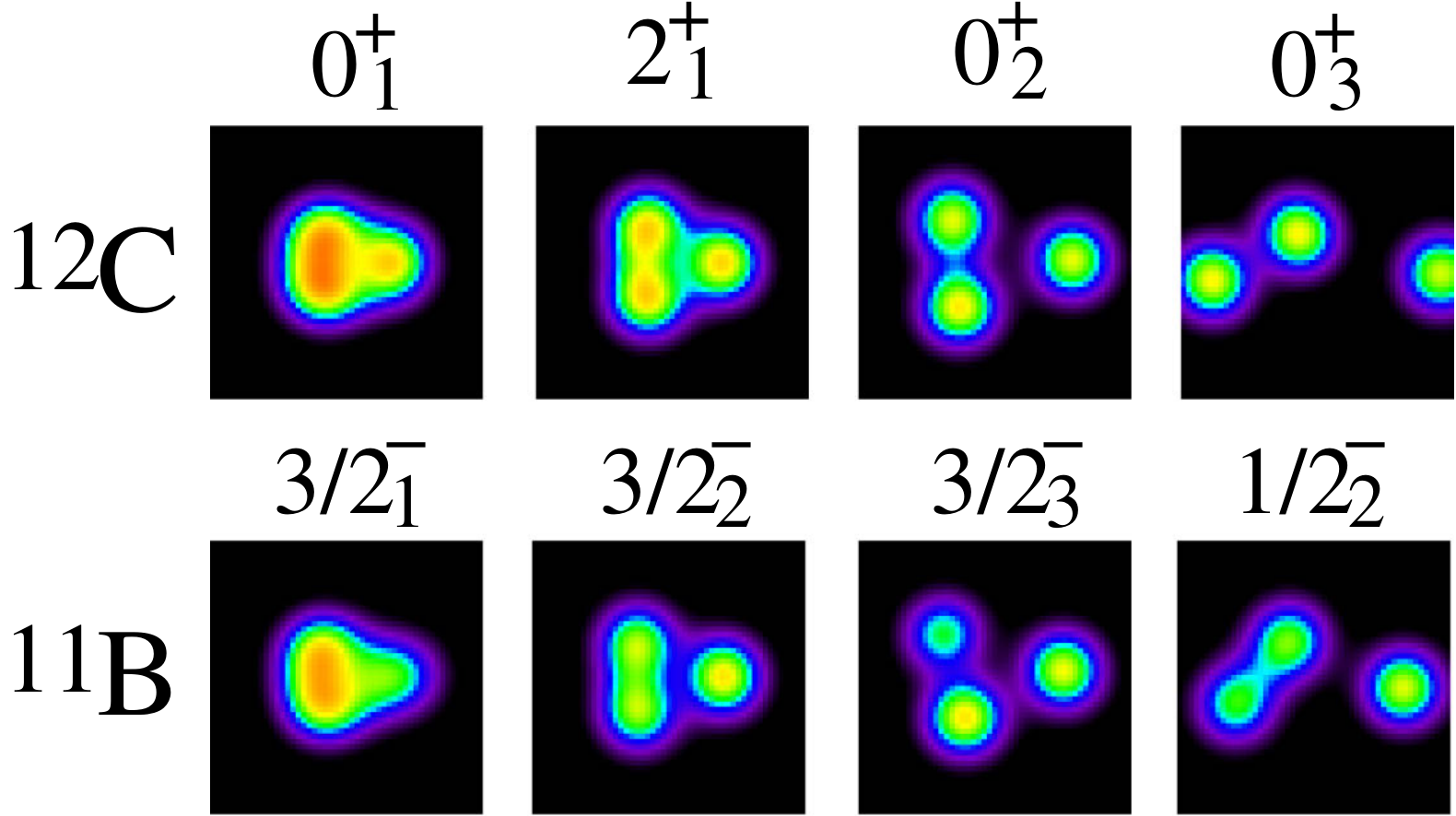}}
\vspace*{8pt}
\caption{\label{fig:b11-c12-dense}
Density distributions of intrinsic states for the ground and excited states of $^{12}$C and $^{11}$B
calculated by AMD-VAP \cite{Enyo-c12v2,Enyo-c11}.}
\end{figure}

To investigate excited states of $^{12}$C we performed the AMD-VAP calculations.
As already mentioned, in the AMD model, all nucleons are independently treated 
without assuming existence of any clusters. 
The formation or breaking of shell structure and clusters 
is expressed in the twelve-nucleon dynamics after the energy variation.
Even though any clusters are not {\it a priori} assumed, $\alpha$ clusters are formed in many excited states 
of $^{12}$C. Indeed, a variety of 3$\alpha$ cluster states were obtained in the AMD results.
The experimental and calculated energy levels of $^{12}$C are shown in Fig.~\ref{fig:c12-b11-spe}, and 
density distribution of intrinsic wave functions for the ground and excited states 
is shown in the upper row of Fig.~\ref{fig:b11-c12-dense}. 
It should be stressed that this is the first calculation that succeeded in simultaneously reproducing the energy spectra of the ground band 
and those of excited states having developed $3\alpha$ cluster structures in $^{12}$C. The success owes to 
flexibility of AMD wave functions which can describe both shell-model and cluster structures. 
The calculation also reproduces 
well transition properties such as $E2$, monopole, and GT transitions. 
In the result, the ground state shows the $p_{3/2}$-shell closure
configuration with a mixing of the 3$\alpha$-core component, 
while the second $0^+$ state has a well-developed $3\alpha$ cluster structure.  
Since the $0^+_2$ wave function has large overlap with various $3\alpha$ configurations
(amplitudes of the superposed wave functions do not concentrate on a specific AMD wave function but
they fragment largely into various AMD wave functions), 
this state is regarded as the $3\alpha$ gas-like state, where three $\alpha$ clusters are rather freely moving in 
a dilute density.  
The 3$\alpha$ chain-like state was suggested in the third $0^+$ state around 10 MeV in the 
AMD result.
It is not a "linear" chain but it shows an obtuse triangle 3$\alpha$ configuration as seen in the figure. 
These results are quite similar to those calculated with FMD with the unitary correlation operator method (UCOM) \cite{Chernykh:2007zz}.

\subsubsection{Cluster structures of $^{11}$B}

We applied the same method, AMD-VAP, to $^{11}$B and its mirror nucleus $^{11}$C,
and investigated the structures while focusing on cluster aspect.
The energy levels of negative-parity states of $^{11}$B were reasonably reproduced by 
the calculations as shown in Fig.~\ref{fig:c12-b11-spe}.

The $3/2^-_3$ states at the excitation energy $E^*\sim$8 MeV, 
$^{11}$B$(3/2^-_3$, 8.65 MeV) and $^{11}$C$(3/2^-_3$, 8.10 MeV)
are experimentally known to have abnormal properties such as 
weak GT and $M1$ transitions compared with normal low-lying states. 
It is also interesting that the $3/2^-_3$ state is missing in 
shell-model calculations for $^{11}$B. These facts suggest that the $3/2^-_3$ 
may not be ordinary shell-model-like state but may have a developed cluster structure.
The AMD calculations for $^{11}$B give good results for the energy levels including 
the $3/2^-_3$ state (Fig.~\ref{fig:c12-b11-spe}) and reproduce well  
the experimental values of transition strengths.
For the $3/2^-_3$ state, the quenchings of GT and $M1$ transitions 
are understood because of the developed cluster structure of the $3/2^-_3$ state
which has small overlap with the low-lying shell-model-like states.
Indeed, the $3/2^-_3$ states of $^{11}$C and $^{11}$B 
exhibit the remarkably developed $2\alpha$+$^3$He and $2\alpha$+$t$ 
clustering (Fig.~\ref{fig:b11-c12-dense}). 

\subsubsection{Analogy of cluster aspects of  $^{11}$B to $^{12}$C}
Comparing the results for $^{11}$B with those for $^{12}$C, we found good analogies of cluster aspects 
between $^{11}$B and $^{12}$C. As shown in Fig.~\ref{fig:b11-c12-dense},
the ground state of $^{11}$B is described by the $p_{3/2}$-shell configuration with a mixing of 
cluster structure as well as that of $^{12}$C. The development of the $2\alpha$+$t$-cluster core in the  $^{11}$B($3/2^-_2$) 
shows a good analogy to that of the $3\alpha$-cluster core in the $^{12}$C($2^+_1$). 
The remarkably developed $2\alpha$+$t$-cluster structures in  the $^{11}$B($3/2^-_3$)  and 
$^{11}$B($1/2^-_2$) can be associated with the developed $3\alpha$ cluster in  the $^{12}$C($0^+_2$)
and $^{12}$C($0^+_3$), respectively.

Particular attention is paid to analogy of  $^{11}$B($3/2^-_3$) to $^{12}$C($0^+_2$). 
Similarly to the case of $^{12}$C($0^+_2$), 
the $^{11}$B$(3/2^-_3)$ wave function has large overlap with various $2\alpha$+$t$ configurations
indicating that the state has no geometric cluster configuration but it should be
regarded as a weakly interacting $2\alpha$+$t$-cluster state.
The root-mean-square radius (r.m.s.r.) 
of the $^{11}$B$(3/2^-_3)$ state is 3.1 fm and it is remarkably large compared
with that of the ground state (2.5 fm). 
Considering amplitudes of the wave function fragmented on various configurations
and the large radius, 
the $3/2^-_3$ state may be a $2\alpha$+$t$-cluster state
with a dilute density like a gas, where clusters are rather freely moving. 
It should be noted that, for the $^{11}$B$(3/2^-_3)$, the energy position relative to 
the three-cluster break-up threshold is lower and the nuclear size is smaller than those
for $^{12}$C($0^+_2$), and therefore the gas-like feature of the $^{11}$B$(3/2^-_3)$
might be weaker than the $^{12}$C($0^+_2$).

Another analogy between $^{11}$B$(3/2^-_3)$  and $^{12}$C($0^+_2$)
is remarkable monopole transition strengths from the ground state. 
The calculated iso-scalar monopole strength $B(IS0)$ for the transition
$3/2^-_1\rightarrow 3/2^-_3$ is 94 fm$^4$ and the value 
is in good agreement with the experimental value 94$\pm$16 fm$^4$\cite{Kawabata06}.
The $B(IS0)$ value is as large as that for the monopole transition
$0^+_1\rightarrow 0^+_2$ in $^{12}$C.

Recently, structures of $^{11}$B has been investigated also by
$2\alpha$+$t$ orthogonality condition model (OCM)  \cite{yamada-b11}.
In the results of $2\alpha$+$t$ OCM, different cluster features 
between $^{11}$B$(3/2^-_3)$ and $^{12}$C($0^+_2$) have been pointed out from the point of view of 
$\alpha$ condensation. Cluster gas features of $^{11}$B are under discussion.

\subsubsection{Cluster structures of $^{14}$C}
As mentioned, the straight-line chain structure of three $\alpha$ clusters 
may not be stable in $^{12}$C even though the linear-chain-like
$3\alpha$ state with the obtuse triangle configuration
might exist in the $0^+_3$ state.
We here consider cluster structures of $^{14}$C, in which
3$\alpha$-core structures with additional two neutrons are expected.

\begin{figure}
\centerline{\includegraphics[width=10cm]{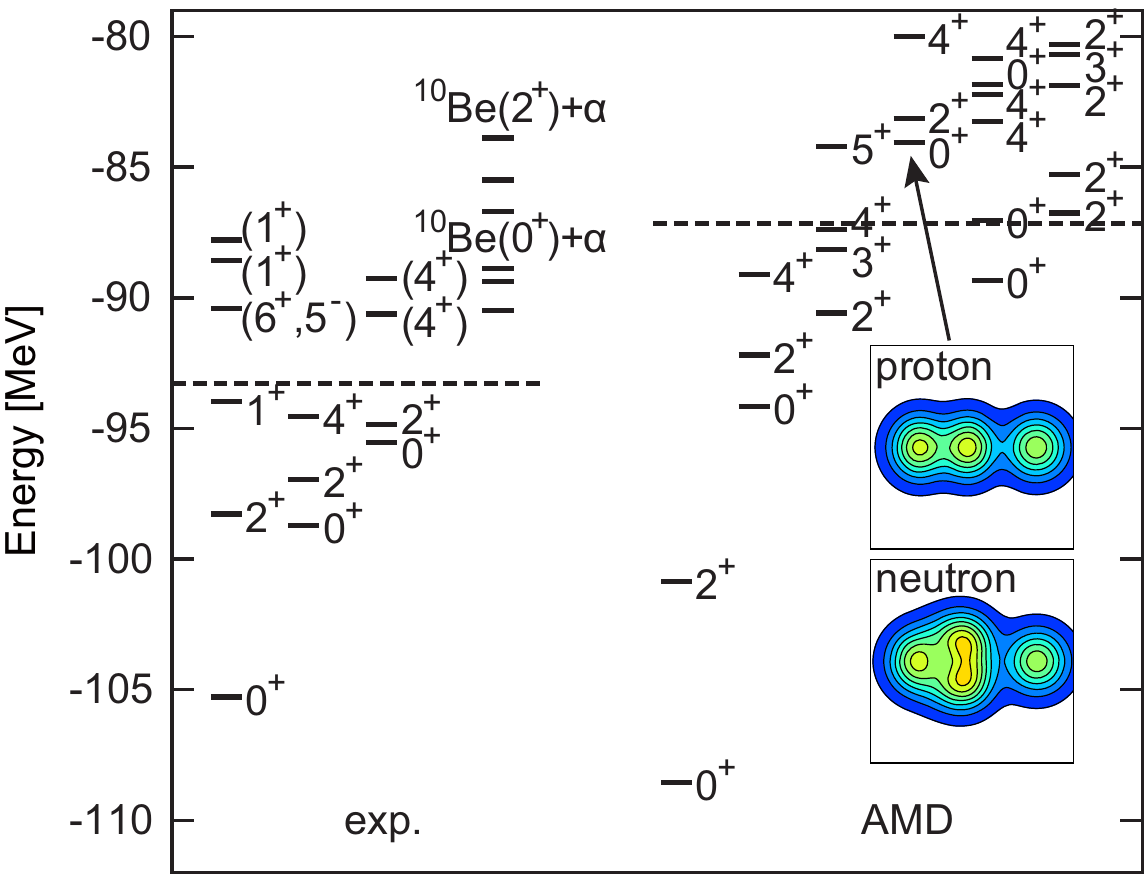}}
  \caption{Energy levels of $^{14}$C. The theoretical data are those in Ref.~\citen{Suhara-c14} 
calculated with $\beta-\gamma$ constraint AMD with GCM by using the Volkov No.2 force with $B=H=0.125$,
$M=0.60$ and the G3RS-type spin-orbit force with $u_I=-u_{II}=1600$ MeV. 
Proton and neutron density distributions for the linear-chain $K^\pi=0^+$ band are also shown.
The dotted lines are the experimental and theoretical $^{10}$Be+$\alpha$ threshold energies.
The figure is taken from Ref.~\citen{Suhara-private}.
}
\label{fig:c14}
\end{figure}

Structures of excited states of $^{14}$C were investigated with a method of $\beta$-$\gamma$ 
constraint AMD in combination with GCM by Suhara and one of the authors (Y.~K-E.) \cite{Suhara-c14}.
We stress again that existence of clusters was not assumed in the model but
dynamics of fourteen nucleons was solved in the AMD model space.
The results suggested a variety of developed 3$\alpha$-cluster core structures 
in excited states. One of the new findings is that a
3$\alpha$ linear-chain structure with valence neutrons can be stabilized in $^{14}$C and
may construct a $K^\pi=0^+$ rotational band above the $^{10}$Be+$\alpha$ threshold energy (Fig.~\ref{fig:c14}).
As shown in the density distributions of protons and neutrons, the linear-chain state indicates 
a strongly coupling $^{10}$Be+$\alpha$ cluster structure, where an $\alpha$ cluster is sitting on the head
of a deformed $^{10}$Be cluster. It was found that additional neutrons play an important role to stabilize the
linear-chain configuration.
Unfortunately, there is no experimental evidence for the linear-chain state.
The $^{10}$Be+$\alpha$ decay observations \cite{price07,Haigh:2008zz} would be helpful 
to identify it.

\subsection{Spectroscopy and exotic phenomena in the island of inversion}
Neutron-rich Ne and Mg isotopes around the $N\sim 20 $ region called ``island of
inversion'' are known to have anomalous properties \cite{War90}. These anomalies originate
in the quenching of the $N=20$  shell gap in 
the neutron-rich nuclei and it leads to the breakdown of the magic number $N=20$ and large
nuclear deformation.  This drastic change of nuclear shell structure has been  intensively
investigated mainly focusing on the spectral properties of yrast states 
\cite{Thi75,Hu78,Mot95,Pov86,Fuk92,Uts99}. Recently
thanks to the development of the experimental technique, the information of the non-yrast
states \cite{Tak09,Sch09,Wan10,Wim10} is rapidly increasing to reveal exotic phenomena
peculiar to the island of inversion. For example,  ``the coexistence of spherical and
deformed shapes'' and ``the coexistence of normal and intruder configurations'' have been
discussed based on the finding of the second excited $0^+$ states of $^{30}$Mg \cite{Sch09}
and $^{32}$Mg \cite{Wim10}.   

AMD combined with GCM is one of the powerful theoretical approaches to investigate the
non-yrast states of nuclei in the island
  of inversion as well as the yrast states
\cite{AMDrev,AMDrev2, Kim02,Kim04,Kim07,Kim11}.
 We here introduce some recent AMD studies for shape coexistence, one-neutron halo and molecular
structure in the island of inversion. 

\subsubsection{Many-particle and many-hole states and shape coexistence in $^{31}$\rm Mg}

  The neutron orbits and the coexistence of normal and intruder
  configurations in neutron-rich Ne and Mg isotopes are most
  sensitively probed by the low-lying yrast and non-yrast states of
  odd-mass isotopes.  The last neutron's orbit, which determines the
  spin and parity of the ground state, is quite sensitive to the
  nuclear deformation.

\begin{figure}[th]
\centerline{\includegraphics[width=\hsize]{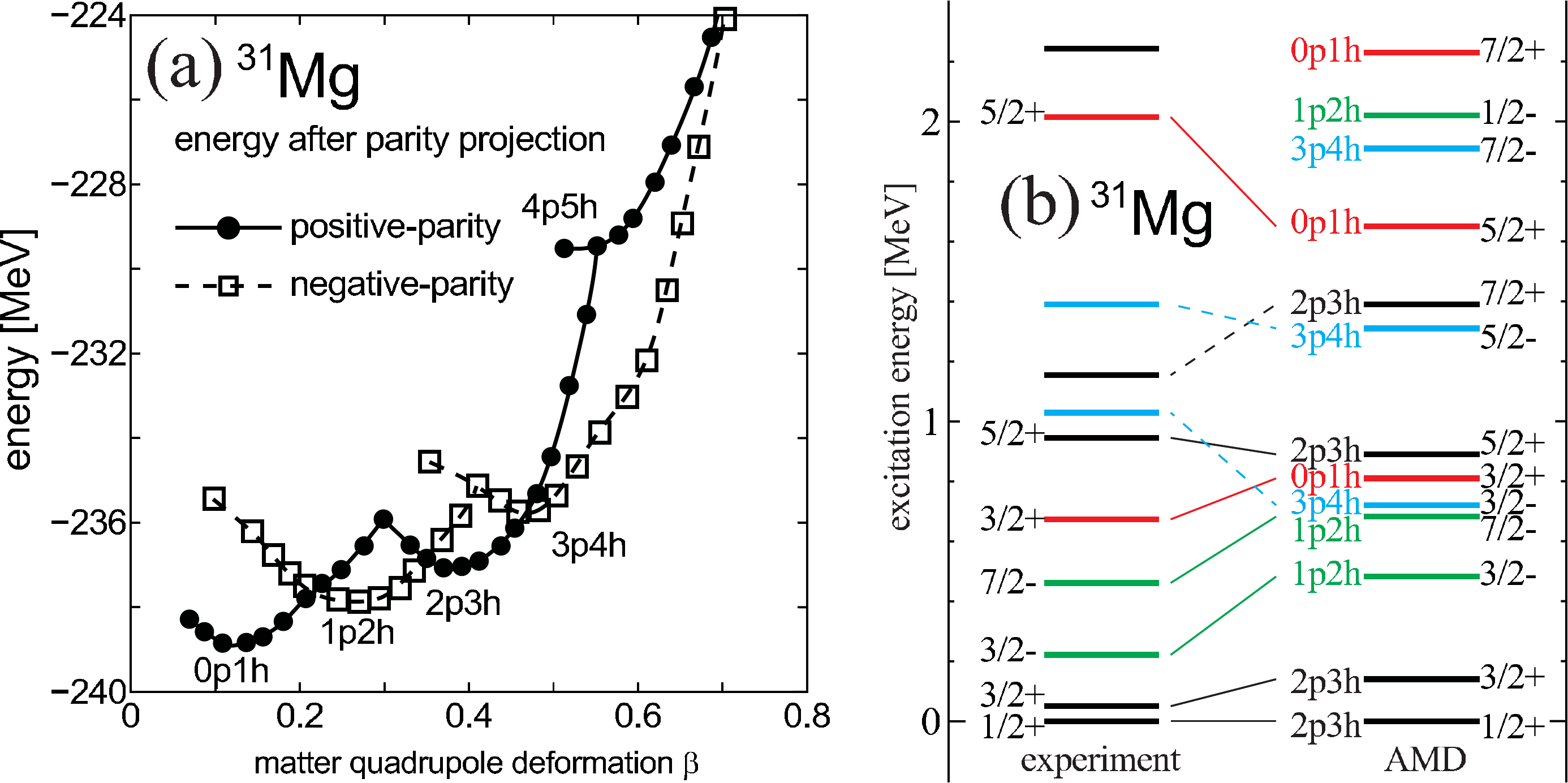}}
\caption{ (a) Energy surfaces of the positive- and negative-parity states of $^{31}$Mg as
 function of quadrupole deformation parameter  $\beta$ calculated by AMD. The AMD results are
 taken from Ref.~\citen{Kim07}. (b) Low-lying  spectrum of $^{31}$Mg calculated by
 AMD+GCM compared with the experimental assignment suggested in Ref.~\citen{Ney11}. 
}\label{fig:Mg31lev}
\end{figure}

 As one of such examples, Fig.~\ref{fig:Mg31lev} shows the calculated energy surface, and predicted
 \cite{Kim07} and observed  spectrum of $^{31}$Mg
 \cite{Ney05,Klo93,Mac05,Mat07,Sei11,Mil09,Ter08,Ney11}.  The energy surface of  
 $^{31}$Mg (Fig.~\ref{fig:Mg31lev} (a)) has  the low-lying local minima with different
 nuclear deformations. 
   By calculating the occupation probabilities of single-particle
   orbits, these local minima are found to
 correspond to the  $0p1h$, $1p2h$, $2p3h$ and $3p4h$ neutron
 configurations in ascending order of  deformation. Here the particle-hole  configurations
 are labeled relative to the $N=20$  shell closure and nuclear deformation becomes larger as
 numbers of particles in $pf$ shell  and holes in $sd$ shell
 increase. 
To predict the ground state configuration, it is essential to perform
a GCM calculation.  In fact,
AMD+GCM calculation
 (Fig.~\ref{fig:Mg31lev} (b)) shows 
 that the ground state is strongly deformed and an almost pure  $2p3h$ configuration in which two 
neutrons are promoted into $pf$ shell across the $N=20$ shell gap to break down the magic
 number.  The calculated magnetic moment of the ground state is   $-0.91\mu_N$, while the
observed value is $-0.88\mu_N$ \cite{Ney05}. Since the spherical $0p1h$  states in which the
 $N=20$ magicity is retained give positive magnetic moment,  the breakdown of the $N=20$ magic
number is confirmed  without ambiguity.   AMD predicts that the ground 
state is followed by $3/2^+_1$, $5/2^+_1$ and $7/2^+_1$ states with $2p3h$ configuration to
constitute the rotational ground band due to the strong deformation. Furthermore,  due to
the quenching of the $N=20$ shell gap, coexistence of three different configurations at small
excitation energy is also predicted. The $1p2h$ configuration appears as the $3/2^-_1$ and
$7/2^-_1$ states at very small excitation energies, the strongly deformed  $3p4h$
configuration constitutes the $K^\pi=3/2^-$ rotational band starting from 720 keV, and the
normal $0p1h$ configuration appears as the $5/2^+_2$ state at 1.6 MeV.  
 Most of these excited states have been observed, in good agreement
 with AMD predictions, by the measurements of $\beta$-decays
 \cite{Klo93,Mac05,Mat07}, one proton or neutron knockout reactions
 \cite{Ter08,Mil09} and Coulomb excitation \cite{Sei11}.
 Thus, the coexistence of various
 neutron configurations and  deformed states is now established.  For the recent 
 discussions on other nuclei,  readers are directed to Ref.~\citen{Kim11}.   

\subsubsection{Neutron-halo with a deformed core in the island of inversion}
As one of the fascinating phenomena in the island of inversion, we focus on the one-neutron halo
structure of $^{31}$Ne. Recent experiments at RI Beam Factory in RIKEN 
have revealed the large Coulomb breakup
cross section \cite{Nak09} and interaction cross section \cite{Tak10-1,Tak10-2} of
$^{31}$Ne, and the $p$- or $s$-wave neutron-halo structure has been suggested from the analysis
of the Coulomb breakup \cite{Nak09,Hor10}. Usually neutron-halo structure 
has been discussed based on the ``spherical inert core + weakly bound neutron'' models
as done for $^{6}$He and $^{11}$Li.
However, in this case the core nucleus $^{30}$Ne is located in the middle of the
island of inversion and the assumption of the spherical inert core is inadequate. The last
neutron of $^{31}$Ne may be coupled to the strongly deformed core with broken magic number. Therefore, the analysis based on a full microscopic theory is more suitable and
necessary.  
\begin{figure}[th]
\centerline{\includegraphics[width=\hsize]{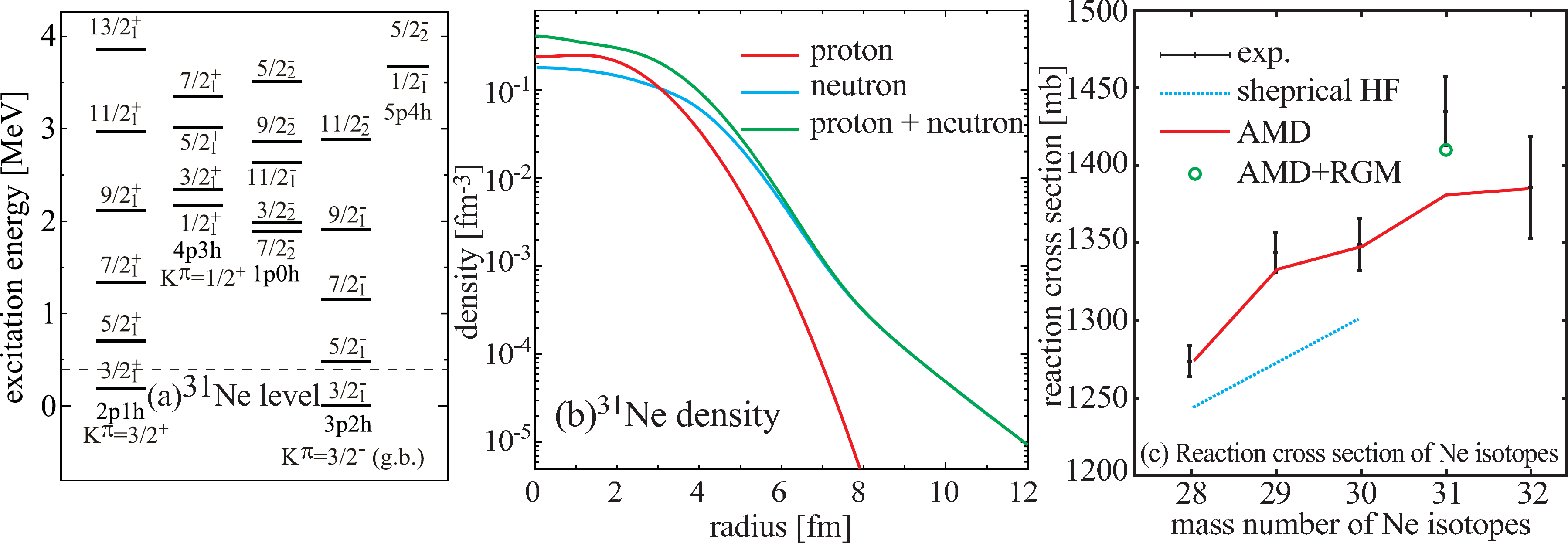}}
\caption{ (a) Low-lying spectrum of $^{31}$Ne calculated by AMD+RGM. Dashed line shows one
neutron threshold energy. (b) Proton, neutron, and proton + neutron densities of
$^{31}$Ne ground state ($J^\pi = 3/2^-_1$) calculated by AMD+RGM. (c) Reaction cross section
of Ne isotopes calculated by double-folding model using the density distribution obtained by
 AMD (solid line), AMD+RGM (open circle for $^{31}$Ne) compared with the observed data
 \cite{Tak10-1,Tak10-2}. The results are taken from Refs.~\citen{Kim11,Min11}.}\label{fig:Ne31dens}
\end{figure}
Since a single-particle wave function is represented by a Gaussian wave packet, AMD cannot
describe properly the tail part of halo nucleus which shows the exponential damping. This
shortage is overcome by combining AMD with resonating group method  (AMD+RGM). We use the 
RGM-type wave function for $^{31}$Ne, 
\begin{align}
 \Psi_{JM}&=\sum_{\alpha l}c_{\alpha l}{\cal A}\left[\chi_{\alpha l}(r)Y_{lm}(\hat r)\phi_n\phi_{^{30}\rm
 Ne}(\alpha)\right]_{JM}, \label{eq:RGM}
\end{align}
where $\phi_n$ and  $\chi_{\alpha l}(r)Y_{lm}(\hat r)$ are the
spin wave function of the valence
neutron and the relative motion between the valence neutron and $^{30}$Ne,
respectively. $\phi_{^{30}\rm Ne}(\alpha)$ is the  internal wave function of 
$^{30}$Ne solved by AMD+GCM \cite{Kim04}, and $\alpha$ labels the ground and excited states of
$^{30}$Ne. Since the AMD wave function of $^{30}$Ne is a superposition of Slater determinant of
Gaussians, the coefficients $c_{\alpha l}$ and the relative wave function $\chi_{\alpha l}(r)$
can be solved by transforming Eq.~(\ref{eq:RGM}) into a
GCM type wave function.  Recent development of high-performance computing enabled this CPU demanding calculation.

Figure \ref{fig:Ne31dens} (a) shows the predicted spectrum of $^{31}$Ne by AMD+RGM. The
ground state is the $3/2^-$ state and the calculated one neutron separation energy is
$S_n=0.45$ MeV, while the observed value is 0.29$\pm$1.64 MeV \cite{Jur07}. Due to this small
separation energy,  the density distribution of $^{31}$Ne calculated by AMD+RGM shows the
long tail at large distance indicating a $p$-wave halo ($1p_{3/2}$-neutron coupled to
$^{30}$Ne) as shown in Fig.~\ref{fig:Ne31dens} (b).  The effect of the 
deformed core appears as the core excitation. From the coefficients $c_{\alpha l}$, it is found
that the $p_{3/2}\otimes {}^{30}{\rm Ne}(2^+)$ configuration amounts to 41\%, that is larger
than  the $p_{3/2}\otimes {}^{30}{\rm Ne}(0^+)$ configuration which amounts to  37\%. These
values suggest that the $p_{3/2}$ neutron is coupled to the deformed and rotating ground
band of $^{30}$Ne. Using AMD and AMD+RGM wave functions, the reaction cross sections of
$^{31}$Ne and other Ne isotopes are analyzed based on the double-folding
model \cite{Min11, Min12}. Figure \ref{fig:Ne31dens} (c) compares the calculated and observed
reaction cross section. We can see that AMD wave function shows overall agreement with the
observation except for $^{31}$Ne, and anomalous large cross section of $^{31}$Ne is
reasonably described by employing the AMD+RGM wave function. Thus, with a help of
high-performance computing,  AMD combined with RGM and reaction theory is a promising method
to investigate neutron-halo nuclei and their reactions in heavier mass region. We can find
many other candidates of weakly bound system such as $^{35}$Mg and $^{37}$Mg from the
systematics of the binding energy, and analysis of them is now ongoing. 

\subsubsection{Molecule-like states in the island of inversion}
Another example of exotic phenomena in the island of inversion is the molecule-like
structure at highly excited region. As discussed in the subsection 3.1, several
excited states of O, F and Ne isotopes have been predicted to have molecule-like structure
analogous  to Be isotopes \cite{k:kimurane,Fur08}. Especially, the candidates of
molecule-like states have been recently observed in $^{18\sim 20}$O
\cite{Milin09,Oer10-1,Oer10-2} and found to qualitatively agree with the AMD 
predictions \cite{Fur08}. Thus the exploration of molecule-like structure is expanding to
heavier neutron-rich systems.  In the case of F and Ne
isotopes, since the neutron drip line is
farther extended than for O isotopes, we can expect  molecule-like structure with more
valence neutrons and more exotic phenomena. 
\begin{figure}[th]
\centerline{\includegraphics[width=0.8\hsize]{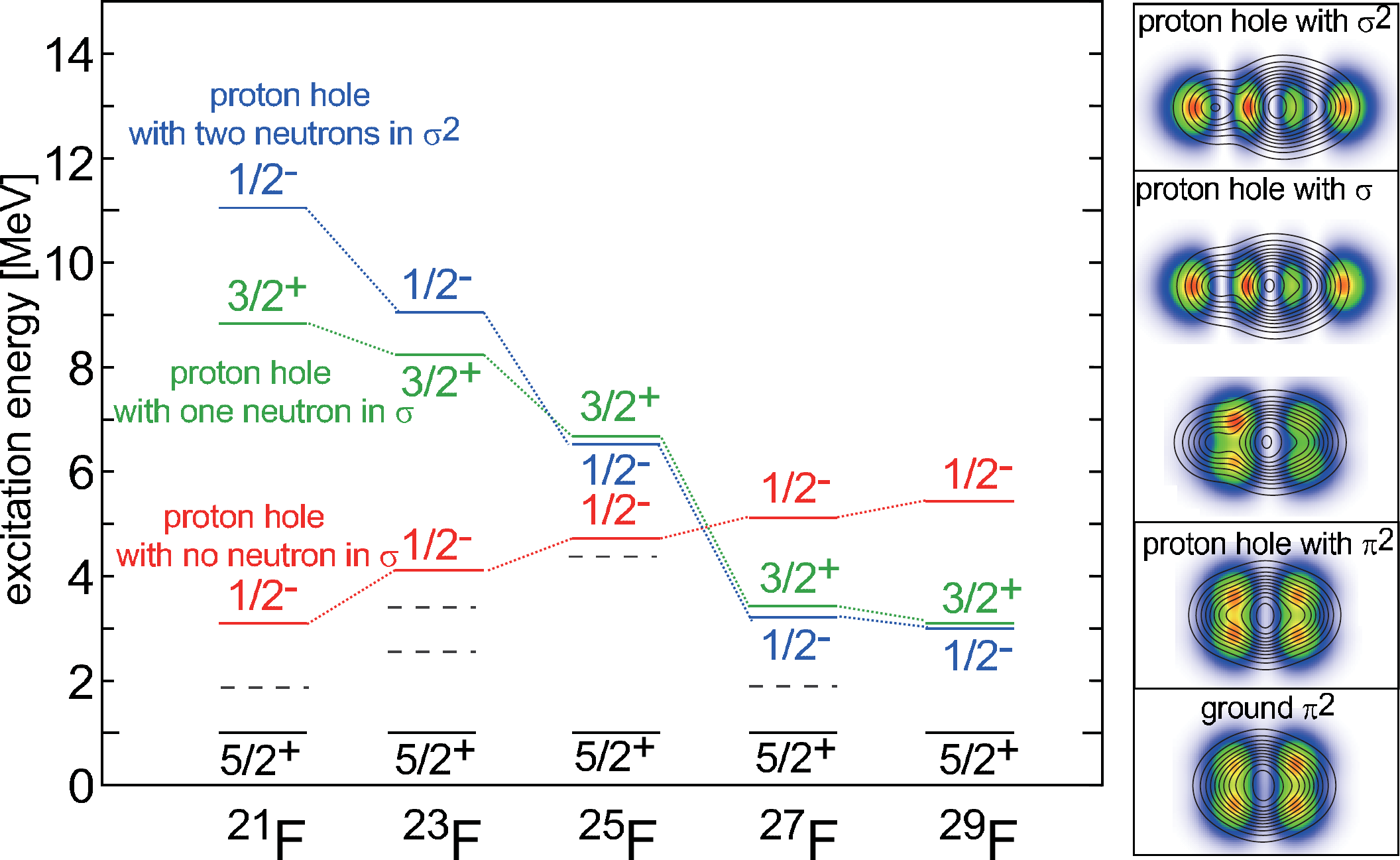}}
\caption{Left: Band-head energies of proton hole states of F isotopes. Red lines show the
 states with a proton excitation from $p$ to $sd$ shell and no neutron in $pf$ shell, while
 green (blue) lines show those with one (two) neutron(s) in $pf$ shell that have
 molecule-like structure. Right: Density distribution of the core ($^{19}$F, contour lines)
 and two valence neutrons (color plot)  of $^{21}$F. From top to bottom, each panel
 corresponds to the states with two, one and no neutrons in $pf$ shell with a proton hole and
 the ground state, respectively.
This figure is taken from Ref.~\citen{AMDrev2}.}\label{fig:Fiso}
\end{figure}

Figure \ref{fig:Fiso} shows the energies of the F isotopes obtained by AMD+GCM \cite{KimF10} as an
 example. All of these F isotopes have molecule-like states in their excited
states and there are always two kinds of molecular bands
 (green and blue lines in Fig.~\ref{fig:Fiso} left).  In these states,
 one or two valence neutrons are excited into $\sigma$ orbital ($pf$
 shell) together with a proton excitation from $p$ to $sd$ shell, and
 the  $\alpha$+N cluster structure develops
 (Fig.~\ref{fig:Fiso} right). It is notable   that the drastic reduction of their excitation
 energies toward the neutron drip line is  predicted.  To understand this reduction, readers are
 reminded following points.  (1) As mentioned in section 3.1, the $\sigma$ orbital
 originates in the $pf$ shell, and its energy is lowered in the island of inversion due to the
 quenching of the $N=20$ shell gap.  (2) If the core has cluster structure, it induces the
 deformation of the system and it further lowers the energy of $\sigma$ orbital as in the case
 of Be isotopes. Therefore, the neutron excitation into $\sigma$ orbital and $\alpha$
 clustering of the core work in a cooperative way to reduce the excitation energies of
 molecular states in the island of inversion. Up to now, several candidates for the molecular states
in lighter
 F isotopes are experimentally known \cite{Kas70,Mai81,Ele04}.
More  data  for F isotopes near the drip line will  be
 experimentally  available in near  future.

\subsection{Superdeformation in sd-shell nuclei}
A recent development in high-spin physics is the discovery of the superdeformed band at very
small mass region ($A\sim 40$) such as $^{36}$Ar \cite{Sve01}, $^{40}$Ca  \cite{Ide01} and
$^{44}$Ti  \cite{Ole03}.In contrast with heavier nuclei, these superdeformed bands are assigned
from high-spin states down to very low-spin states ($J^\pi=0^+\sim 4^+$). From this discovery, the
low-lying $0^+$ states known for a long time are now identified as the band-head of the
superdeformed bands. On the other hand, $\alpha$ clustering of low-lying states and
molecular structure of highly excited states have long been discussed. AMD studies have been
made to investigate properties of superdeformed states and to reveal the relationship
between the superdeformed states and cluster states. 
 
\subsubsection{Dual nature of superdeformed states and its evolution to the molecular states}
The strongly deformed excited state of $^{32}$S is of  particular interest because of the
following reasons. 1) $^{32}$S ($N=Z=16$) is a double magic nucleus of superdeformation. A
couple of mean-field calculations \cite{Rod00,Ina03,Ben03} have predicted the superdeformed state
around $E^*\sim 10$ MeV and  it has $4\hbar\omega$ excited configuration relative to the
ground state.  2) Using unique optical potential for $^{16}$O-$^{16}$O scattering
\cite{Nic99,Oer03}, 
$^{16}$O+$^{16}$O cluster model \cite{Ohk02} showed the presence of three molecular
bands. Among them the lowest energy band is located a few MeV below the $^{16}$O+$^{16}$O
threshold energy and coincides with the predicted superdeformed band mentioned above, while
the highest energy band nicely reproduces the well-known $^{16}$O+$^{16}$O molecular
resonances \cite{Gai81,Mor85}.  Thus the two different theoretical approaches give
qualitatively the same result and suggest the relationship between the superdeformed states
and cluster states. 

\begin{figure}[th]
\centerline{\includegraphics[width=\hsize]{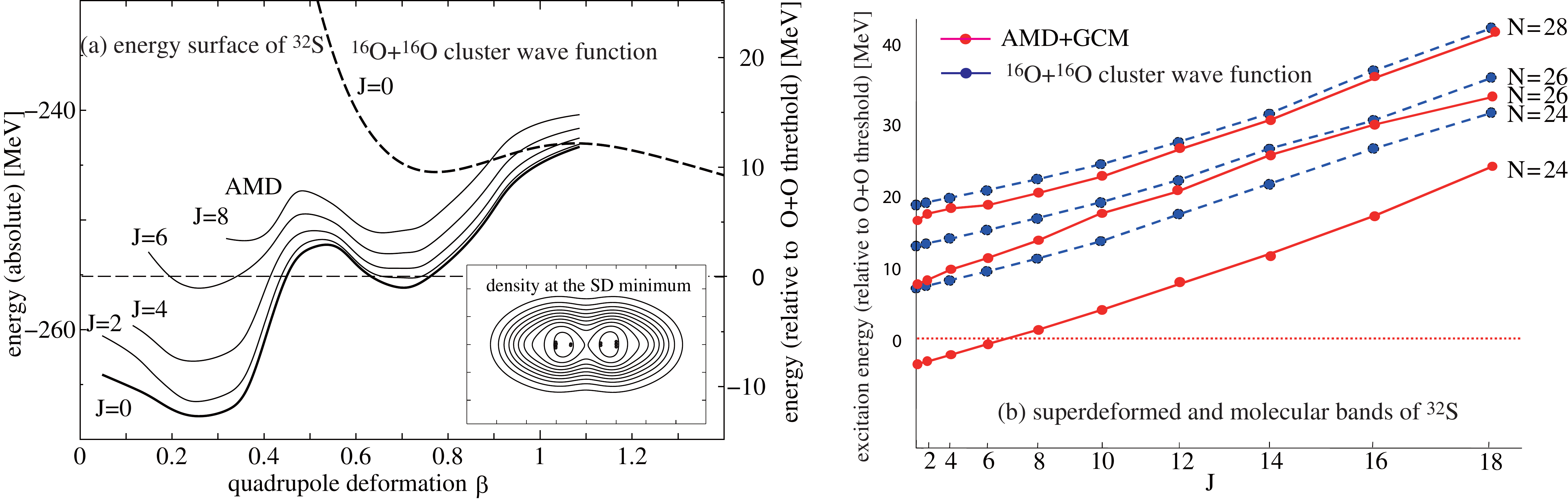}}
\caption{(a) Energy surface of $^{32}$S as function of quadrupole deformation $\beta$.
 Solid lines show the energy surfaces for each angular momentum obtained by AMD, while the
 dashed line shows that obtained by $^{16}$O+$^{16}$O cluster wave function. (b) Three
 rotational bands built around the superdeformed minimum obtained by the AMD+GCM (solid
 lines) and  $^{16}$O+$^{16}$O cluster model wave functions. The principal quantum number
 $N$ for each  rotational band is also shown. The figures are taken from 
Ref.~\citen{Kim04-2}.}\label{fig:S32} 
\end{figure}

AMD gives a unified understanding for the results of the mean-field and cluster model studies
and shows that the superdeformed band evolves to the $^{16}$O+$^{16}$O molecular
bands as the excitation energy increases \cite{Kim04-2}. Similar to the mean-field  
calculations, AMD predicts a well pronounced superdeformed minimum
(Fig.~\ref{fig:S32} (a)). The density distribution of the superdeformed wave function  clearly
shows $^{16}$O+$^{16}$O molecule-like structure and the distance between $^{16}$O+ $^{16}$O
increases as deformation becomes larger. 
Note that the superdeformed wave function is smoothly connected
to the $^{16}$O+$^{16}$O cluster wave  function around the Coulomb barrier
($\beta\sim$1.1) in AMD. 
Around the superdeformed
  minimum, the  $^{16}$O clusters are distorted by the formation of the
mean-field and by the spin-orbit interaction to gain more binding energy, while the pure $^{16}$O+$^{16}$O cluster structure is restored as inter-cluster
distance (deformation) becomes larger.  The AMD+GCM calculation has shown that three
rotational bands appear as the superposition of the wave functions around this superdeformed
minimum (Fig.~\ref{fig:S32} (b)). The lowest energy band coincides 
with the superdeformed band predicted by the mean-field calculations. The $^{16}$O+$^{16}$O
cluster component in this band amounts to 57\% which means the superdeformed band has the
$^{16}$O+$^{16}$O molecule-like structure, but $^{16}$O clusters are considerably
distorted. This shows dual nature of this band (mixing of mean-field and cluster nature) and
explains why the mean-field  calculations and cluster model give qualitatively the same
results. The degree-of-freedom of inter-cluster motion embedded in the superdeformed band
shows up as two excited rotational bands.  The cluster components of these two bands amount
to more than 90\% and these bands are  interpreted as the excitation mode of the
superdeformed band in which the relative motion between two $^{16}$O clusters is excited by 2 and 4
$\hbar\omega$, that are confirmed by the analysis of the inter-cluster motion of the AMD
wave function. The highest band with $4\hbar\omega$ excitation of relative motion plausibly
agrees with the observed $^{16}$O+$^{16}$O molecular band \cite{Gai81,Mor85}. Thus the superdeformed
band and the molecular band can be regarded as a series of the $^{16}$O+$^{16}$O cluster
bands, and the superdeformed band evolves to the $^{16}$O+$^{16}$O molecular band
as the inter-cluster motion is excited.

\subsubsection{Superdeformation and clusters in $A\sim 40$ region}
\begin{figure}[th]
\centerline{\includegraphics[width=\hsize]{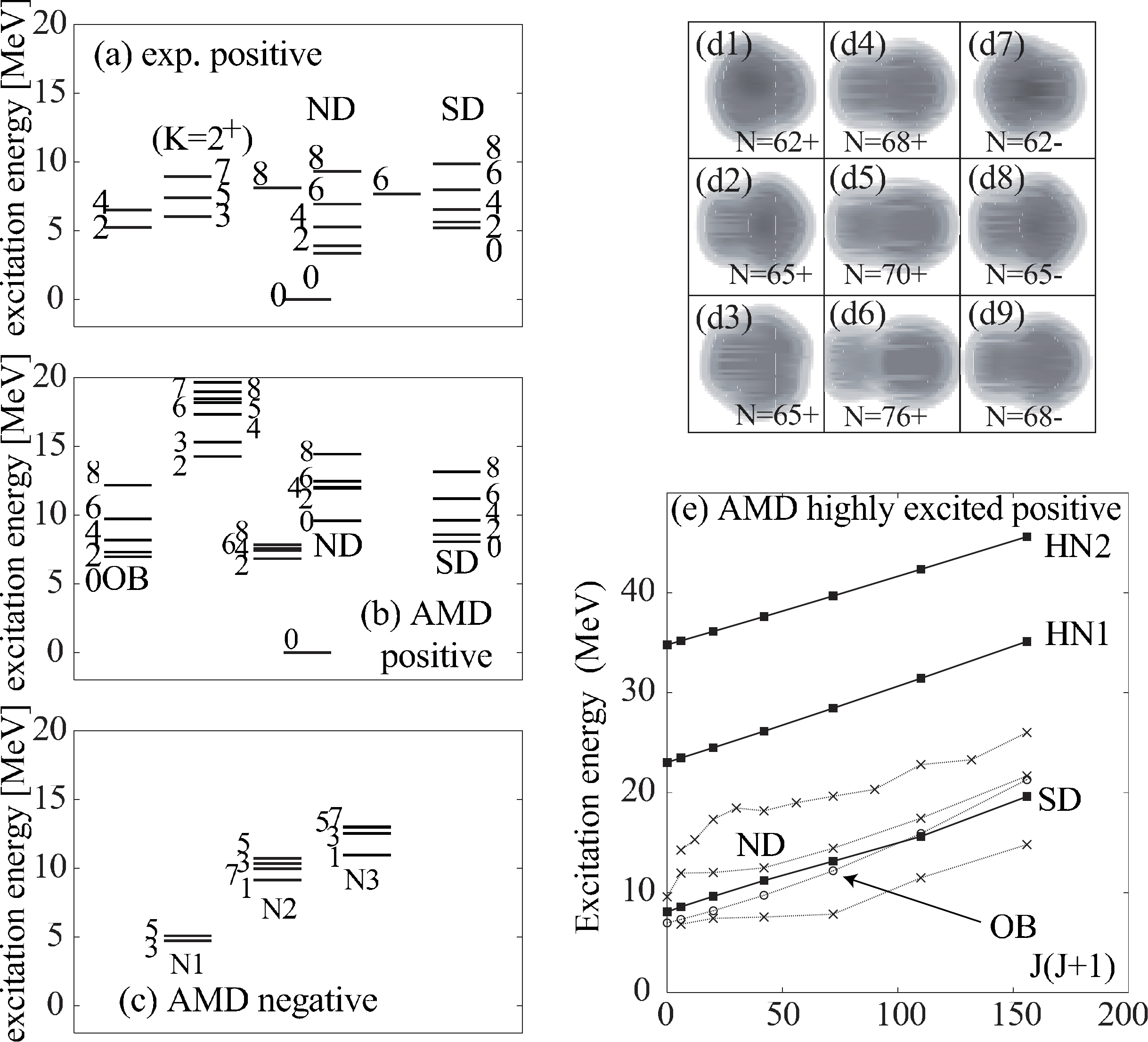}}
\caption{(a) Observed positive-parity states of $^{40}$Ca.  (b) and (c) Calculated positive-
 and negative-parity spectra by AMD+GCM \cite{Eny05}. 'OB', 'ND', 'SD' and
 'N3' denote the oblate and prolate deformed bands, superdeformed band and parity doublet
 partner of the superdeformed band,  respectively. (d1)-(d9) Intrinsic density  distribution
 obtained by the constraint  variational calculation (see text). (e) Calculated  spectra of
 the highly excited  positive-parity states. HN1 and HN2 denote the higher nodal
 $^{12}$C+$^{28}$Si cluster  bands in which the inter-cluster  motion is excited. The 
AMD results are taken from Ref.~\citen{Eny05}.}\label{fig:Ca40} 
\end{figure}

The superdeformed state and $^{16}$O+$^{16}$O clustering of $^{32}$S shed  light on the
relationship between the (super)deformed states and clustering.  As an illustrative and
interesting example, Fig.~\ref{fig:Ca40} (a) shows the observed spectra of doubly magic
nucleus $^{40}$Ca. Two deformed rotational bands start from  the $0^+$ states at 3.35 and
5.21 MeV. It has been suggested that the former band is  predominated by a $4\hbar\omega$ excited
configuration and the latter is a $8\hbar\omega$ configuration \cite{Ger67,Ger69}.
The discovery of the high-spin states \cite{Ide01} identified the latter band as a
superdeformed band, and several theoretical studies based on the mean-field models have been
performed \cite{Ina03,Ben03}. On the other hand, based on the cluster model,
$\alpha$+$^{36}$Ar clustering of the  band starting from 3.35 MeV has been suggested
\cite{Oga77, Sak94} and experimentally confirmed by  the $\alpha$ transfer reaction on $^{36}$Ar
 \cite{Yam93, Yam94}.  In the
following, we denote the band starting at 3.35
  MeV as ND band and the band at 5.21 MeV as SD band.   

AMD studies have been performed to clarify nature of the ND and SD bands\cite{Eny05,Tan07} to
reveal the coexistence of the different deformed bands and their relationship to the cluster
structure. Here we mainly discuss the result reported in Ref.~\citen{Eny05}, in which
 the variational calculation is performed under the constraint on the principal quantum
 number of the harmonic oscillator,  
\begin{align}
 N&=\sum_{i=1}^A\hat{a}^\dagger_i \hat{a}_i 
 = \sum_{i=1}^{A}
 \left[
 \frac{{\bf p}_i^2}{4\hbar^2\nu} + \nu {\bf r}_i^2 - \frac{3}{2}
\right].
\end{align}
Here the lowest Pauli allowed value is $N=60$, and the average number of nucleons promoted from
$sd$ to  $pf$ shell increases for a  larger value of $N$.  Figure \ref{fig:Ca40} (d1)-(d9)
shows the intrinsic density distributions obtained for different values of $N$. For a  small value
of $N$, an almost spherical state is obtained (d1), and prolate (d2) and oblate (d3) deformed
states appear by slight increase of $N$. Further increase of $N$ develops the parity
asymmetric prolate deformed state (d4). The parity asymmetry of this wave function
originates in  $^{12}$C+$^{28}$Si cluster nature of the superdeformed state, since 
it evolves into a prominent $^{12}$C+$^{28}$Si cluster state as $N$ increases
(d4)-(d6). An important point is that similar $^{12}$C+$^{28}$Si cluster-like state (d9)
also appears in the negative-parity state with large $N$, while the negative-parity state
with small $N$ (d7) corresponds to the $1\hbar\omega$  excited state built on the spherical
state (d1).  

Several rotational bands are obtained as the superposition of those wave functions by
AMD+GCM as shown in Fig.~\ref{fig:Ca40} (b) and (c)  
compared with the observed data (Fig.~\ref{fig:Ca40} (a)).  In the positive-parity
states, there are two rotational bands (denoted as ND and OB in Fig.~\ref{fig:Ca40} (b)) that
are dominantly composed of the wave functions (d2) and (d3) respectively, predicting a
prolate and oblate shape coexistence. Though the calculated excitation energy is larger than the
observation, the calculated ND band is assigned to the observed band starting from 3.35 MeV from the
comparison of their $B(E2)$ strengths, while the corresponding oblate deformed band has not 
been experimentally assigned yet.

\begin{figure}[th]
\centerline{\includegraphics[width=0.8\hsize]{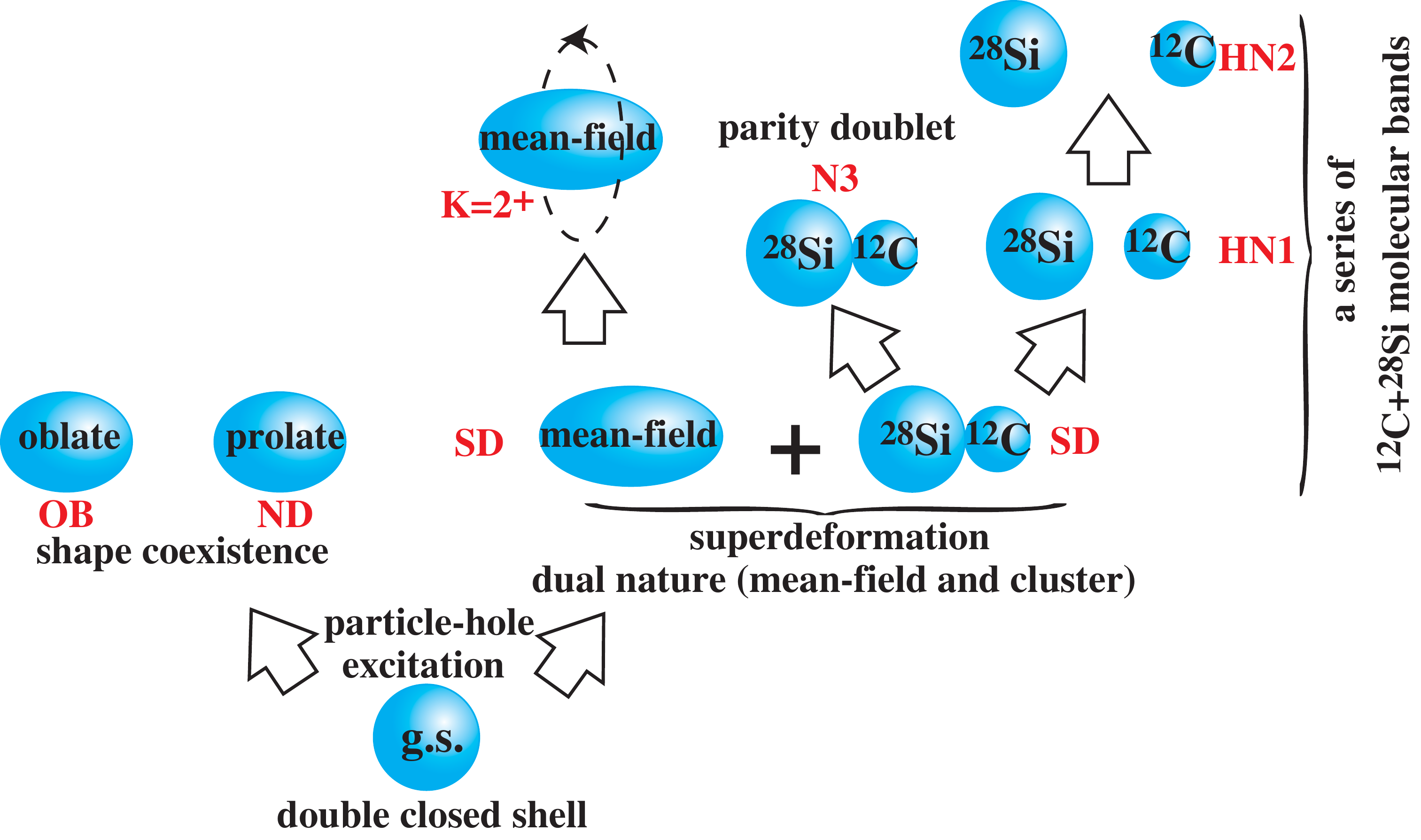}}
\caption{Schematic figure summarizing the excitation modes of $^{40}$Ca. Prolate and
 oblate deformed states coexist at small excitation energy. The superdeformed band has dual
 nature of the mean-field and $^{12}$C+$^{28}$Si clustering. The triaxiality of the
 superdeformed band generates the $K^\pi=2^+$ sideband, while parity asymmetry of the
 $^{12}$C +$^{28}$Si cluster structure generates the  parity doublet partner. Furthermore,
 the nodal excitation of the inter-cluster motion  generates a series of the $^{12}$C +
 $^{28}$Si molecular bands.}\label{fig:Ca40illust}  
\end{figure}

The highlight of the  result  is the superdeformed band and related ones  that are associated
with the parity-asymmetric $^{12}$C+$^{28}$Si clustering 
nature (Fig.~\ref{fig:Ca40} (d4)-(d6) and (d8)-(d9)). The calculated superdeformed band (SD
in Fig.~\ref{fig:Ca40}  (b)) is dominated by the wave functions (d4)-(d6) and starts around
8 MeV, while the observed band is located at 5.21 MeV. The AMD calculation has predicted the
presence of the parity doublet partner of the superdeformed band generated by the parity
asymmetry. The $K^\pi=0^-$ band starting
around 11 MeV in  the negative parity (N3 in Fig.~\ref{fig:Ca40} (c)) is mainly composed of
the wave function shown in Fig.~\ref{fig:Ca40} (d9) that also manifests $^{12}$C+$^{28}$Si
clustering and can be understood as the parity doublet partner of the superdeformed band. 
The clustering nature of the superdeformed band also appears as the two excited
molecular bands that are predicted at high excitation energy region (HN1 and HN2 in
Fig.~\ref{fig:Ca40} (e)). These bands are dominated by the wave
functions with $N$ larger than for
the superdeformed band
dominated by (d4)-(d6), which indicates that they are the nodal
excitation modes of the inter-cluster motion. Similar to the case of $^{32}$S, the superdeformed
band (SD) and nodal excited bands (HN1 and HN2)  can be understood as a series of the
$^{12}$C+$^{28}$Si cluster bands. Thus, the observation of the parity doublet partner and
nodal excited bands will be essential to reveal clustering nature of the superdeformed band.
It is also noted that the triaxial deformation  of the superdeformed band has been also discussed
in Ref.~\citen{Tan07}, and it has been suggested that the $K^\pi=2^+$ band accompanies 
the superdeformed band.

The situation of the
$^{40}$Ca is schematically summarized as shown Fig.~\ref{fig:Ca40illust}. Above the ground
state, a prolate and oblate shape coexistence presents as two rotational bands. The
superdeformed band is also located at the similar energy region and manifests the dual
nature of the strongly deformed mean-field and $^{12}$C+$^{28}$Si clustering. The AMD
calculation has predicted that the $^{12}$C+$^{28}$Si clustering nature of the superdeformed
band generates the parity doublet partner in the negative-parity states from the parity asymmetry
and the nodal excited bands at highly excited region from the excitation of the inter-cluster motion.

\subsection{Decoupling between Proton and neutron deformations}
For $Z\ne N$ unstable nuclei, exotic phenomena concerning nuclear deformation have been discovered.
If the shell effect for proton orbits and that for neutron ones compete with each other,
the shape of proton density may be affected by the neutron structure, or it might be insensitive to
neutron structure. In the latter case, decoupling of deformations 
between proton density and neutron one may occur. The decoupling, i.e., different proton and neutron
deformations is possible in light-mass nuclei and it can be observed in 
the quadrupole transition properties such as the ratio of 
the neutron transition matrix amplitude to the proton one (so-called $M_n/M_p$ ratio).

\begin{figure}[th]
\centerline{\includegraphics[width=12cm]{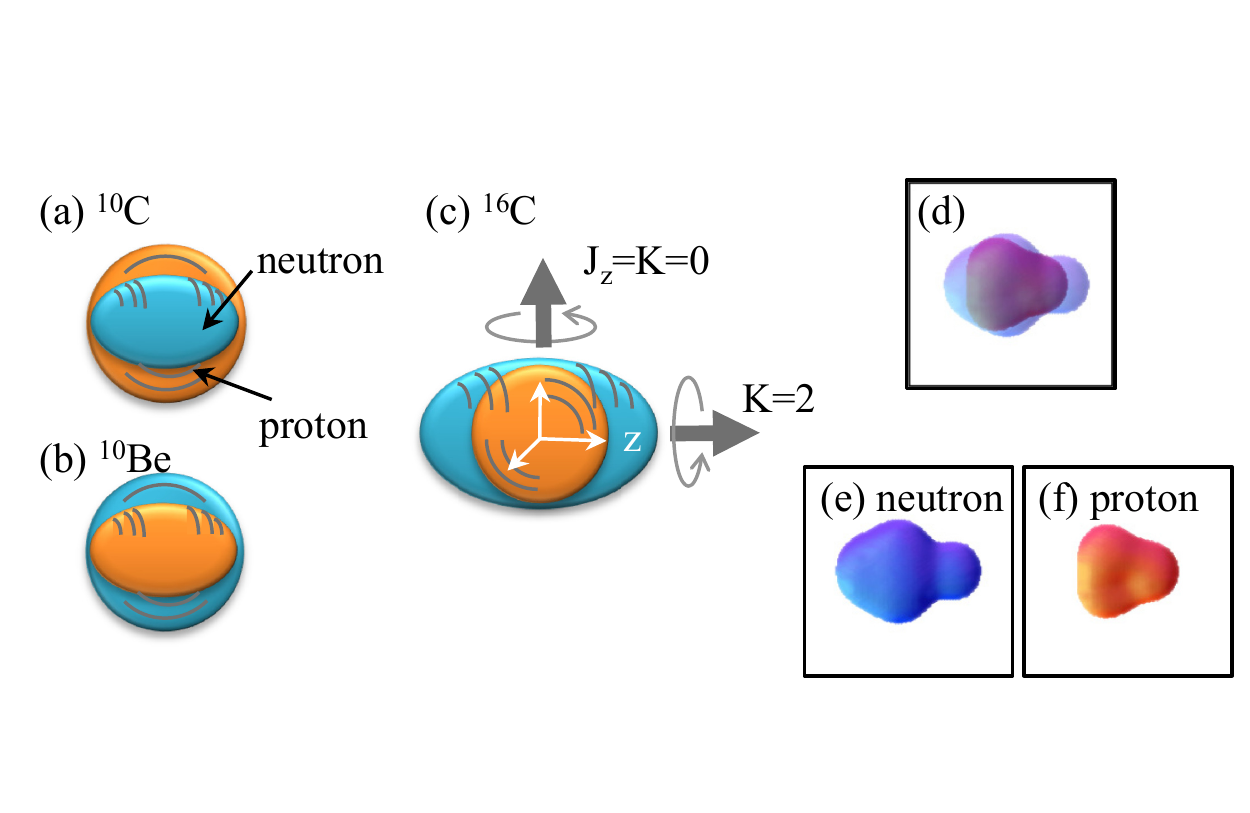}}
\vspace*{8pt}
\caption{\label{fig:c10-c16}
Schematic figures for different shapes of proton and neutron densities in (a) $^{10}$C,  (b) $^{10}$Be, and
(c) $^{16}$C. (d) Surface cut at constant proton and neutron densities of $^{16}$C obtained by the VBP calculations with AMD. 
(e) Prolate neutron density of  $^{16}$C. (f) Oblate proton density of  $^{16}$C. }
\end{figure}

Such a decoupling between proton and neutron shapes has been suggested, for instance, in 
$^{16}$C for which an enhanced $M_n/M_p$ ratio called neutron dominance has been observed 
in the ground-band transition, $2^+_1\rightarrow 0^+_1$, in inelastic scattering \cite{Elekes04}. 
The neutron dominance was described by opposite deformations, i.e., 
an oblate proton shape and a prolate neutron one (Fig.~\ref{fig:c10-c16})
in the study with AMD \cite{KanadaEn'yo:2004bi,Takashina:2005bs}.
The opposite deformations in $^{16}$C has been supported also by
the abnormally small $E2$ transition strength $B(E2;2^+_1\rightarrow 0^+_1)=$2.6 $\pm$0.9 e$^2$fm$^4$
observed by the life time measurement of the $^{16}$C($2^+_1$)
\cite{Ong:2007jb}. 
The strength $B(E2;2^+_1\rightarrow 0^+_1)$ in $^{16}$C is small  
compared with those for other C isotopes, $^{10}$C, $^{12}$C,
and $^{14}$C. The hindrance of the $B(E2;2^+_1\rightarrow 0^+_1)$
implies a small proton deformation, however, it seems to contradict
a large deformation expected from the small excitation energy of the $2^+_1$ state 
if proton and neutron deformations are assumed to be consistent as usual.

\begin{figure}[th]
\centerline{\includegraphics[width=12cm]{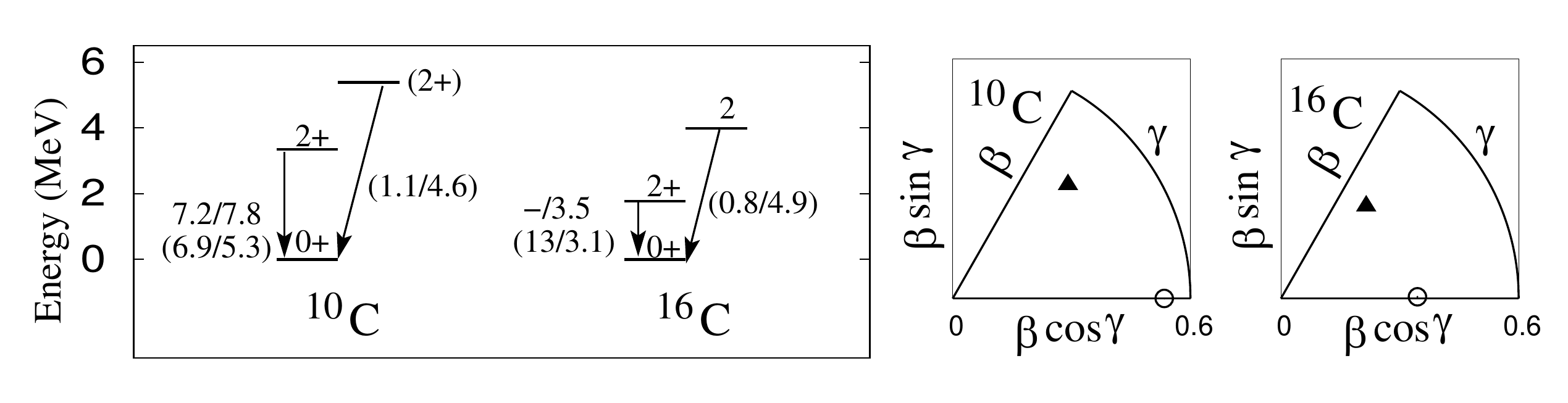}}
\vspace*{8pt}
\caption{\label{fig:c10-c16-spe022}
Left: the experimental energy levels of the $0^+_1$, $2^+_1$, and $2^+_2$ states 
and $M_n/M_p$ ratios for the $2^+_1\rightarrow 0^+_1$ and $2^+_2\rightarrow 0^+_1$
of $^{10}$C and $^{16}$C\cite{nucldata,Ong:2007jb}. 
The experimental values of the neutron matrix amplitude ($M_n$) are deduced from the corresponding 
$B(E2)$ values of the mirror nucleus.
The values in the parentheses are the theoretical values for $M_n/M_p$ of 
the AMD calculation\cite{KanadaEn'yo:2011mj}.
Right: deformation parameters for the intrinsic wave functions of $^{10}$C and $^{16}$C. 
The filled triangles indicate $\beta_p$ and $\gamma_p$ for the proton part
and the open circles are $\beta_n$ and $\gamma_n$ for the neutron part.}
\end{figure}

We studied the structure of even-even C isotopes
with AMD \cite{KanadaEn'yo:2004bi,KanadaEn'yo:1996hi}.
The systematics of the binding energies, radii, and $E2$ transition strengths of C isotopes were 
qualitatively reproduced by a simple version of AMD calculations (VBP). 
The results indicate that the neutron shape drastically changes depending on the neutron number,
while the proton shape is rather stable and insensitive to the neutron structure.
One of the striking features is that 
the difference between proton and neutron shapes was suggested 
in $^{16}$C and $^{10}$C in which prolate neutron shapes are favored. 
In spite of the prolate neutron structure, the proton structure shows an oblate deformation 
resulting in the opposite deformations. 
The deformation parameters for proton and neutron densities of the intrinsic state are 
$(\beta_p,\gamma_p)=(0.41,0.27\pi)$ and $(\beta_n,\gamma_n)=(0.53,0.00\pi)$ for $^{10}$C, and they are 
$(\beta_p,\gamma_p)=(0.32,0.26\pi)$ and $(\beta_n,\gamma_n)=(0.34,0.00\pi)$ for $^{16}$C 
(Figs.~\ref{fig:c10-c16} and \ref{fig:c10-c16-spe022}). 
The reason for opposite proton and neutron deformations is that a $Z=6$ system favors an oblate proton shape 
because of the proton shell effect
while a $N=10$ or $N=4$ nucleus has prolate trends of the neutron shape due to the neutron shell effect.
In other words, the $Z=6$ proton structure is not so much affected by the neutron structure 
but it keeps the oblate tendency.

To discuss the neutron deformation, mirror analysis is useful.  
In the mirror analysis for $^{10}$C and $^{10}$Be, 
the neutron transition matrix $M_n$ for the ground-band transition is evaluated from $B(E2)$ in $^{10}$Be
by assuming mirror symmetry. 
The experimental value of the $M_n/M_p$ ratio in $^{10}$C deduced by the mirror analysis
is described by the AMD calculation, and it can be understood with the opposite deformations between proton 
and neutron densities (Fig.~\ref{fig:c10-c16-spe022}).
The neutron dominance in the ground-band transition is more remarkable in $^{16}$C
as seen in the theoretical results. Unfortunately, there is no direct data of the $E2$ strength 
for the mirror nuclei of $^{16}$C, however, as mentioned before, 
the observed inelastic scattering cross section implies an enhanced $M_n/M_p$ ratio 
indicating the neutron dominance \cite{Elekes04}. 
It is worth to mention that 
microscopic coupled-channel calculations with the transition densities obtained by the AMD calculation
have reproduced the inelastic scattering data successfully \cite{Takashina:2005bs}.

To clarify the oblate shape of the proton structure in $^{16}$C and $^{10}$C, 
observations of possible $K=2$ side bands and their transition properties would be helpful probes.
Such a nucleus having oblate proton and prolate neutron structures may show an
isovector triaxiality. If the case,
a $K=2$ side band can be constructed from the rotation
around the symmetric axis of the prolate neutron part
instead of the rotation around the perpendicular axis for the $K=0$ ground band
(Fig.~\ref{fig:c10-c16-spe022}). 
Since the proton contribution should be dominant while the neutron contribution 
is minor in the rotation for the $K=2$ side band, the inter-band transition $2^+_2\rightarrow 0^+_1$
may show the proton dominance resulting in a small $M_n/M_p$ ratio.
In fact, the calculated $M_n/M_p$ ratios
for the $2^+_2\rightarrow 0^+_1$ in $^{10}$C and $^{16}$C are quenched 
and they indicate the proton dominance.
There is no experimental information of the 
transition strength for $2^+_2\rightarrow 0^+_1$.
Inelastic scatterings of $^{10}$C and $^{16}$C will be good probes to experimentally 
confirm the proton dominance. 

\section{Applications of time-dependent AMD method to nuclear response and reaction}  \label{sec:sec4}

Interesting phenomena, including clustering phenomena, appear in
nuclear many-body systems not only near the ground state but also in
excited states in wide range of excitation energies.  These include
giant resonances of collective modes and stronger expansion of the
system which is often followed by the disintegration into fragment
nuclei.  Not only the collective degrees of freedom but also
non-collective (thermal) excitations play important roles in many
cases.  Since the number of states increases very rapidly as the
excitation energy goes up, it is difficult to describe these phenomena
based on individual quantum states.  The time-dependent version of AMD
has been a powerful approach for such problems of highly excited
systems.

\subsection{Dipole resonances}
Among exotic phenomena in neutron-rich nuclei,
one of the hot subjects is dipole strengths in low energy regions, 
which are expected to enhance in neutron-rich nuclei
due to excess neutrons. For instance, soft resonances in extremely low-energy regions
and pygmy resonances below giant dipole resonance (GDR) energy 
are attracting great interests (see, for example, 
Refs.~\citen{Honma90,Hamamoto96b,Catara,Colo,Nakatsukasa,Inakura:2009vs} and references therein).
To investigate isovector dipole responses of neutron-rich nuclei,
we applied a time-dependent version of AMD without stochastic terms\cite{KanadaEn'yo:2005wd}.

In order to calculate responses to external fields, 
we first solve the static problem and obtain the optimum 
parameter set
${Z}^0$, which gives the energy minimum state $\Phi_{\text{AMD}}({Z}^0)$
in the AMD model space.
Then, we boost the $\Phi_{\text{AMD}}({Z}^0)$ instantaneously at $t=0$
by imposing an external perturbative dipole field,
\begin{eqnarray}
\Psi(t=0+)&=&e^{-i\epsilon F}\Phi_{\text{AMD}}({Z}^0),\\
V_{\rm ext}({\bf r},t) &=& \epsilon F({\bf r})\delta(t),\\
F({\bf r})= {\cal M}(E1,\mu)&=&
\sum_i^A e^{\rm rec} r_iY_{1\mu}(\hat {\bf r_i}),
\end{eqnarray}
where $\epsilon$ is an arbitrary small number and 
$e^{\rm rec}$ is the $E1$ recoil charge, $Ne/A$ for protons and 
$-Ze/A$ for neutrons.
Note that the initial state $\Psi(t=0+)$ after imposing the dipole field is written with 
a single AMD wave function $\Phi_{\text{AMD}}({Z}(t=0+))$.

Following the time-dependent AMD method, we can calculate the time evolution of the system, 
$\Psi(t)=\Phi_{\text{AMD}}({Z}(t))$, from the initial
state $\Psi(t=0+)$ by using the equation of motion Eq.~(\ref{eq:eqmotion}).
Once the wave function $\Psi(t)$ is obtained as a function of time, 
the transition strength can be obtained by 
Fourier transform of the expectation value of ${\cal M}(E1,\mu)$ as follows,
\begin{eqnarray}\label{eq:BE1}
\frac{d B(\omega;E1,\mu)}{d\omega}  &\equiv& \sum_n 
|\langle n|{\cal M}(E1,\mu)|0 \rangle|^2 \delta(\omega-\omega_n) \\ 
&=& 
-\frac{1}{\pi\epsilon} {\rm Im}\int^\infty_0 dt \langle \Psi(t)| {\cal M}(E1,\mu) 
|\Psi(t) \rangle e^{i\omega t},
\end{eqnarray}
where $|0\rangle$ is the ground state and 
$|n\rangle$ is the excited state with the excitation energy $\hbar\omega_n$.

In the present framework, 
$d B(\omega;E1)/d\omega$ consists of 
discrete peaks in principle, because the present AMD method is a bound state
approximation and continuum states are not taken into account.
We introduce a smoothing parameter $\Gamma$ by hand in the 
Fourier transform in Eq.~(\ref{eq:BE1}) which may simulate
the escape and spreading widths of resonances.

One of the advantages of time-dependent AMD is that
we can obtain intuitive interpretations for each mode 
by analyzing time evolution of Gaussian centers ${\bf Z}_i(t)$ of single-particle wave packets.
Another advantage is that 
the present method is free from the spurious center-of-mass motion because  
center-of-mass motion can be exactly separated from $\Psi(t)$. 

\begin{figure} 
\centerline{\includegraphics[width=14.0 cm] {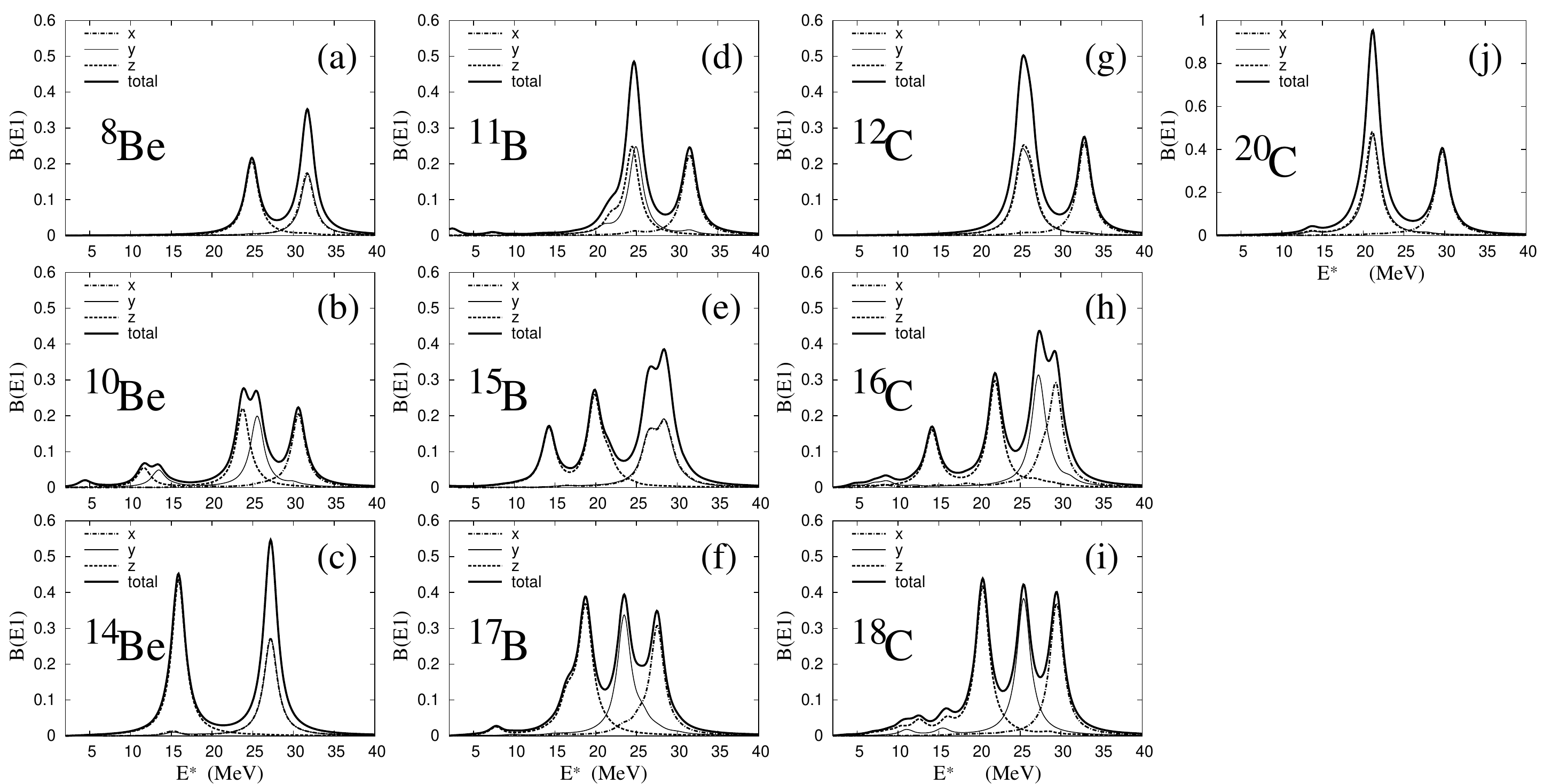}}
 
   \caption{$E1$ transition strengths ($e^2$fm$^2$/MeV) 
of Be, B, and C isotopes calculated with the time-dependent AMD method
by using the MV1 ($m=0.576$ and $b=h=0$) + G3RS ($u_{I}=-u_{II}=900$ MeV) force\cite{KanadaEn'yo:2005wd}.
The smoothing parameter is chosen to be $\Gamma=2$ MeV.
Thin dash-dotted, solid, and dotted lines are the contribution of 
vibration for the $x$,$y$, and $z$-directions, respectively.
The total strengths are shown by the thick solid lines.}
   \label{fig:e1strength}
\end{figure}

We applied this method to Be, B, and C isotopes and investigated the $E1$ resonances 
 \cite{KanadaEn'yo:2005wd}.
The $E1$ strengths are shown in Fig.~\ref{fig:e1strength}. 
It was found that remarkable peaks appear in $^{10}$Be, $^{15}$B, 
and $^{16}$C in the $E^*=10-15$ MeV region decoupling from the GDR. 
Those soft dipole resonances arise from the relative motion of
excess neutrons against a core, which is decoupled from the motion inside the
core. In other words, the soft resonances appear due to the
excitation of excess neutrons around the rather hard core.
In fact, the strengths of the soft dipole resonances 
almost exhaust the cluster sum rule values for the core and valence neutrons.
In further neutron-rich B and C 
isotopes with $N>10$, the strengths for the
soft dipole resonances decline
compared with those in $^{15}$B and 
$^{16}$C. The reason for decreasing low-lying strengths is that motion 
of excess neutrons assimilates into neutron motion inside the core
and its decoupling from the core weakens.
As a result, the excitation energies of the GDR decrease 
with the enhancement of the neutron skin.
It is striking that strengths of the soft dipole resonances 
do not necessarily increase with the increase of excess neutrons.
Instead, the feature of the soft resonances rapidly changes depending on the
proton and neutron numbers of the system. The strengths of the soft dipole
resonances depends on how much the coherent motion 
of the excess neutrons decouples from the motion inside the core.

\subsection{Radial oscillations}

Radial breathing oscillations of nuclei are interesting particularly
because they are closely related to the compressibility of nuclei and
nuclear matter.  The AMD approach is, however, not very suitable for
the precise analysis of the monopole strengths since the state boosted
by the monopole operator from the ground state AMD wave function is no
longer an AMD state.  Nevertheless, AMD can be a powerful tool to
explore the radial oscillations at various amplitudes.  The large
amplitude oscillations should be continuously linked to the expansion
of nuclei without restoration at higher energies where many-body
correlations play important roles in forming clusters and fragment
nuclei.  Therefore the possibility of cluster correlations in lower
energy oscillations is an interesting question.

\begin{figure}
\begin{center}
\includegraphics[width=0.5\textwidth]{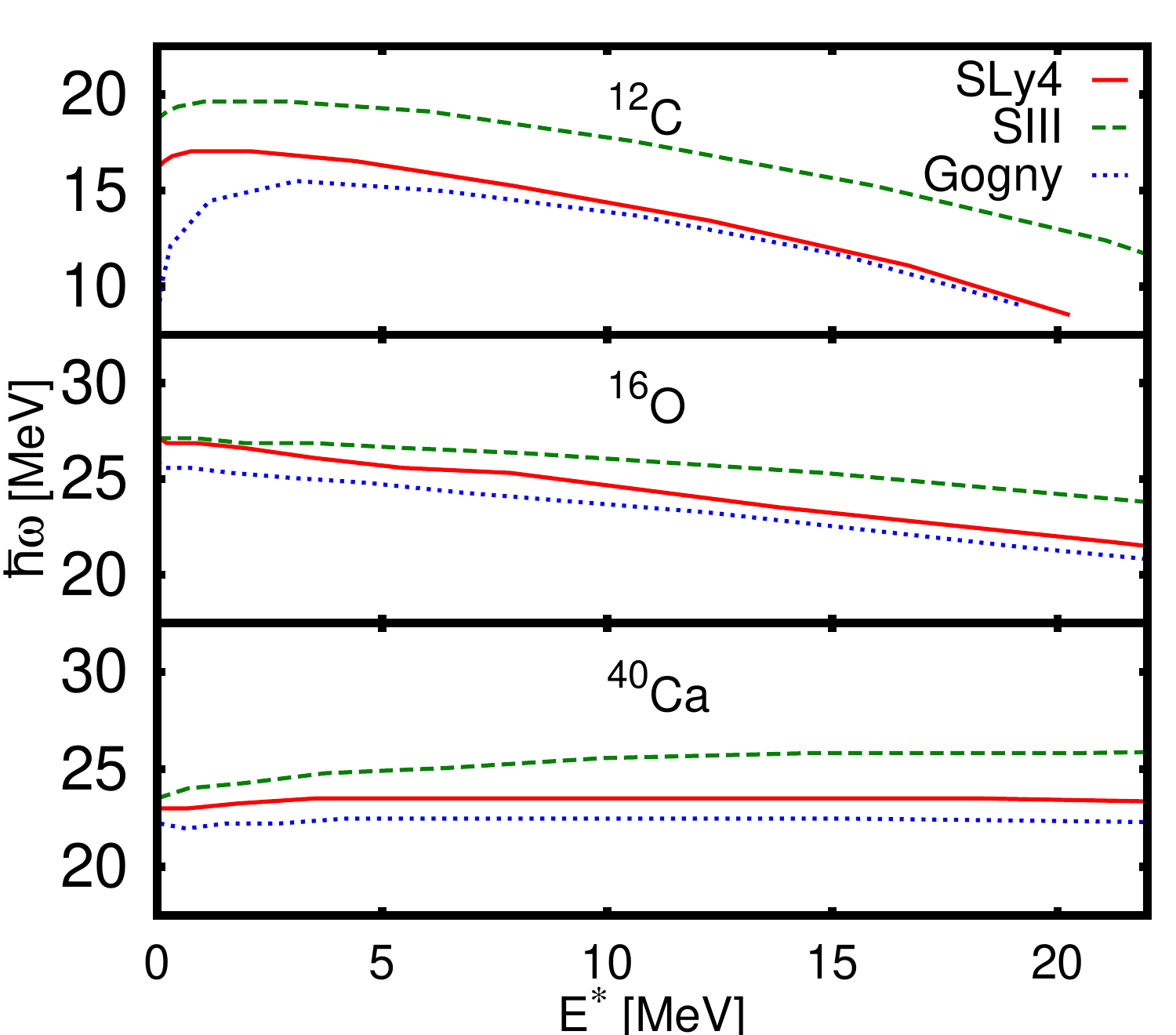}
\end{center}
\caption{AMD prediction for the frequency $\hbar\omega$ of the radial
  oscillation as a function of the oscillation energy for
  ${}^{12}\mathrm{C}$, ${}^{16}\mathrm{O}$ and ${}^{40}\mathrm{Ca}$
  with the SLy4, SIII and Gogny interactions. The nuclear matter
  incompressibilities for these interactions are $K=230$, 355 and 228
  MeV, respectively.  The figure is taken from Ref.\ \citen{Furuta:2010}.}
\label{fig:AMD3force}
\end{figure}

The radial oscillations of the ${}^{12}\mathrm{C}$ and other nuclei
were calculated with AMD [Eq.\ (\ref{eq:eqmotion})] without any
stochastic terms in Ref.\ \citen{Furuta:2010}.  The time evolution is
solved from the initial state prepared by placing three $\alpha$
clusters on a regular triangle in the ${}^{12}\mathrm{C}$ case.  By
changing the size of the initial regular triangle, oscillations with
different amplitudes were studied.  Figure \ref{fig:AMD3force} shows
the dependence of the oscillation frequency on the amplitude or the
excitation energy.  The results for three different effective
interactions are shown.  Dependence on the incompressibility is
clearly observed.  The dependence on the amplitude suggests the
unharmomicity of the oscillation.  It should be noted that the
one-phonon excitation corresponds to the excitation energy of
$E^*=\hbar\omega$ where $\omega$ is the angular frequency.

The calculations of Fig.\ \ref{fig:AMD3force} were carried out with
the width parameters $\nu$ that optimize the ground state energy.
Depending on the chosen width parameter, however, the frequency
sometimes shows anomalous behavior as in the system of a double-well
potential \cite{Furuta:2010}.

\begin{figure}
\begin{center}
\includegraphics[width=0.5\textwidth]{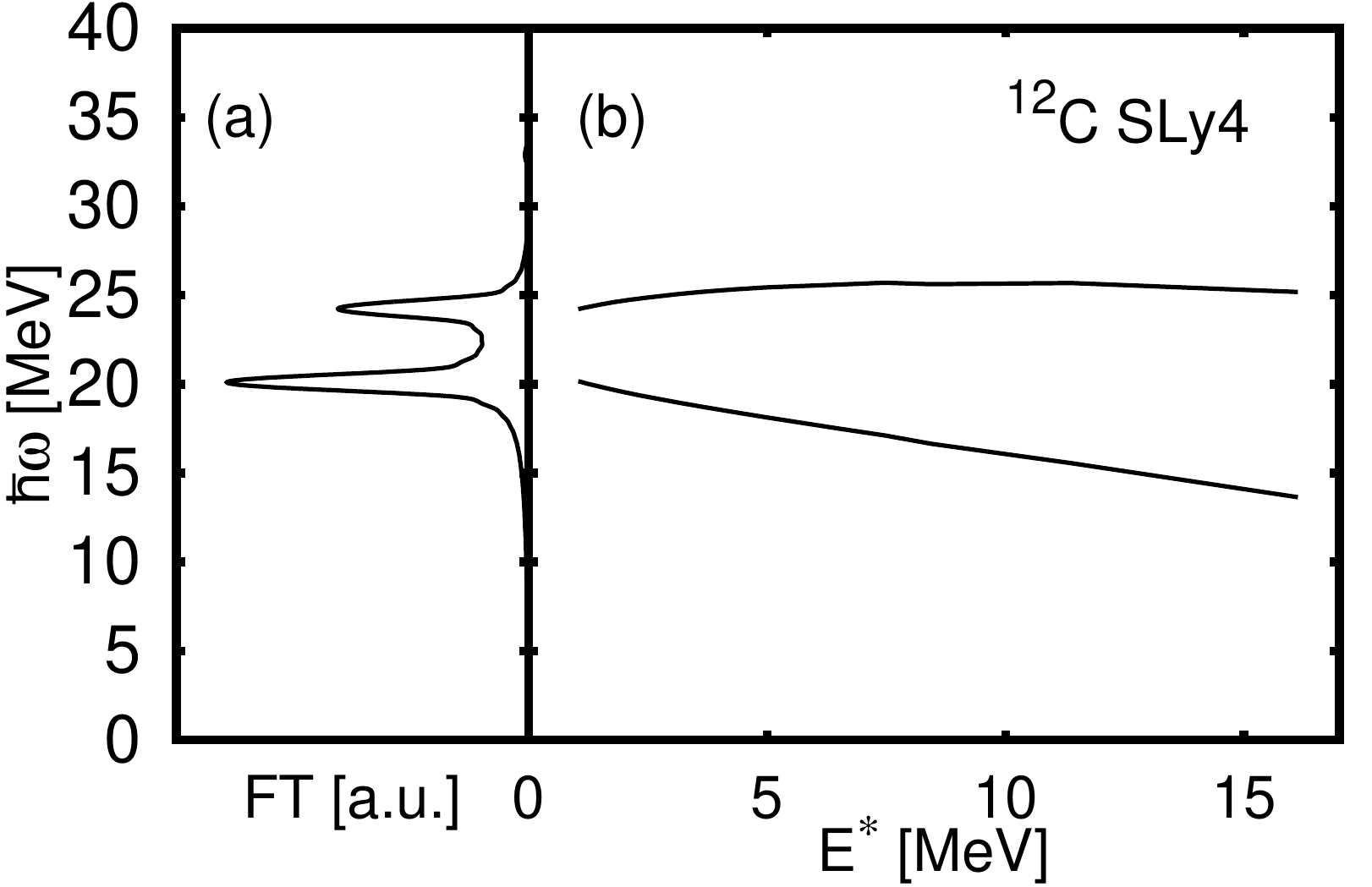}
\end{center}
\caption{(a) Fourier transform $\text{FT} [r](\omega)$ of the FMD
  oscillation pattern in small amplitude, with the same initial
  condition as in the AMD calculation. (b) The peak position
  $\hbar\omega$ as a function of the oscillation amplitude.  The
  figure is taken from Ref.\ \citen{Furuta:2010}.}
\label{fig:FMD_C12}
\end{figure}

The same problem was studied by the fermionic molecular dynamics (FMD)
as well \cite{Furuta:2010}.  In FMD, the width parameters $\nu_i(t)$
of individual wave packets are treated as time-dependent variables as
well as the centroid variables.  In this case, the calculated result
shows an oscillation pattern composed of two modes with different
frequencies as shown in Fig.\ \ref{fig:FMD_C12}.  By analyzing the
motions of the variables, it was found that one of the two modes
corresponds to the change of the wave packet widths and the other
corresponds to the motion of the wave packet centroids.  The latter
mode is the same as that observed in the AMD result and related to the
$\alpha$-clustering degrees of freedom.  The former mode is the
breathing of individual single-particle wave packets.  Thus these
calculations suggest that the single-particle excitation and the
clustering excitation are both important in radially oscillating
systems.  The simplest version of AMD without any stochastic terms can
describe clustering excitations.

\subsection{Multifragmentation in expanding systems}

It has been well known experimentally that a lot of fragment nuclei
are produced in each event of heavy-ion collisions in various
situations if the incident energy is more than ten MeV/nucleon.
Multifragmentation is an interesting problem in excited nuclear
systems, in which quantum many-body correlations play essential roles
as well as the existence of nuclear liquid-gas phase transition.  The
microscopic description of multifragmentation is, in principle, a
highly complicated problem of quantum many-body systems.  Transport
models have been developed for heavy-ion collisions with some
classical approximations.  Compared with other transport models, some
quantum features have been incorporated into AMD by employing fully
antisymmetrized wave functions.

\begin{figure}
\includegraphics[width=\textwidth]{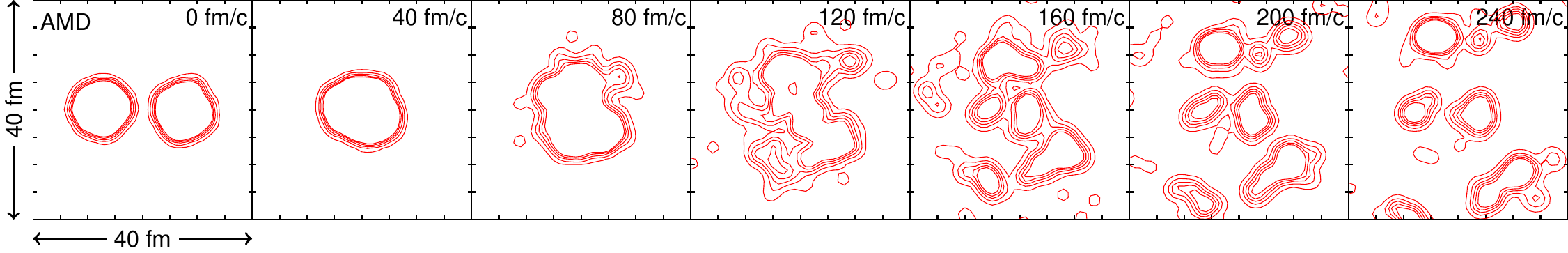}\\
\includegraphics[width=\textwidth]{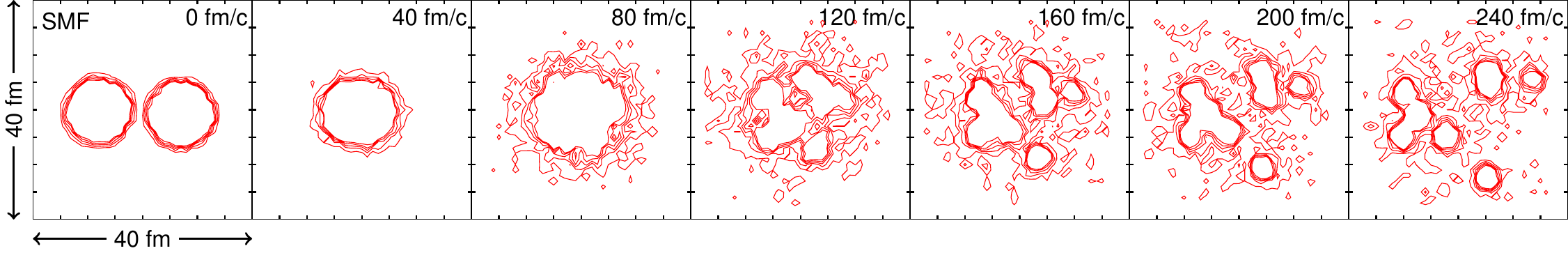}\\
\caption{Time evolution of density profiles obtained by the AMD
  (upper) and SMF (lower) models, for the central collision of
  $^{112}\mathrm{Sn} + {}^{112}\mathrm{Sn}$ at 50 MeV/nucleon.}
\label{fig:dens_SMF_AMD}
\end{figure} 

Figure \ref{fig:dens_SMF_AMD} shows a typical example of the time
evolution of ${}^{112}\mathrm{Sn}+{}^{112}\mathrm{Sn}$ central
collisions at 50 MeV/nucleon \cite{Rizzo:2007,Colonna:2010}.  The
calculation by AMD in the upper row is compared with the calculation
by the stochastic mean-field (SMF) model in the lower row.  In both
cases, the system compressed at an early stage starts to expand almost
spherically.  As the expansion proceeds, the density fluctuation
develops to form many fragments.

The AMD calculation was performed with the coherence time
$\tau=\tau_{\text{NN}}$.  The SMF model, which is based on the
single-particle motion in the mean field, takes into account the
two-nucleon collisions and fluctuations \cite{Colonna:1998}.
Therefore these two models are conceptually similar, but the results
can be different due to the different approximate treatments of
fluctuations.  In fact, it is observed that the density fluctuation
(among different events) is already developing in AMD at the
relatively early stage of $50\lesssim t\lesssim 100$ fm/$c$, while the
fluctuation develops in SMF only at a later stage $t\sim100$ fm/$c$
suggesting a fragmentation mechanism by spinodal decomposition
\cite{Chomaz:2004}.  Thus the many-body correlations are stronger in
AMD.  This difference can be interpreted as the origin of the
differences in the expansion velocity, the nucleon emission and so on
predicted by these models \cite{Rizzo:2007,Colonna:2010}.

\begin{figure}
\begin{center}
\includegraphics[width=0.5\textwidth]{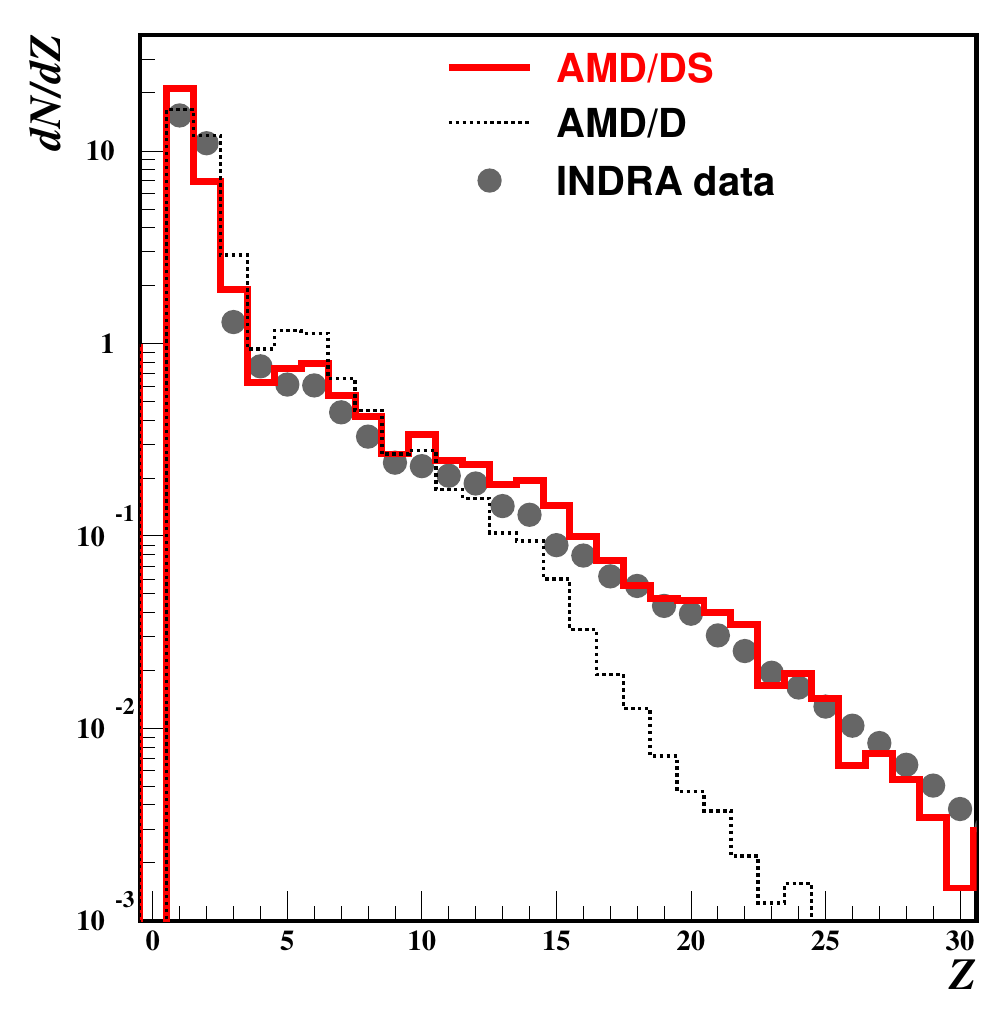}
\end{center}
\caption{\label{fig:XeSn-zmulti} The charge distribution of the
  produced clusters in ${}^{129}\mathrm{Xe}+\mathrm{Sn}$ collisions at
  50 MeV/nucleon with the impact parameter $0<b<4$ fm, after
  calculating the secondary decay of excited clusters and applying the
  experimental filter for the detector setup.  Solid histogram
  (labeled AMD/DS) shows the result of AMD with the coherence time
  $\tau=\tau_{\text{NN}}$, while the dotted histogram (labeled AMD/D)
  shows the result with the strongest decoherence $\tau\rightarrow0$.
  The INDRA experimental data are shown by solid points.  The figure
  is taken from Ref.\ \citen{ONOj}.}
\end{figure}

The final results of these models for the fragment charge distribution
have been compared with the experimental data with reasonable
successes \cite{Frankland:2001,ONOj}.  Figure \ref{fig:XeSn-zmulti}
shows the AMD results compared with data.  The result depends very
much on the choice of the coherence time.  In this reaction system,
the fragment yields for $Z\gtrsim 3$ are well reproduced by AMD when
the wave packet splitting with the coherence time
$\tau=\tau_{\text{NN}}$ is introduced.  It is often convenient to
define the liquid and gas parts of the system as the parts composed of
$Z\ge3$ fragments and $Z\le2$ particles, respectively.  The comparison
shows that the total charge of the liquid part, $Z_{\text{liq}}$, in
the AMD result (with $\tau=\tau_{\text{NN}}$) is consistent with the
experimental data, and therefore the total charge of the gas part,
$Z_{\text{gas}}=Z_{\text{system}}-Z_{\text{liq}}$, is also consistent.

However, a problem is found in the composition of the gas part in the
result with $\tau=\tau_{\text{NN}}$.  The $\alpha$-particle
multiplicity $M_\alpha\approx7$ is too small and the proton
multiplicity $M_p\approx20$ is too large compared with the
experimental data $M_\alpha\approx M_p\approx 10$.  It should be noted
that only about 10\% of the total protons in the system is emitted as
free protons in this reaction at 50 MeV/nucleon.  It is also known
experimentally that still a half of the protons are bound in clusters
even at 1 GeV/nucleon \cite{Reisdorf:2000}.  Thus the experimental
data have been suggesting the importance of cluster correlations.  The
comparison with the data indicates that the AMD approach with the wave
packet splitting, which is largely based on the single-particle motion
in the mean field, does not include sufficient cluster correlations in
the dynamics.  The problem of the gas composition may influence on the
liquid part of the system because the energy balance and the number of
the effective degrees of freedom will change as the gas composition
changes.  Therefore the proper treatment of cluster correlations in
dynamical approaches is an urgent issue.

\subsection{Fragmentation in collisions of light nuclei}

\begin{figure}
\begin{center}
\includegraphics[width=0.5\textwidth]{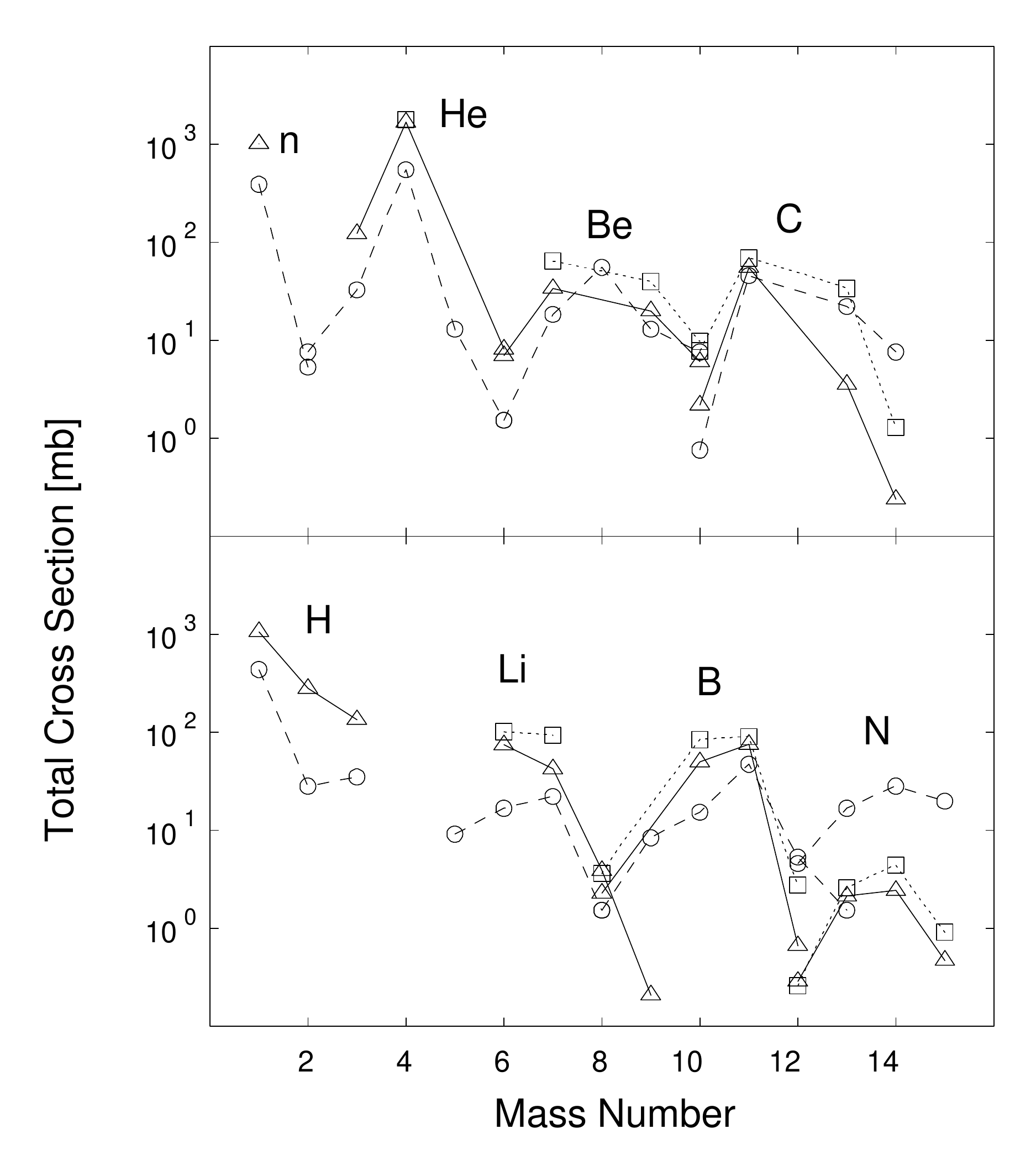}
\end{center}
\caption{\label{fig:CC-stdi} Isotope distribution in
  ${}^{12}\mathrm{C}+{}^{12}\mathrm{C}$ reaction at 28.7
  MeV/nucleon. Circles are the yields before the secondary decay,
  while triangles are those after the secondary decay. Squares are the
  experimental data. Lines connect isotopes. The figure is taken from
  Ref.\ \citen{Ono:1992uy}.}
\end{figure}

In an early study with AMD, the fragmentation in the
${}^{12}\mathrm{C}+{}^{12}\mathrm{C}$ reaction was studied at 28.7
MeV/nucleon \cite{Ono:1991uz,Ono:1992uy}.  The fragment isotope
distribution is reproduced well by AMD as shown in Fig.\
\ref{fig:CC-stdi}.  Especially the large production cross section of
$\alpha$ particles is well reproduced in this case.  The calculation
was done with the stochastic two-nucleon collisions but without wave
packet splitting.  Thus the situation here is different from the
multifragmentation in heavier systems as seen in the previous
subsection where the wave packet splitting is very important.

\begin{figure}
\begin{center}
\includegraphics[width=0.45\textwidth]{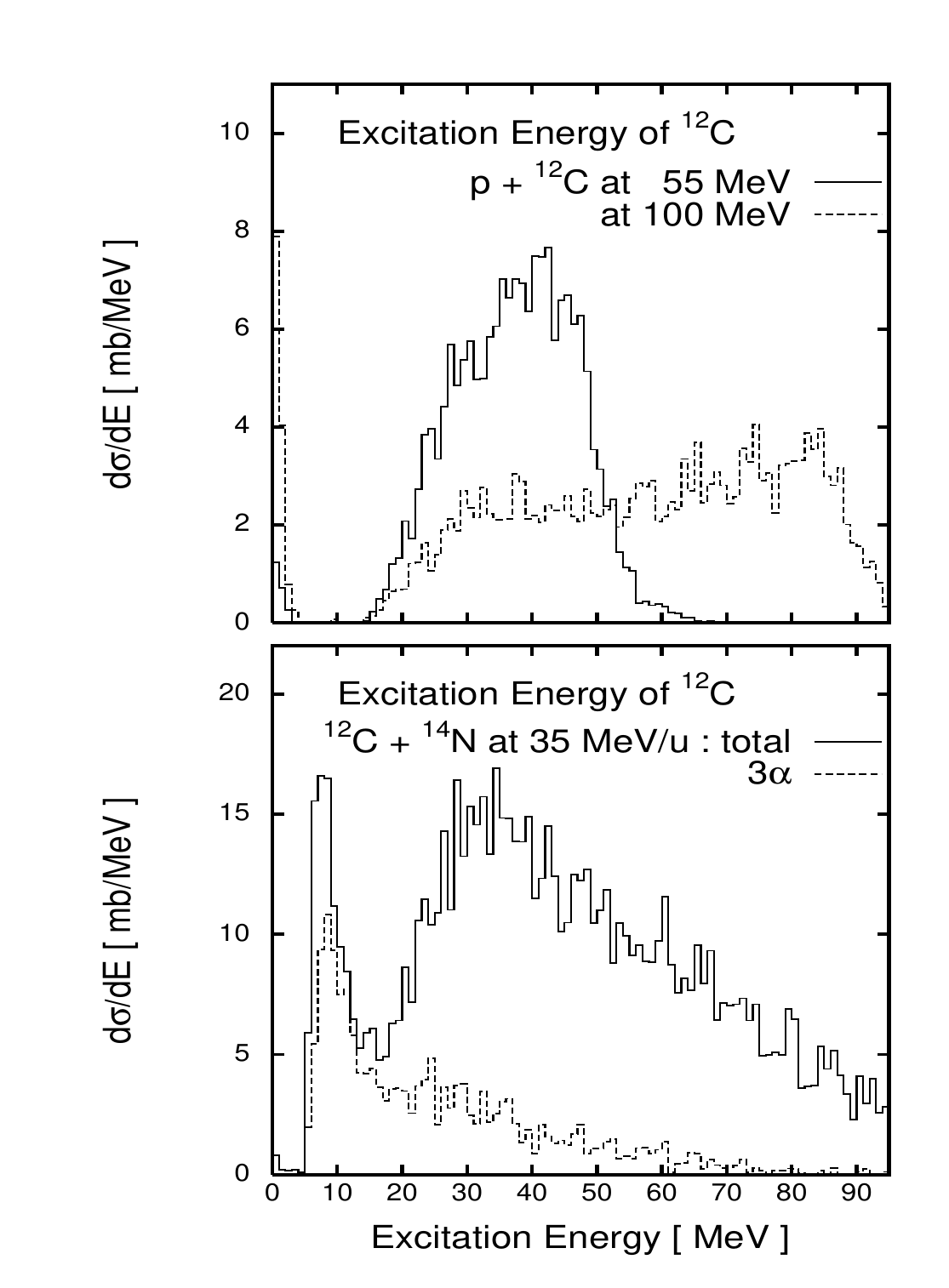}
\end{center}
\caption{\label{fig:takemoto-12C-break} The distribution of excitation
  energy of ${}^{12}\mathrm{C}$ projectile before its breakup. Upper
  figure is for $p+{}^{12}\mathrm{C}$ collisions, while the lower
  figure is for ${}^{12}\mathrm{C}+{}^{14}\mathrm{N}$ collision at 35
  MeV/nucleon.  The figure is taken from Ref.\ \citen{Takemoto:1996}.}
\end{figure}
The detailed mechanism of the fragmentation of the ${}^{12}\mathrm{C}$
projectile was studied in Ref.\ \citen{Takemoto:1996} by Takemoto
\textit{et al}. The solid line of the lower part of Fig.\
\ref{fig:takemoto-12C-break} shows the distribution of the excitation
energy of the ${}^{12}\mathrm{C}$ projectile after the interaction
with the target in ${}^{12}\mathrm{C}+{}^{14}{\rm N}$ reaction at 35
MeV/nucleon.  In addition to the big bump around $E^*\sim 30$ MeV,
there is a sharp peak at $E^*\sim 10$ MeV.  The $\alpha$-clustering
states of ${}^{12}\mathrm{C}$ in this energy region are strongly
excited by the heavy-ion reaction, and they contribute to the large
yield of $\alpha$ particles. In fact, the dashed line shows the
contribution from the events where ${}^{12}\mathrm{C}$ broke up into
three $\alpha$ particles. On the other hand, in the proton induced
reactions, the excitation of ${}^{12}\mathrm{C}$ is mainly of single
particle nature and the $\alpha$ clustering states are not excited at
all as shown in the upper part of Fig.\
{}\ref{fig:takemoto-12C-break}.

\begin{figure}
\begin{center}
\includegraphics[width=0.5\textwidth]{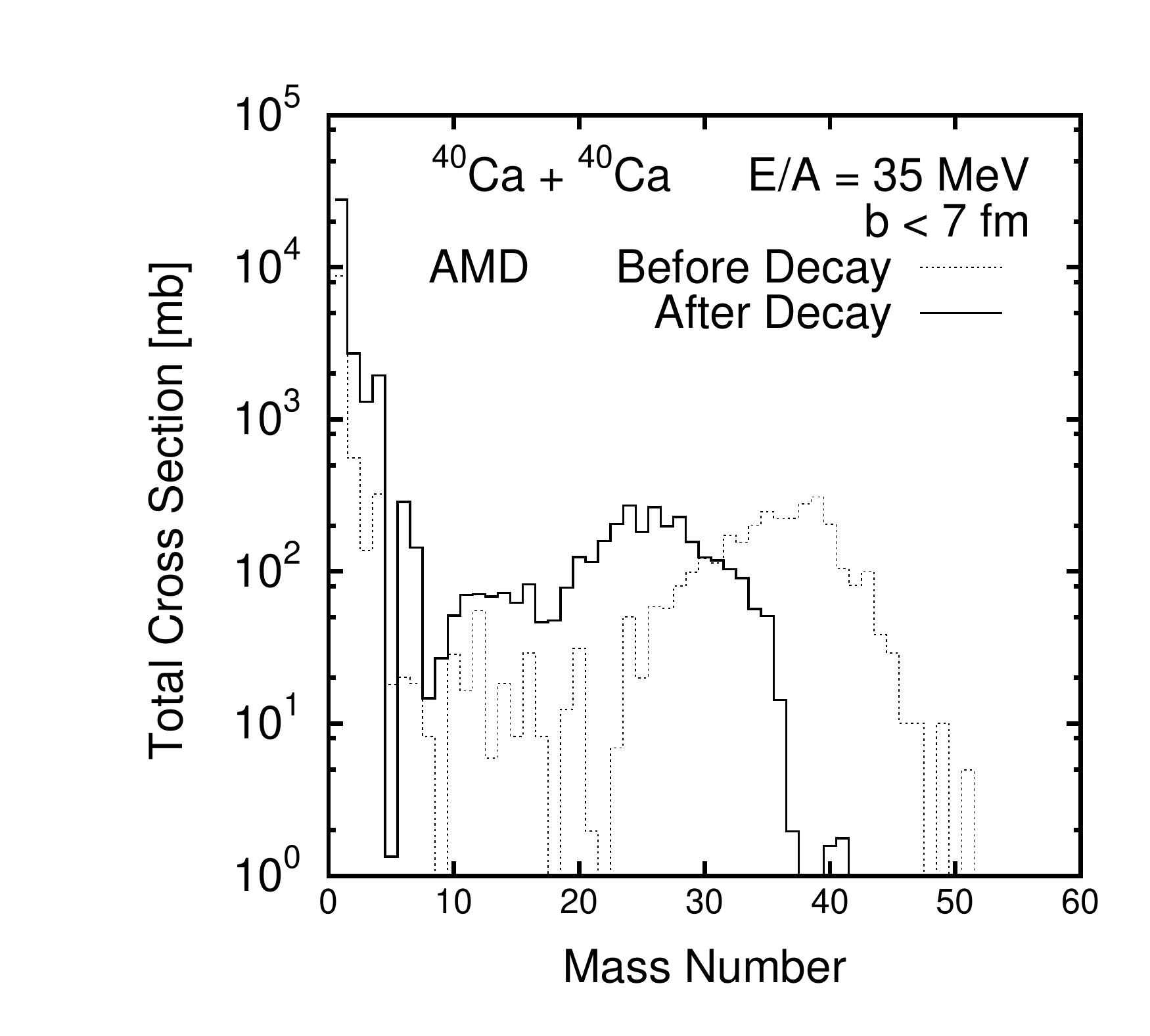}%
\hspace{-0.05\textwidth}
\includegraphics[width=0.5\textwidth]{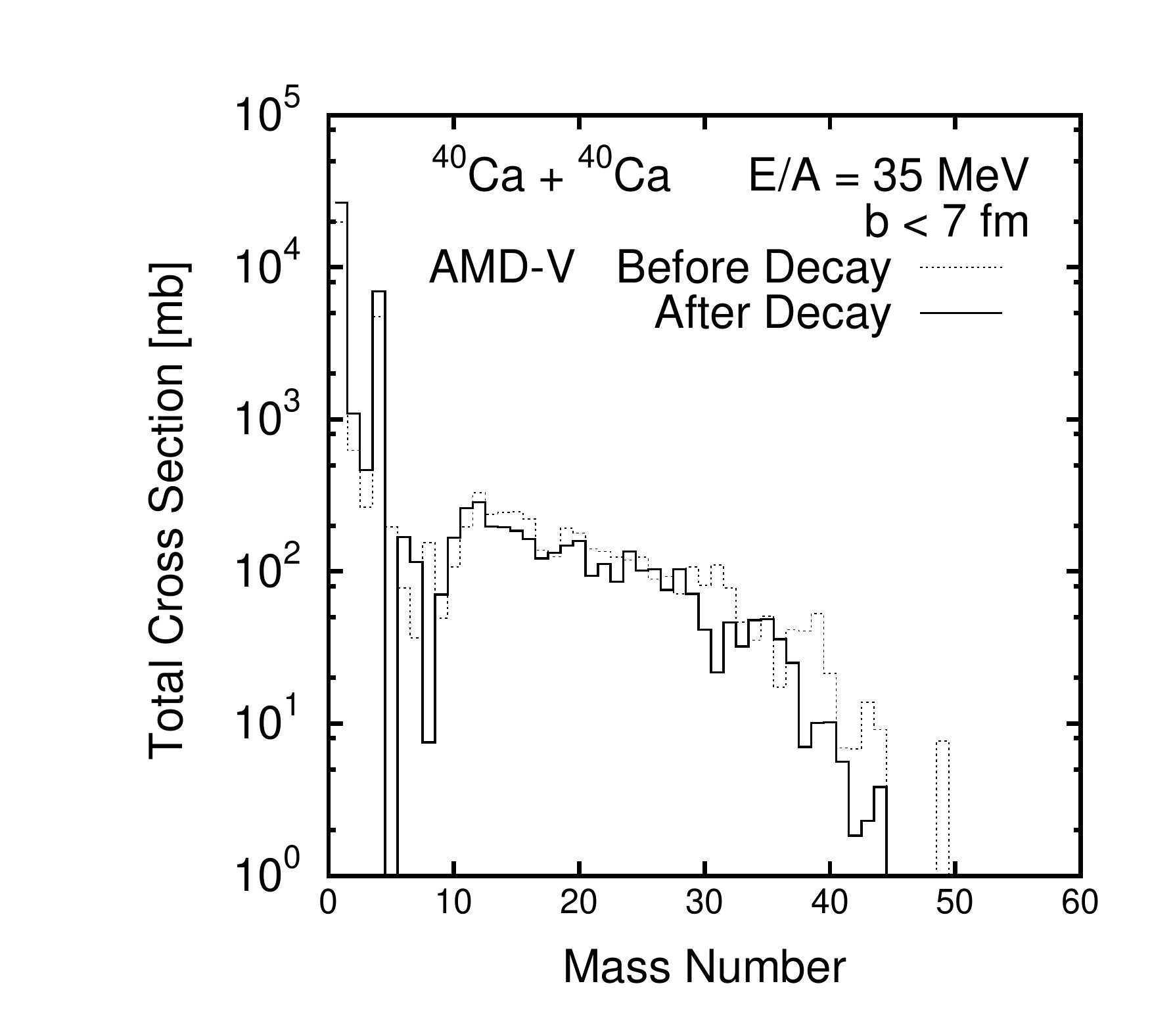}
\end{center}
\caption{\label{fig:CaCa-msdst} Mass distribution in
  ${}^{40}\mathrm{Ca}+{}^{40}\mathrm{Ca}$ collision at 35
  MeV/nucleon. Calculated results of AMD before and after the
  secondary decay are shown by dotted and solid histograms,
  respectively.  The left figure is the result by AMD without wave
  packet splitting, and the right figure is the result with the
  strongest decoherence $\tau\rightarrow0$.  The figure is taken from
  Ref.\ \citen{Ono:1996rk}.}
\end{figure}

For the multifragmentation in ${}^{40}\mathrm{Ca}+{}^{40}\mathrm{Ca}$
collisions at 35 MeV/nucleon, the wave packet splitting plays
important roles to enable the breakup of the system into small pieces.
Figure\ \ref{fig:CaCa-msdst} compares the results without wave packet
splitting and with the wave packet splitting in the limit of the small
coherence time $\tau\rightarrow0$ (i.e., the strongest decoherence).
The latter result is consistent with the experimental data.  Without
wave packet splitting, the two nuclei go through each other without
forming fragments of intermediate size.  The wave packet splitting
allows the mixing of the two nuclei or the neck formation, so that
more than two fragments can be formed from the system expanding in the
beam direction.  The $\alpha$-particle multiplicity also depends on
the wave packet splitting very much.

In short, the strength of the wave packet splitting (viewed from AMD)
or the decoherence of the single-particle states (viewed from the
mean-field theory) is a key ingredient for the description of
fragmentation.  For the stronger splitting, the system tends to expand
strongly and to break into small fragments and many $\alpha$
particles.  Unfortunately, the appropriate strength seems to depend on
the size of the system and/or the incident energy.  A more consistent
understanding may be possible if the cluster correlations in dynamical
systems are more explicitly treated.

\subsection{Statistical properties of excited systems}

One of the aims of the study of multifragmentation has been to extract
the information of excited nuclear matter in which the liquid-gas
phase transition is expected as in the system of the Van der Waals
equation of state.  The concept of phase transition in finite
many-body systems, as in the heavy-ion collision systems, has been
improved very much in recent studies \cite{Gross:1997,Gross:2002}.
Phase transition is clearly defined in finite systems by considering
microcanonical ensembles.

By solving the time evolution of a many-body system in a container of
a finite volume for a very long time, it is possible in principle to
generate a microcanonical ensemble for the given energy and volume.
However, it is a non-trivial question whether dynamical models such as
AMD can produce a correct statistical ensemble.  In fact, the
introduction of wave packet splitting into AMD was first motivated for
the purpose to get a proper statistical properties with the fermionic
caloric curve $E^*=aT^2$ at low temperature \cite{ONOf,ONOg}.  From a
different point of view, Ohnishi and Randrup also introduced
stochastic terms into molecular dynamics for quantum statistics
\cite{OhnishiNPA565}.  The caloric curves in the region of liquid gas
phase transition was calculated by Sugawa and Horiuchi by employing
AMD with an implementation of wave packet splitting
\cite{SugawaPRC60,SugawaPTP105}.  Fermionic molecular dynamics was
also applied to the caloric curves \cite{SchnackPLB409}.  The caloric
curves obtained in these studies were drawn under the condition of
fixed volumes or for a system under a confining potential.

\begin{figure}
\begin{center}
\includegraphics[width=0.5\textwidth]{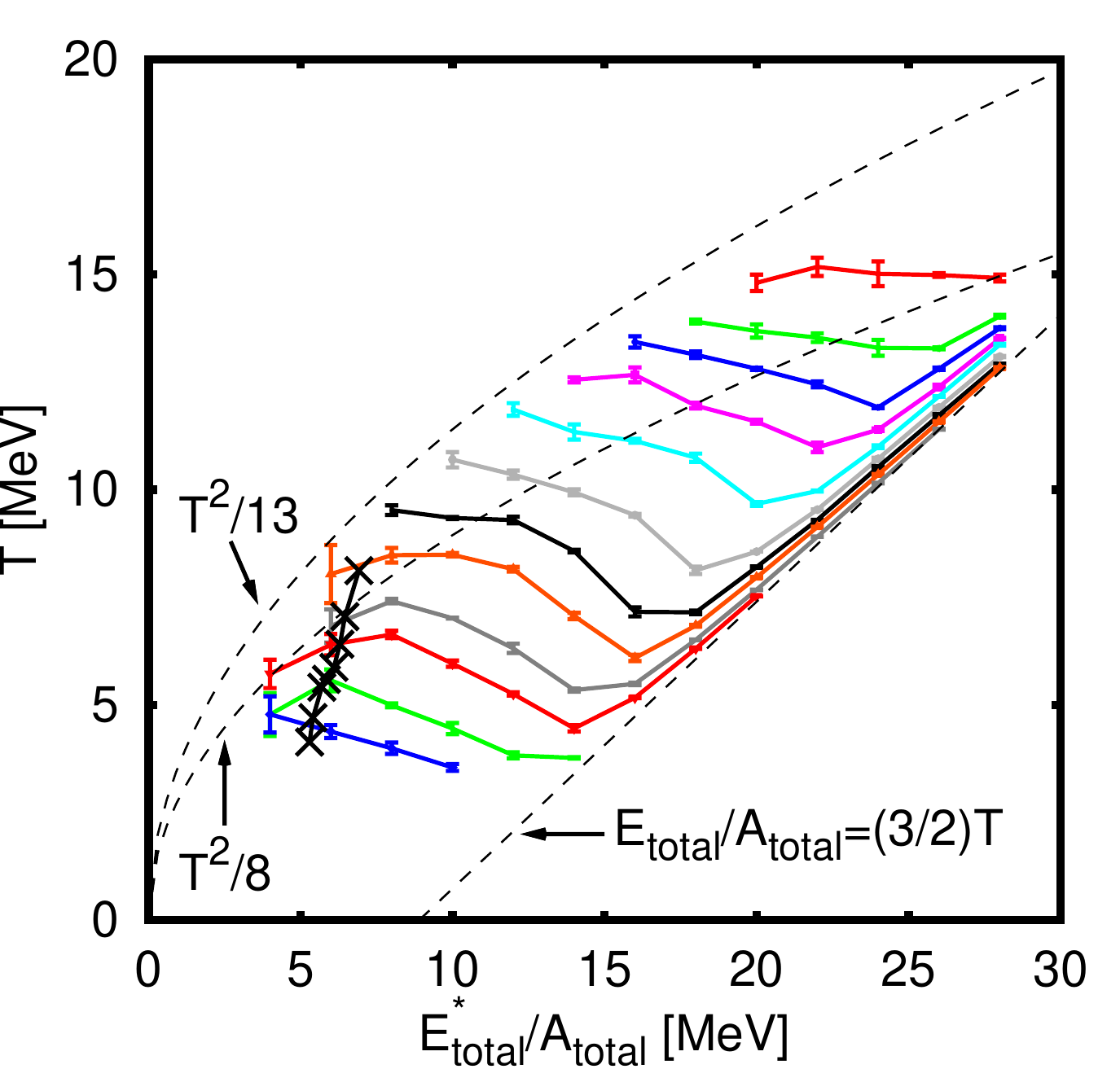}
\end{center}
\caption{\label{fig:pconst-path} Constant pressure caloric curves
  calculated with AMD for the $A=36$ system.  The cross symbols
  indicate the equilibrium state corresponding to the reaction system
  from $t=80$ to 300 fm/$c$ for
  ${}^{40}\mathrm{Ca}+{}^{40}\mathrm{Ca}$ central collisions.  The
  figure is taken from Ref.\ \citen{Furuta:2009}.}
\end{figure}

More recently, in Refs.\ \citen{Furuta:2006,Furuta:2009}, Furuta and Ono
performed AMD calculation to obtain constant-pressure caloric curves
in which phase transition should be clearly identified.  Wave packet
splitting was considered with a density-dependent coherence time
$\tau(\rho)$ in this study.  For a calculated microcanonical ensemble
of a given energy $E$ and a volume $V$, the temperature $T$ is defined
by using the kinetic energies of gas-like nucleons.  The pressure $P$
is obtained from the information of the reflections of particles at
the boundary of the container.  Then caloric curves are drawn as
$T(E)$ for different values of $P$.  The result for the system of
$N=18$ and $Z=18$ is shown in Fig. \ref{fig:pconst-path}.  Nuclear
liquid-gas phase transition, as a first order phase transition, is
clearly seen in this result as the back-bending of caloric curves.

\begin{figure}
\begin{center}
\includegraphics[width=0.75\textwidth]{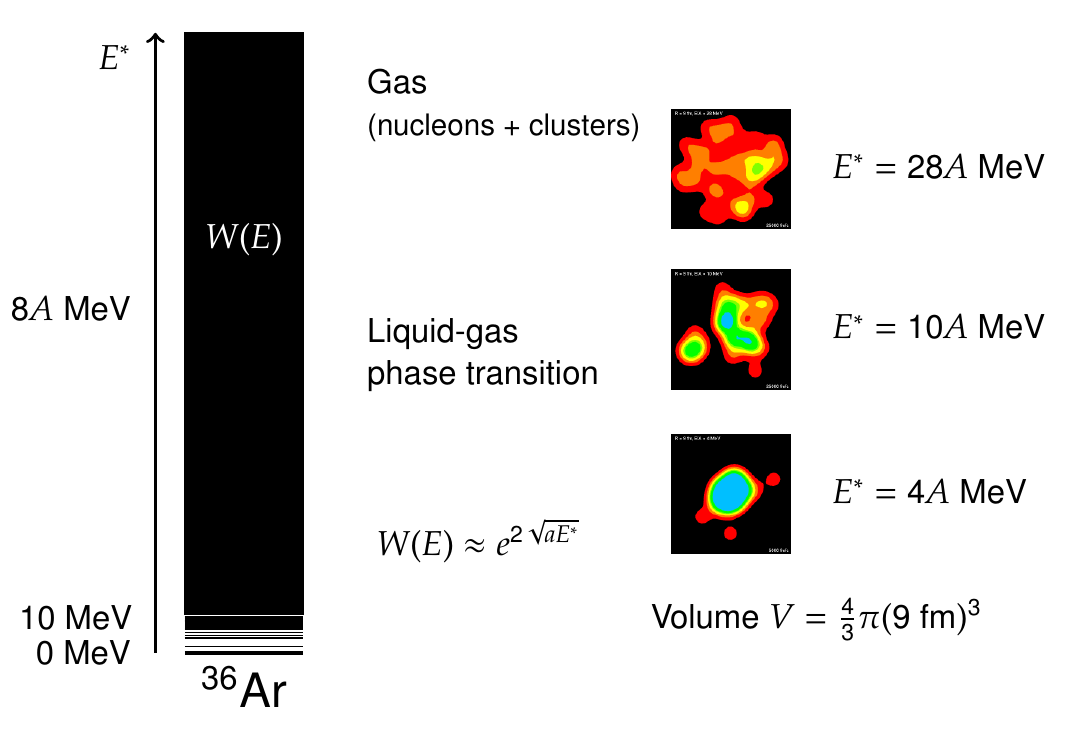}
\end{center}
\caption{\label{fig:excitednucleus} Excited states of a many-nucleon
  system with $N=Z=18$ confined in a virtual spherical container with
  a radius of 9 fm.  The density distribution at each excitation
  energy ($E^*/A=4, 10$ and 28 MeV) shows a snapshot taken from the
  AMD equilibrium calculation of Ref.\ \citen{Furuta:2006}.}
\end{figure}
The construction of microcanonical ensembles can also be regarded as a
way to explore the many-body states at various excitation energies, as
illustrated in Fig.\ \ref{fig:excitednucleus}.  Above the ground state
and low-lying excited states of the nucleus, there should be quite a
lot of states which are characterized by the density of states $W(E)$.
The density profile at each of the excitation energies $E^*/A=4$, 10
and 28 MeV is a sample taken from the calculated microcanonical
ensemble \cite{Furuta:2006}.  At low excitation energies, a single large
nucleus is usually observed.  The energy $E^*/A= 10$ MeV is in the
region of liquid-gas phase transition where $W(E)$ shows an anomalous
behavior and each density profile typically shows several nuclei being
related to multifragmentation.  It should be noted that the system
confined in a container does not become a gas of nucleons even though
the excitation energy is higher than the binding energy of the
nucleus.  When the energy is further raised to 28 MeV/nucleon, for
example, the state may be regarded as a gas state but the gas is
composed of clusters as well as nucleons.

Thus it is now a great advantage of AMD that it can describe both
dynamical reactions in heavy-ion collisions and virtually equilibrated
systems reasonably well.  In Ref.\ \citen{Furuta:2009}, we
investigated the question whether equilibrium is really relevant in
multifragmentation by comparing the details of reaction calculations
and equilibrium calculations performed by the same AMD model.  The
calculations show that there exists an equilibrium ensemble which well
reproduces the reaction ensemble at each reaction time $t$ for the
investigated period $80\leq t\leq300$ fm/$c$ in
${}^{40}\mathrm{Ca}+{}^{40}\mathrm{Ca}$ central collisions at 35
MeV/nucleon, as far as fragment observables (fragment yields and
excitation energies) are concerned.  Thus the corresponding
temperature and excitation energy (or the volume and pressure) can be
identified at each reaction time.  In Fig.\ \ref{fig:pconst-path}, the
path of the reaction form $t=80$ to 300 fm/$c$ is drawn by the cross
symbols on the caloric curve figure.  It is also important to note
that there are some other observables which show discrepancies between
the reaction and equilibrium ensembles \cite{Furuta:2009}.  These may
be interpreted as dynamical effects in the reaction. In particular,
the usual static equilibrium at each instant is not realized since any
equilibrium ensemble with the same volume as that of the reaction
system cannot reproduce the fragment observables.

\subsection{Symmetry energy effects in heavy-ion collisions}

The AMD simulations for heavy-ion collisions are useful not only to
explain the experimental data but also to know what kind of
information is reflected in the fragment formation.  In particular, as
demonstrated in the previous subsection, we may expect that
statistical properties such as the equation of state can be extracted
from the fragment observables even in dynamical collisions.

In heavy-ion collisions with unbalanced neutron and proton numbers,
the difference between the neutron and proton motions is an
interesting new degrees of freedom.  The difference of flow pattern
between neutrons and protons has been predicted to be sensitive to the
density dependence of symmetry energy.  From the viewpoint of
liquid-gas phase transition, neutron-rich systems are quite
interesting because new characters as two-component systems are
expected.  Namely, the gas part of the system is more neutron-rich
than the liquid part, which can be called fractionation or distillation.
This effect of isospin fractionation/distillation should be observable
in the neutron-to-proton ratio of produced fragments.

In Refs.\ \citen{ONOk,ONOl}, the fragment yields were analyzed in
the AMD simulations for multifragmentation reactions of the central
collisions of Ca isotopes at 35 MeV/nucleon, in order to see how the
fragment isospin composition is related to the symmetry energy term of
the effective interaction adopted in the calculation.  For the
fragment yields $Y_i(N,Z)$ in the reaction $i$ at $t=300$ fm/$c$, it
was found that the isoscaling relation
\begin{equation}
Y_j(N,Z)/Y_i(N,Z)\propto e^{\alpha_{ij}N+\beta_{ij}Z}
\end{equation}
is satisfied for any two reaction systems $i$ and $j$ which are
different in the proton-to-neutron ratios.  Isoscaling is expected
under an equilibrium assumption and has been observed in the
experimental data \cite{XU-isoscale}.  Isoscaling is equivalent to the
statement that the fragment yields are expressed as
\begin{equation}
Y_i(N,Z)=\exp[-K(N,Z)+\alpha_iN+\beta_iZ+\gamma_i]
\end{equation}
by using a function $K(N,Z)$ that is independent of the reaction
system $i$.  If the equilibrium is relevant to the reaction, $K(N,Z)$
should contain a term $(C_{\text{sym}}/T)(N-Z)^2/(N+Z)$, where
$C_{\text{sym}}$ is a kind of symmetry energy which may depend on $A$
or $Z$ in principle.  Based on this assumption, the isoscaling
parameter is related to the symmetry energy by
\begin{equation}
\alpha_{ij}=\frac{4C_{\text{sym}}}{T}\Bigl[
(Z/\bar{A}_i(Z))^2-(Z/\bar{A}_j(Z))^2\Bigr],
\end{equation}
where $Z/\bar{A}_i(Z)$ represents the mean isospin asymmetry of the
fragments for each given $Z$ in the reaction $i$.

\begin{figure}
\begin{center}
\includegraphics[width=0.48\textwidth]{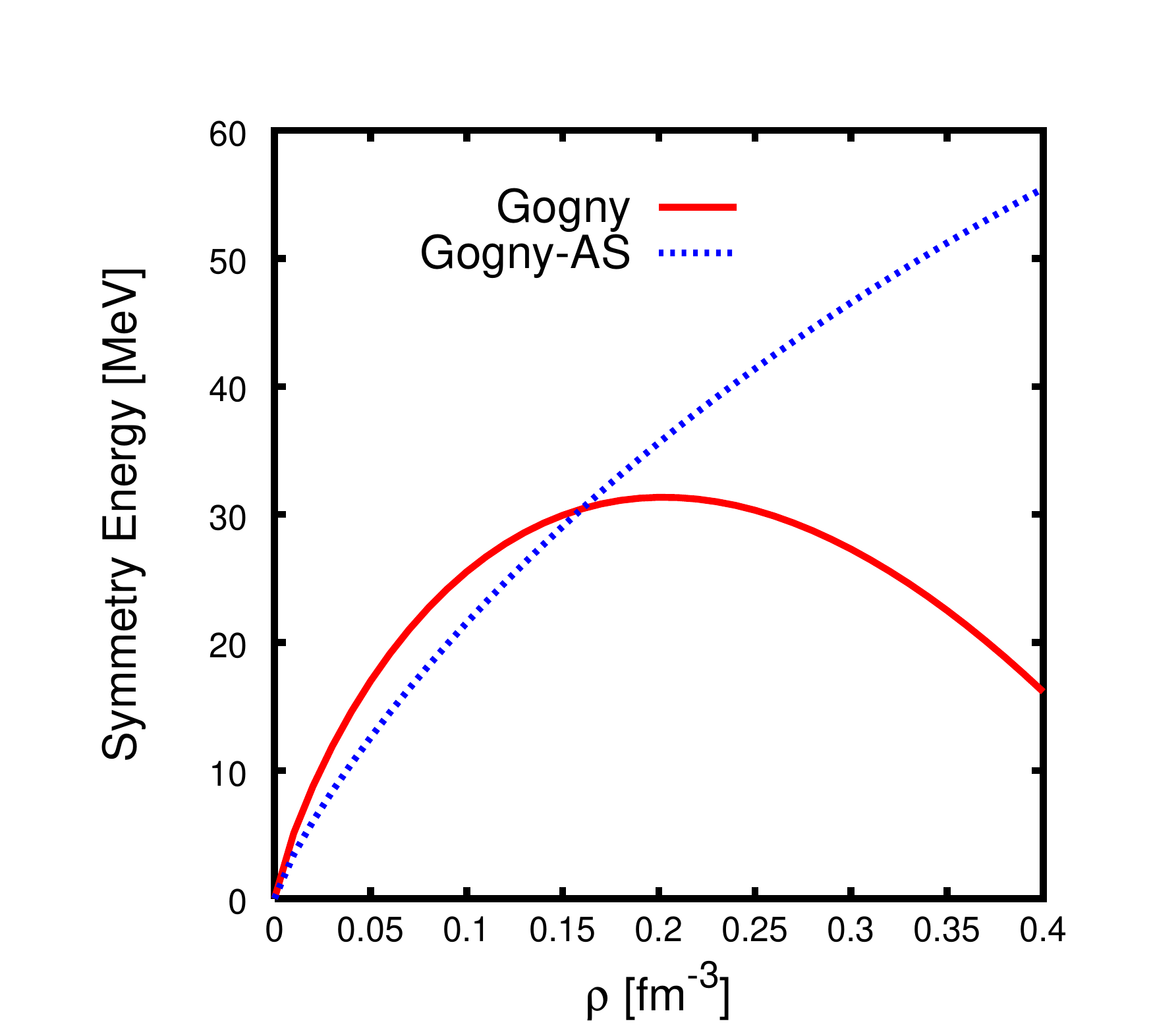}
\end{center}
\caption{\label{fig:symeng} Density dependence of the symmetry energy
of nuclear matter for the Gogny force (solid line) and for the
Gogny-AS force (dashed line).}
\end{figure}
\begin{figure}
\begin{center}
\includegraphics[width=0.47\textwidth]{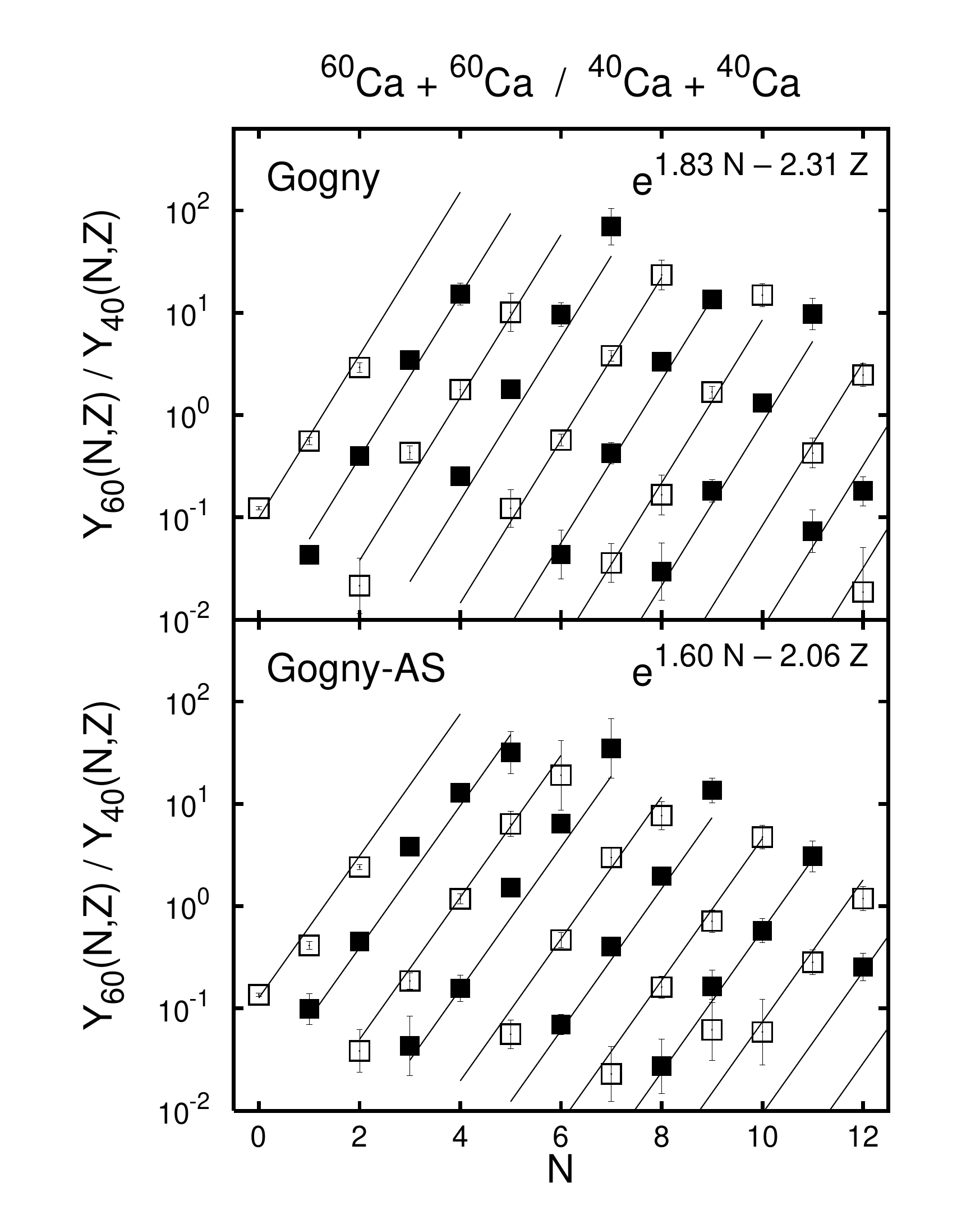}
\end{center}
\caption{\label{fig:CaCa-isoscale2-6040} The fragment yield ratio
  between the AMD simulations of central
  ${}^{60}\mathrm{Ca}+{}^{60}\mathrm{Ca}$ and
  ${}^{40}\mathrm{Ca}+{}^{40}\mathrm{Ca}$ collisions at 35
  MeV/nucleon, at time $t=300$ fm/$c$.  The top and bottom panels
  show, respectively, the results obtained using the Gogny and
  Gogny-AS forces.  The extracted isoscaling parameters are
  $\alpha=1.82\pm0.06$ and $\beta=-2.23\pm0.08$ for the Gogny force,
  and $\alpha=1.64\pm0.05$ and $\beta=-2.09\pm0.07$ for the Gogny-AS
  force.  The figure is taken from Ref.\ \citen{ONOk}.}
\end{figure}
AMD calculations were carried out with the Gogny force and the
Gogny-AS force which are different in the density dependence of the
symmetry energy as shown in Fig.\ \ref{fig:symeng}.  The result shows
that the isoscaling parameter $\alpha$ actually depends on the density
dependence of the symmetry energy as shown in Fig.\
\ref{fig:CaCa-isoscale2-6040} \cite{ONOk}.  Furthermore, the value of
$C_{\text{sym}}/T$ extracted from the simulation result is found to be
almost independent of the fragment size $A$ or $Z$ \cite{ONOl}, which
suggests that bulk properties are reflected in the fragment isotope
yields rather than the symmetry energy for the ground state binding
energies that depends on $A$ due to the surface effect.  To explain
the obtained result, $C_{\text{sym}}$ is identified with the bulk
symmetry energy at about $\frac{1}{2}\rho_0$, with $\rho_0$ being the
saturation density of nuclear matter, and the temperature should be
$T\approx3.4$ MeV.  These values of the density and the temperature
are reasonable as the condition for fragmentation.

It should be noted that the above analysis has been done for the
fragment yields at $t=300$ fm/$c$.  In order to compare with
experimental data, the effects of the decay of these primary fragments
should be carefully considered.

\section{Summary and perspective} \label{sec:summary}
In nuclear systems, cluster aspect is one of the essential features as well as mean-field aspect. 
Coexistence of cluster and mean-field aspects brings a variety of phenomena 
as functions of excitation energy and isospin degrees of freedom.
For usual theoretical models, it is not easy to describe  
both behaviors of independent single nucleons in a mean field and spatially 
correlating nucleons in clusters. 
The AMD method is one of the theoretical approaches that can describe
those two kinds of nature. In AMD, single-particle wave functions are
written by localized Gaussian wave packets whose dynamics expresses 
assembling and disassembling of nucleons. 
The method has been applied to investigate nuclear reactions and structures 
and it has been proved to successfully describe a variety of phenomena in general nuclei.

In this paper, we reviewed the AMD approach and its application to nuclear 
structure and reaction studies. To show applicability of AMD 
some topics studied with time-independent and time-dependent versions of AMD 
were explained. 
In the applications of time-independent AMD to nuclear structure studies, 
structures of neutron-rich nuclei such as Be, C, F, Ne, and Mg isotopes were described
focusing on cluster aspects. 
The results suggested a variety of structures appear in
unstable nuclei as well as in stable nuclei.
Deformation and cluster phenomena in $Z \sim N$  nuclei in $p$- and $sd$- shell regions
were also described.  
The applications of time-dependent AMD contain various topics such as 
fragmentation in heavy-ion collisions, nuclear
responses and virtual systems in thermal equilibrium. 
AMD calculations successfully described multifragmentation which is one of the remarkable phenomena 
in heavy-ion collisions. 
  Comparisons with predictions by other models and/or experimental
  data suggest the important balance between the single-particle
  motions and the many-body correlations to form clusters and fragments
  in these phenomena.
The AMD approach is suited to link the reaction observables to the
equilibrium properties of nuclear matter such as liquid-gas phase
transition and the equation of state of asymmetric nuclear matter.

Success of those studies using the AMD approach greatly owes to characteristics of the AMD model, for instance, 
advantages listed below.
\begin{itemize}
\item It is able to describe cluster and mean-field aspects without assuming 
existence of clusters nor mean fields.
\item It is applicable to both static and dynamical problems.
\item It it applicable to general nuclei with given proton and neutron numbers.
\item Center-of-mass motion can be exactly extracted.
\end{itemize}
In addition to the above advantages, the model can be easily extended 
because of flexibility of AMD wave functions.
For instance, parity and angular-momentum projections and superposition of wave functions are performed in structure calculations, 
and stochastic collisions are incorporated in calculations of heavy-ion collisions. 

Present structure studies with the AMD method cover light-mass regions of nuclear chart 
up to $pf$-shell nuclei. It is a future problem to apply the method to further heavy-mass regions
and progress systematic studies covering wide regions of the nuclear chart. 

Origins of cluster formation and breaking should be clarified from the point of view of 
nuclear force. 
Unfortunately, the present AMD framework is model calculation and 
it requires phenomenological effective nuclear interactions. 
To calculate nuclear systems 
based on realistic nuclear forces is one of the important issues in nuclear physics. 
In fact, there are many attempts of {\it ab initio} calculations though practical {\it ab initio} 
calculations are still limited in very light systems. 
For systematic study covering a wide mass number region, 
model calculations are efficient. 
To achieve model calculations starting from realistic nuclear forces 
a main problem is how to deal with complicated many-body correlations.
One of the promising methods is the unitary correlation operator method (UCOM) \cite{ucom98,ucom03,ucom10} 
recently proposed to incorporate efficiently 
short-range and tensor correlations in structure 
models such as FMD.
To understand nuclear structure from fundamental point of view,
more sophisticated versions of the AMD method would be required. 

Another issue of the AMD approach would be wide application to reaction phenomena. 
Although the AMD method has been extensively applied to
violent reactions above ten MeV/nucleon, 
its applications to lower-energy reaction phenomena are limited. 
Motivated by recent progress of experimental studies,
low-energy reactions such as resonances, fusion/capture, and transfer reactions are interesting problems 
to be solved concerning cluster phenomena in unstable nuclei.

\section*{Acknowledgments}
We would like to thank A. Dote, T. Furuta, N. Furutachi, H. Horiuchi, T. Suhara, and Y. Taniguchi, 
for fruitful discussions and collaborations. 
Parts of numerical calculations of this work were performed by using 
supercomputers at RCNP in Osaka University, those in High Energy Accelerator Research Organization, 
and those at YITP in Kyoto university.
This work was supported by Grant-in-Aid for Scientific Research from Japan Society for the 
Promotion of Science (No.~21540253, No.~22540275).


%

\end{document}